\documentclass[sigconf, nonacm, pdfa]{acmart}

\newcommand\vldbdoi{10.14778/3749646.3749666}
\newcommand\vldbpages{3951 - 3964}
\newcommand\vldbvolume{18}
\newcommand\vldbissue{11}
\newcommand\vldbyear{2025}
\newcommand\vldbauthors{\authors}
\newcommand\vldbtitle{\shorttitle} 
\newcommand\vldbavailabilityurl{https://github.com/mageirakos/crack-ivf-vldb}
\newcommand\vldbpagestyle{empty}

\usepackage{graphicx}
\usepackage{subcaption}
\usepackage{booktabs} 
\usepackage{placeins} 
\usepackage{afterpage} 
\usepackage{multirow}
\usepackage{enumitem}

\usepackage{pdfpages}  

\usepackage{algorithm}
\usepackage{algorithmic}  

\begin{document}

\title{Cracking Vector Search Indexes}

\author{Vasilis Mageirakos}
\affiliation{%
  \institution{Systems Group, ETH Zurich, Switzerland}
}
\email{vmageirakos@inf.ethz.ch}

\author{Bowen Wu}
\affiliation{%
  \institution{Systems Group, ETH Zurich, Switzerland}
}
\email{bowen.wu@inf.ethz.ch}

\author{Gustavo Alonso}
\affiliation{%
  \institution{Systems Group, ETH Zurich, Switzerland}
}
\email{alonso@inf.ethz.ch}

\newcommand{\note}[1]{\begingroup\color{red}\textbf{NOTE: } #1\endgroup}
\newcommand{\todo}[1]{\begingroup\color{red}\textbf{TODO: } #1\endgroup}

\definecolor{ETHc}{RGB}{18,105,176}
\newcommand{\new}[1]
{\begingroup\color{blue}\textbf{} #1\endgroup}



\begin{abstract}

Retrieval Augmented Generation (RAG) uses vector databases to expand the expertise of an LLM model without having to retrain it. The idea can be applied over data lakes, leading to the notion of embedding data lakes, i.e., a pool of vector databases ready to be used by RAGs. The key component in these systems is the indexes enabling Approximated Nearest Neighbor Search (ANNS). However, in data lakes, one cannot realistically expect to build indexes for every dataset. Thus, we propose an adaptive, partition-based index, CrackIVF, that performs much better than up-front index building. CrackIVF starts answering as a small index, and only expands to improve performance as it sees enough queries. It does so by progressively adapting the index to the query workload. That way, queries can be answered right away without having to build a full index first. After seeing enough queries, CrackIVF will produce an index comparable to those built with conventional techniques. CrackIVF can often answer more than 1 million queries before other approaches have even built the index, achieving 10-1000x faster initialization times. This makes it ideal for cold or infrequently used data and as a way to bootstrap access to unseen datasets. 

\end{abstract}

\maketitle

\pagestyle{\vldbpagestyle}
\begingroup\small\noindent\raggedright\textbf{PVLDB Reference Format:}\\
\vldbauthors. \vldbtitle. PVLDB, \vldbvolume(\vldbissue): \vldbpages, \vldbyear.\\
\href{https://doi.org/\vldbdoi}{doi:\vldbdoi}
\endgroup
\begingroup
\renewcommand\thefootnote{}\footnote{\noindent
This work is licensed under the Creative Commons BY-NC-ND 4.0 International License. Visit \url{https://creativecommons.org/licenses/by-nc-nd/4.0/} to view a copy of this license. For any use beyond those covered by this license, obtain permission by emailing \href{mailto:info@vldb.org}{info@vldb.org}. Copyright is held by the owner/author(s). Publication rights licensed to the VLDB Endowment. \\
\raggedright Proceedings of the VLDB Endowment, Vol. \vldbvolume, No. \vldbissue\ %
ISSN 2150-8097. \\
\href{https://doi.org/\vldbdoi}{doi:\vldbdoi} \\
}\addtocounter{footnote}{-1}\endgroup

\ifdefempty{\vldbavailabilityurl}{https://github.com/mageirakos/crack-ivf-vldb}{
\vspace{.3cm}
\begingroup\small\noindent\raggedright\textbf{PVLDB Artifact Availability:}\\
The source code, data, and/or other artifacts have been made available at \url{\vldbavailabilityurl}.
\endgroup
}

\section{Introduction}

Large language models (LLMs) \cite{brown2020language} can be complemented with retrieval-augmented generation (RAG) \cite{lewis2020retrieval}. In RAG, external data is represented as \textit{vector embeddings} (i.e., learned representations of data that results in a multidimensional vector). It is indexed and accessed through approximate nearest neighbor (ANN) search so that LLMs can answer queries on information they were not trained on. The effectiveness of RAG is tied to that of the ANN index structures, making them a critical component.

\begin{figure}[t]
    \centering
    \includegraphics[width=0.7\linewidth]{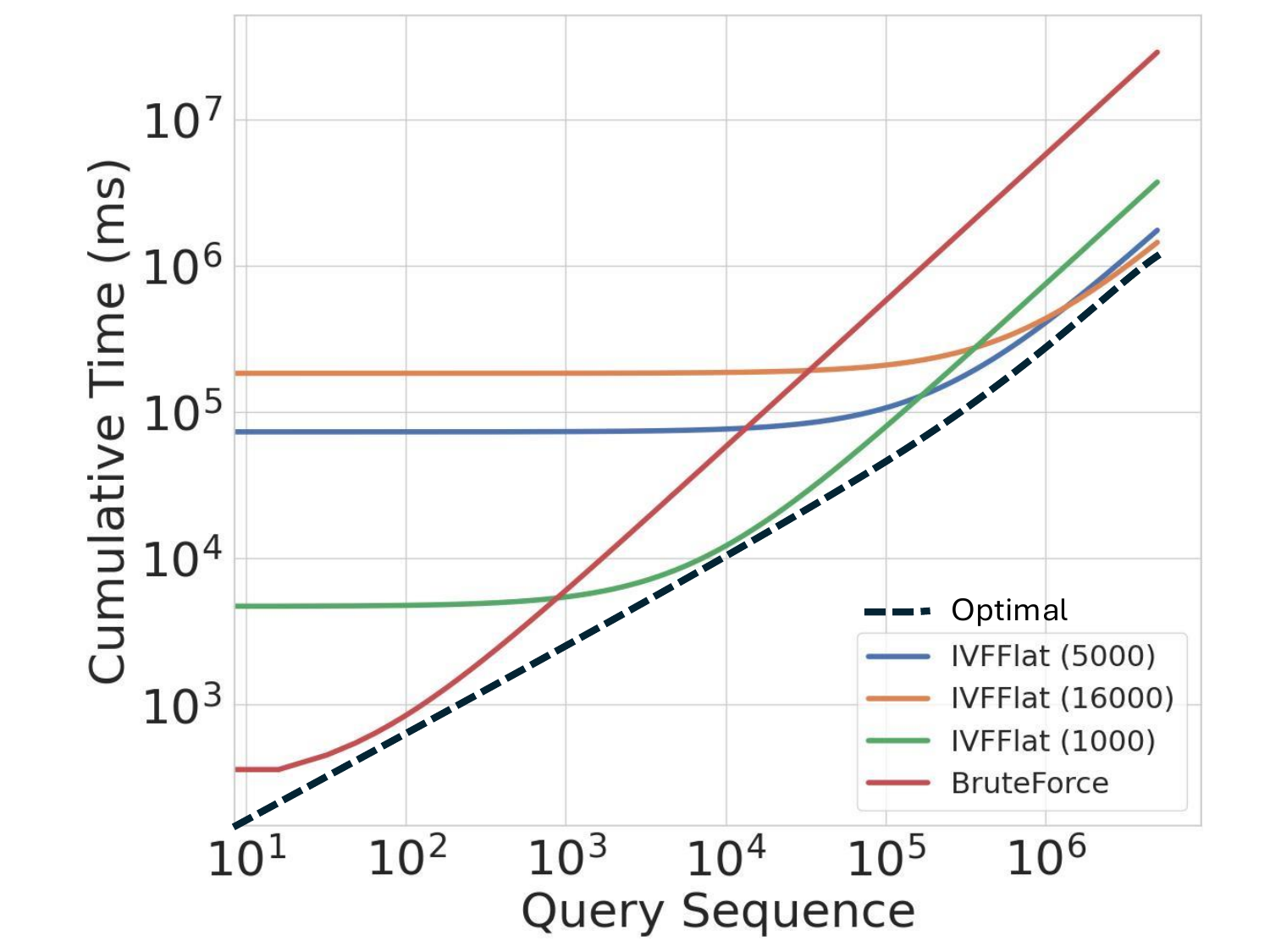}
    \caption{Total time to answer the queries submitted for different indexing strategies vs the number of queries submitted}
    \label{fig:optimal_index_is_workload_dependent_example}
\end{figure}

In practice, the vast majority of private, unstructured data remains untapped. It is estimated to be up to 80-90\% of all data \cite{MerrillLynch1998, Harbert2021Tapping}, and is often stored in data lakes on open data formats \cite{miller2018open}. As much as 70\% of it remains unused \cite{Gualtieri2016Hadoop}, classified as ``dark'' data \cite{heidorn2008shedding, zhang2016extracting}. 
While the datasets can be made discoverable \cite{chapman2020dataset}, the actual underlying data in them often remains unindexed. As of March 2020, Google Research’s Dataset Search \cite{brickley2019google} had indexed around 28 million structured and unstructured datasets \cite{benjelloun2020google} based on their metadata. Embedding-based vector search and RAG techniques could however help us go a step further by enabling direct retrieval and question answering on the underlying data itself. This requires embedding the data, as well as the creation of approximate nearest neighbor indexes so as to expose the data to RAG systems. We refer to such an approach as \textit{embedding data lakes} (EDL), where unstructured data is stored alongside its vector representations and is queried by a pool of vector databases. Efforts in this direction are already ongoing. Databricks uses vector search \cite{Databricks2024} over their Lakehouse \cite{Armbrust2021Lakehouse}, and new custom storage formats \cite{lance_format} are being used to store embeddings on data lakes. Researchers are also exploring RAG-based techniques to query multi-modal data lakes \cite{tangsymphony, chen2023symphony}, as well as answering queries over unstructured data \cite{liupalimpzest, anderson2024design}. 

\textit{The issue we address in this paper is \textbf{index selection at scale}}. Embedding data lakes, in principle, would require building millions of indexes across diverse datasets, modalities, workloads, and embedding models, which is neither feasible nor cost efficient. Some datasets may receive the majority of queries while others are rarely accessed. When selecting an index for embedding lakes, there is a trade-off. On the one hand, building an index requires significant upfront costs that might not pay off if the data is rarely queried. On the other hand, skipping the indexing process and relying on a brute-force search is not scalable. Hence, data must be indexed, which raises two questions: \textbf{when and how to index the data}. 

An optimal ANN index depends on the workload patterns and future usage, which are often unknown. As shown in Figure~\ref{fig:optimal_index_is_workload_dependent_example}, the index that minimizes total cumulative time spent on building and searching varies with the query volume. Larger indexes take longer to build and require many queries to justify their upfront cost (e.g., the large IVFlat configurations (5000, 16000 clusters) in the figure). Brute-force search, and smaller indexes (IVFlat 1000 clusters), minimize time-to-query for fast discovery but struggle at scale due to their higher query response times.

To tackle index selection at scale, we adopt a well-established strategy: carefully avoiding the need to build the full index upfront. We draw inspiration from deferred data structuring \cite{karp1988deferred} and database cracking \cite{idreos2007database,halim2012stochastic,schuhknecht2016experimental}. Similar approaches have been extended to multidimensional data \cite{holanda2018cracking, zardbani2023adaptive, pavlovic2018quasii, nerone2021multidimensional}. However, prior work does not scale to the embedding dimensionalities nor supports the k-Nearest Neighbour (k-NN) queries typical in RAG deployments. Recently, the AV-Tree was the first adaptive index for high-dimensional k-NN search \cite{lampropoulos2023adaptive}. AV-Tree targets short-lived data, with workloads of up to one thousand queries, and is a tree-based Exact Nearest Neighbor (ENN) adaptive index. This makes it impractical for RAG-scale workloads, which rely on Approximate Nearest Neighbor index structures to trade off a bit of accuracy for lower query response time and higher scalability. Nonetheless, the idea of dynamically building an index tailored to the workload is compelling. In this paper, we demonstrate how to do this efficiently in the context of RAGs by employing the same data structures \cite{sivic2003video} and systems \cite{douze2024faiss} used for ANN search workloads \cite{aumuller2020ann, simhadri2022results}.

We introduce \textbf{CrackIVF}, an incrementally built, partition-based ANN index that dynamically improves itself to adapt to increasing workload demands, minimizing both time-to-query and cumulative cost associated with suboptimal static index selection. CrackIVF evolves as a side effect of query execution. Each query is a candidate around which we can add a new partition, or crack. Likewise, its local region visited during search is a candidate for local refinement, allowing the index to grow and improve as more queries arrive. Efficiency is maintained through two control mechanisms: one deciding where to apply cracking and refinement in the search space, and another controlling when, balancing indexing and search times. In this work we focus on the static setting (i.e., no data or query distribution shifts) and evaluate CrackIVF on multiple standard open source datasets for ANN search \cite{Bertin-Mahieux2011,pennington2014glove,jegou2011searching,babenko2016efficient}. 

For partition-based indices, index selection means deciding upfront how many partitions the index will use, which has an impact on the ANN search performance. More partitions increase the upfront construction cost but improve query performance. This decision has to be made without knowing the future query workload. On the other hand, CrackIVF waits to build the index as it sees enough queries and decides on the clusters based on the query distribution over the search space. As more queries arrive, the clusters are redefined, and their number is increased to adapt to the increasing query workload. CrackIVF is an incrementally built index that asymptotically matches or surpasses the performance of pre-built indexes, while it has been able to answer queries along the way.

Across benchmarks, CrackIVF consistently outperforms pre-built partition-based indexes. It converges to near-optimal query response time, yet achieves several orders of magnitude lower startup time than other indexes. It efficiently scales to large workloads and, in some cases, can process 1 million queries before the baseline indexes have finished building. It is the only index that consistently remains near the Pareto frontier of minimum cumulative time across the number of queries posed to the system.

\section{Background}
\label{sec:background_related_work}
\textbf{Nearest Neighbor Search}
\label{sec:nn_search}
In the \emph{k-nearest neighbor} (\emph{k-NN}) search problem, we are given a set of points 
$P \in \mathbb{R}^{|P|\times d}$ and a query $q \in \mathbb{R}^d$. The aim is to find the $k$ 
points in $P$ that are closest to $q$ under some notion of distance or similarity (e.g., minimize the Euclidean distance (L2) or maximize the inner product (IP)). A straightforward Exact Nearest Neighbor approach is a linear scan of every point, a.k.a \textit{brute-force search}, incurring $O(|P|\cdot d)$ complexity. There exist more sophisticated \emph{exact indexing} structures (e.g., KD-trees \cite{bentley1975multidimensional}, R-trees \cite{guttman1984r}). However, due to the curse of dimensionality \cite{DBLP:conf/vldb/WeberSB98}, performance degrades to near-linear-scan in higher dimensions, making them impractical when $|P|$ or $d$ are large. With embeddings of hundreds to thousands of dimensions ~\cite{openai_embedding2023} and real-world datasets at the scale of billions of data points~\cite{simhadri2022results}, RAG systems turn to Approximate Nearest Neighbor search techniques. They allow a small retrieval error $\varepsilon$ in exchange for better runtime or memory usage. The approximation quality can be evaluated by the fraction of exact top-$k$ vectors recovered.

\begin{figure*}[t]
    \centering
    \begin{subfigure}[t]{0.24\linewidth}
        \centering
        \includegraphics[width=\linewidth]{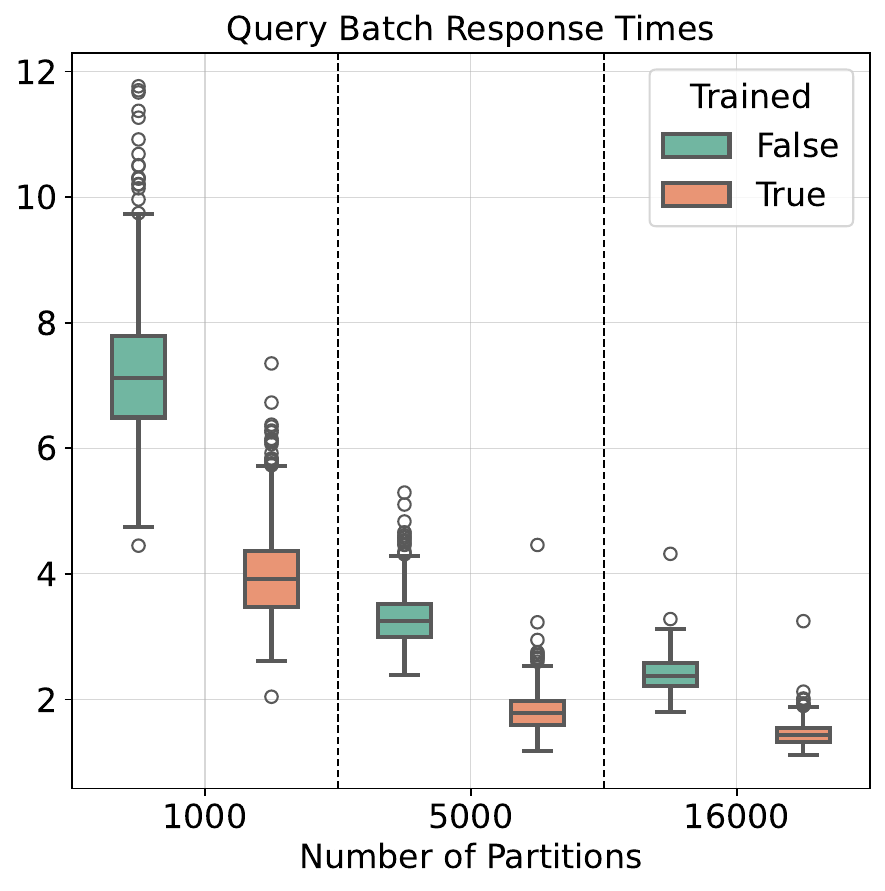}
        \caption{Response Time with increasing partitions and training K-Means.}
        \label{fig:fig1a_response_time}
        \label{fig:improving_response_time_example}
    \end{subfigure}
    \hfill
    \begin{subfigure}[t]{0.38\linewidth}
        \centering
        \includegraphics[width=1\linewidth]{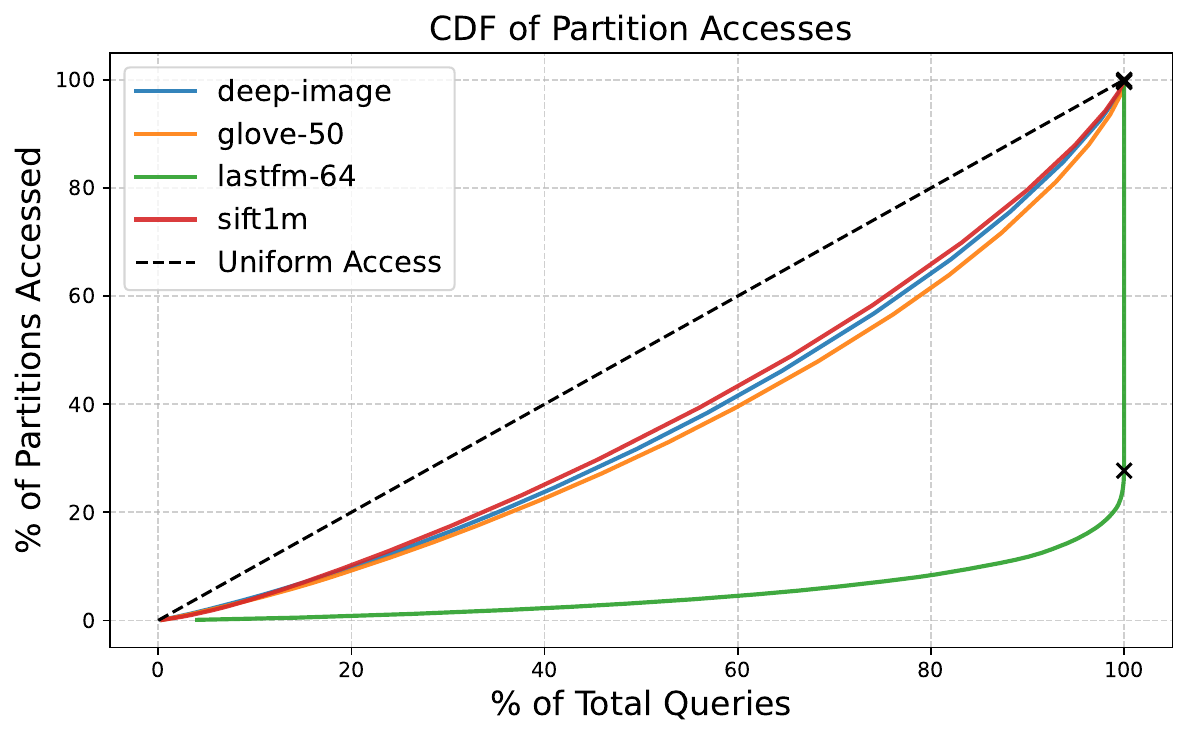}
        \caption{Queries exhibit a skewed access pattern over the search space that is often far from uniform.}
        \label{fig:plot_skew_cdf}
        \label{fig:all_datasets_have_skew}
    \end{subfigure}
    \hfill
    \begin{subfigure}[t]{0.24\linewidth}
        \centering
        \includegraphics[width=\linewidth]{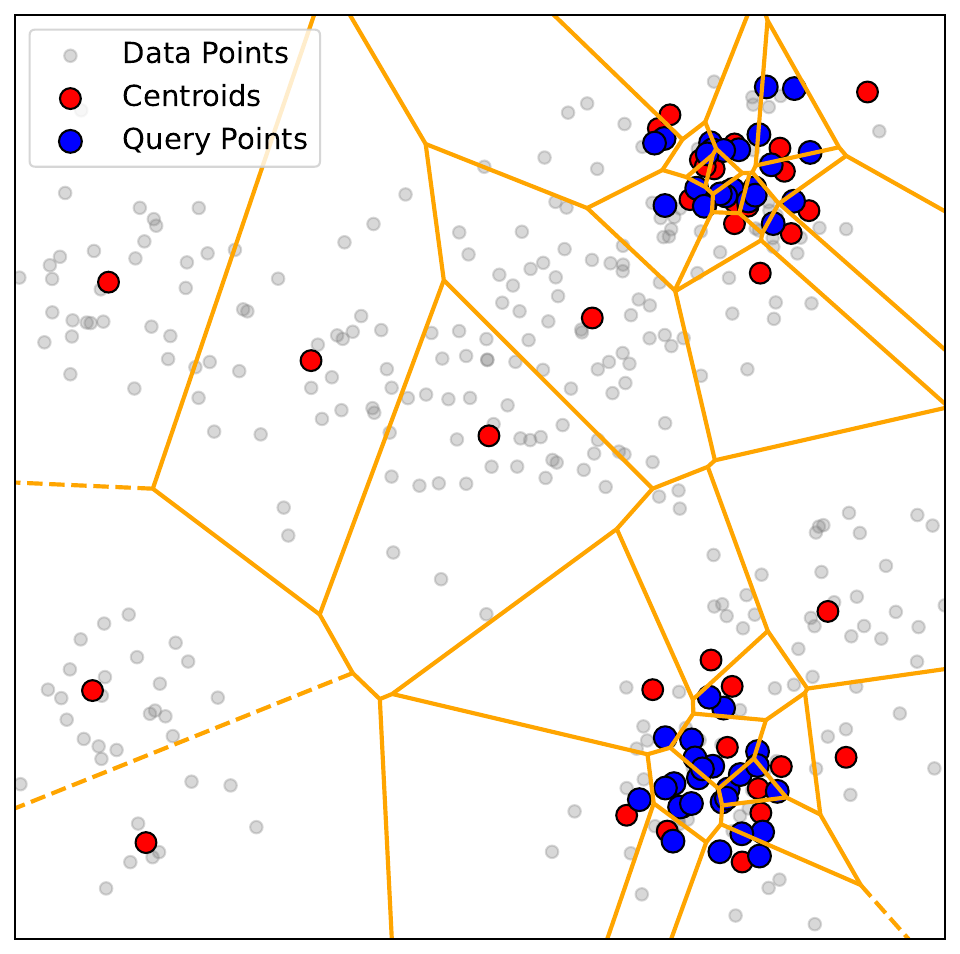}
        \caption{Finer partitioning of the Vector Space based on queries received.}
        \label{fig:cracking_on_queries_voronoi_cell_example}
    \end{subfigure}
    \caption{Observations behind the design: (a) separating the number and refinement of partitions; (b) access to the index is generally skewed towards certain regions; (c) regions queried often can be clustered and refined at a much lower granularity    
    }
    \label{fig:approach_intuition}
\end{figure*}

\textbf{Partition based IVF methods}
\label{sec:partition-based-ann}
One of the most common ANN search structures is \emph{partition-based IVF indexes} \cite{sivic2003video, chen2021spann, sun2024soar}. Instead of scanning all points in $P \subset \mathbb{R}^d$, they divide the vector space into $n_{\text{list}}$ disjoint groups, or \emph{partitions}. At query time, the search process first identifies the $n_{\text{probe}} \leq n_{\text{list}}$ nearest partitions to $q$, whose representative vectors are closest (or most similar) to the query. Once these partitions are selected, the points within them are exhaustively scanned. Since the remaining partitions are skipped,  some nearest neighbors may be missed, which is why the result is an \emph{approximate} solution.  A partition-based index is built by first selecting $n_{\text{list}}$ \emph{representative vectors} for each partition, typically using $k$-means on a training subset $P_{\text{train}} \subset P$. Each data point is then assigned to the nearest representative. These assignments are typically stored in an \emph{inverted file} (IVF) structure \cite{sivic2003video}. A widely adopted implementation of IVF indexes is the FAISS library from Meta \cite{douze2024faiss}, which we built CrackIVF on and use as a baseline. See Section~\ref{sec:related_work} for other ANN index types.

\section{Solution Overview}
\label{sec:solution_overview}

Embedding lake deployments involve constructing ANNS indexes for each data set of interest. The goal is to minimize index construction and memory overhead while also ensuring high queries per second (QPS). Although other index types excel in specific areas, IVF performs well in all three. For example, HNSW \cite{malkov2018efficient}, a graph-based index, can achieve higher QPS, but at the cost of much higher memory. For both IVF and HSNW, the memory-QPS trade-offs, Figure 5 in \cite{aumuller2020ann}, ``perform nearly indistinguishably'' and better than alternatives, but as mentioned IVF is a much smaller index. The build times of IVF indexes have been shown to be 100$\times$ lower than any other index and 8000$\times$ lower than HNSW (Figure 10 in \cite{aumuller2020ann}). Thus, looking at indexing speed versus QPS trade-offs \cite{soar_scann_blogpost}, graph-based methods struggle, where IVFs are the best (faiss-ivfpqfs\cite{douze2024faiss}, and ScaNN\cite{guo2020accelerating}).  Focusing on a single index type doesn't solve the index selection problem. While IVF indexes reduce compute and storage costs, they are still paid upfront before the query workload is known. We show how to extend IVF to enable incremental construction, scaling performance based on the query volume, and amortizing build and storage costs across time.

\subsection{Cracking in high dimensions}

An approach to avoid the upfront cost for indexing is \textit{cracking}. In the original database cracking \cite{idreos2007database}, the final state is where a single column is sorted. Over time, fewer points are physically reorganized reducing the overhead. In high-dimensional spaces, the having a final sorted state no longer holds, and space filling curves, like Z-order \cite{samet1984quadtree}, do not scale past a few dimensions. Additionally, in a naive implementation, the physical reorganization happens with every incoming query. In high-dimensions, this becomes increasingly difficult. High-dimensional vectors require moving multiple bytes per dimension, increasing data movement costs. In IVF, moving points between inverted lists incurs random access overhead, with no eventual locality benefits. Furthermore, inverted lists are often implemented as memory-aligned contiguous arrays to allow for efficient SIMD-based operations, and they may require linear-time compaction even for a single vector to move. These factors contribute to the latency overhead, making physical reorganization after every query difficult. The challenge is as follows: A cracking-based IVF index must efficiently manage the significantly higher overhead of physical reorganization in high dimensions, while minimizing time-to-query, continuously improving the index, and ensuring minimum cumulative time across all scales.

\subsection{Outline of the approach }
Index construction can be divided into two operations that independently improve index performance, \emph{i}) increasing the number of partitions and \emph{ii}) refining partitions by using K-means, which helps to distribute points more evenly (Figure~\ref{fig:fig1a_response_time}). The best results consistently come after reaching a maximum index size, even with randomly chosen representatives, followed by K-means training to optimize their placement. Thus, the optimal approach is to first increase the number of partitions up to a limit and then also refine their locations using K-means.

\textbf{Observation 1:} \textit{Index construction can be decoupled into two distinct build operations, which can improve the index independently and be \textbf{separated in time}}.

Existing partition-based IVF vector search indexes apply build operations evenly across the vector space, with partitions sampled uniformly and K-means refinements performed globally. While simple and intuitive, this approach assumes equal importance across the space, leading to larger than necessary index build times. However, vector search workloads can exhibit highly skewed access patterns, where certain regions are significantly more important than others. For instance, in an industrial workload~\cite{mohoney2023high, mohoney2024incremental}, up to 85\% of partitions are never accessed during search, which would make partitioning or training efforts in those regions unnecessary. As we show in Figure~\ref{fig:plot_skew_cdf}, skewed query distributions also appear in well known open-source vector search datasets \cite{Bertin-Mahieux2011,pennington2014glove,jegou2011searching,babenko2016efficient}. Even on open source datasets showcasing moderate amounts of skew, where $80\%$ of the total queries access $60\%$ of the partitions, the disparity between the most frequently and least frequently accessed partitions is upwards of 50 times. There is a fundamental inefficiency in current uniform indexing approaches. A large portion of the vector space may rarely accessed, yet the indexing effort is evenly distributed across the space, and always paid upfront. 

\textbf{Observation 2:} \textit{Just as index construction can be separated in time, it can also be \textbf{separated in space}. Index build operations do not have to be applied uniformly across the vector space. Instead, they can be local to regions that are accessed. Adaptive index construction that follows the query distribution is one strategy to optimize for efficiency.}

In partition-based IVF indexes,  operations performed during search and those performed during index construction are very similar. During search the distances of the query to centroids and nearby points are computed, and during index construction the distances of points to centroids are computed. This means that by treating the query as a potential new partition representative ``centroid'', we can both identify good locations to add new cracks, and reuse the intermediate computations of search to perform incremental index construction. This naturally leads to a cracking-based approach for IVF, where index construction can occur as a side effect of query execution. 

\textbf{Observation 3:} \textit{We can further amortize the cost of specific build operations, such as assigning vectors to partitions and identifying the local regions to crack and refine, Figure~\ref{fig:cracking_on_queries_voronoi_cell_example}. We can \textbf{reuse computations from search operations} to perform index construction as a side effect of query execution.
}

\subsection{Outline of the solution}

We propose a partition-based crackable IVF index for ANN search. It is designed to remain near the minimum cumulative time across the number of queries submitted, be constructed as the query load increases, and avoid indexing unused parts of the vector space.

At initialization, CrackIVF starts as a small IVF index, with coarse partitioning of the entire dataset. As queries come in, more partitions are added in the accessed regions related to the queries (Figure~\ref{fig:cracking_on_queries_voronoi_cell_example}). The index employs two localized and independent build operations: \textit{CRACK} and \textit{REFINE}. The former introduces new partitions, while the latter applies a localized K-means to refine them. These operations run independently across both time and region of the vector space. Queries and their local visited regions act as \textit{CRACK} and \textit{REFINE} candidates, determining where new partitions should be introduced. Using the queries to create partitions, rather than selecting random ones, allows the distances computed during search to be used for stealing points from the local region, that are closer to the query than their current assignment. Thus, intermediate results from search operations are reused to amortize the cost of construction. Additionally, using the queries as candidates for \textit{CRACK} and \textit{REFINE} enables CrackIVF to follow the query distribution in terms of when and where to execute build operations. Partitions that receive a larger number of queries will be proportionally more likely to be selected for cracking and refinement.

Finally, to mitigate the high overhead of storage reorganization in high-dimensional IVF indexes, we do not perform build operations after every query. Instead, \textit{CRACK} operations are buffered and committed only when it is beneficial, while \textit{REFINE} operations are executed eagerly, but infrequently, to improve local regions that queries are visiting. We introduce two control mechanisms to make these decisions. The first is a set of heuristic rules that determine the decision boundaries for where in the vector space cracks should be buffered or refinements executed. The second is a cost-based approach that estimates the latency overhead of a build operation and constrains it relative to the cumulative runtime of the index since initialization. As more queries arrive and the index is used more frequently, the budget for build operations increases proportionally, allowing the index to improve as the query workload demands increase.

\begin{table}[t]
\caption{Notation used in algorithms and cost model.}
\label{tab:notation}
\footnotesize
\centering
\begin{tabular}{p{0.34\columnwidth} p{0.55\columnwidth}}
\hline
\textbf{Symbol} & \textbf{Description} \\
\hline
$|X|$ & Cardinality, e.g., $|X| = N$ if $X \in \mathbb{R}^{N \times d}$ \\
$metric(a, b)$ & Similarity metric, where in our work \\
& $metric \in \{\text{Eucl. Distance (L2)}, \text{Inner Product (IP)}\}$ \\
$P \in \mathbb{R}^{|P| \times d}$ & Data points, $d$-dimensional \\
$C \in \mathbb{R}^{|C| \times d}$ & Committed cracks \\
$C_{\text{local}} \in \mathbb{Z}^{nprobe}$ & Local region crack IDs  \\
$P_{\text{local}}$ & Local region points  \\
$P_{\text{train}}$ & K-means training points, where $|P_{\text{train}}|=\min(|P_{\text{local}}|, |C_{\text{local}}| \times \max_{\text{points}})$ \\
$n_{\text{iter}}$ & K-means iterations \\
$\max_{\text{points}}$ & K-means max training sample per crack \\
$nlist$ & Total inverted lists storing $P$, where $nlist=|C|$ \\
\hline
$Q \in \mathbb{R}^{b_s \times d}$ & Query set per search \\
$D \in \mathbb{R}^{b_s \times |r|},\, I \in \mathbb{Z}^{b_s \times |r|}$ & kNN distances and point indices from \textit{SEARCH}, where $|r| \in [k, |P_{\text{visited}}|]$ \\
$bs$ & Total queries batched in single \textit{SEARCH} \\
$nprobe$ & Number of partitions scanned per \textit{SEARCH} \\
$k$ & Number of nearest neighbors to return \\
$\alpha$ & Ratio of total build time to total time \\
$A_P^\ast \in \mathbb{Z}^{|P|}$ & Point assignments to crack IDs \\
$D_P^\ast \in \mathbb{R}^{|P|}$ & Point distances to crack representative \\
$H_C^\ast \in \mathbb{N}^{|C|}$ & Histogram of points per crack \\
\multicolumn{2}{p{0.95\columnwidth}}{\textbf{Note:} $\ast \in \{\text{dyn}, \text{true}\}$, where \textit{dyn} refers to the dynamic state (includes buffered cracks) and \textit{true} refers to the true index state.} \\
\hline
\end{tabular}
\end{table}

\section{Crack-IVF}
\label{sec:crack_ivf}

CrackIVF, is implemented in FAISS~\cite{douze2024faiss}. We describe the core logic of the algorithm and operations in Section~\ref{subsec:algorithm-operations} and Algorithm \ref{alg:search_and_crack}; the control mechanisms, including heuristic rules and cost models, in Sections \ref{subsec:where-heuristics} and \ref{subsec:when-to-operate}. These aid in making accurate decisions about where and when to perform the build operations, \textit{CRACK} and \textit{REFINE} (Algorithms \ref{alg:crack} and \ref{alg:refine}). The notation used is in Table \ref{tab:notation}.

\subsection{Algorithm and Operations}
\label{subsec:algorithm-operations}

CrackIVF performs index construction and storage reorganization operations as a side-effect of query execution to incrementally improve the index. It has three core operations, $SEARCH$ finds the k-NN results to return to the user, while $CRACK$ and $REFINE$ operations reorganize the IVF structure and change the index state. 

\textbf{\textit{SEARCH}}: During the search procedure of IVF, the local region to the query is identified by finding the $nprobe$ nearest partitions to the query. Subsequently, the point assignments to these local partitions are scanned to find the $k$ nearest neighbors. The result includes the nearest point ids $I$, their distances $D$ to the query, and the $C_{visited}$ ids of the local region cracks that the query visited.

\begin{algorithm}[t]
\footnotesize
\caption{\textsc{SearchAndCrack}}
\label{alg:search_and_crack}
\begin{algorithmic}[1]
\REQUIRE $k, Q, \alpha , State_{dyn}, State_{true}, T_{buld}, T_{search} $
\STATE \textbf{\texttt{\textbackslash\textbackslash} Step 1: SEARCH}
\STATE $T_{\text{start\_search}} \gets \textsc{CurrentTime}()$
\STATE $(D, I, C_{visited}) \gets \text{\textit{SEARCH}}(\textit{Q}, k)$
\STATE $T_{\text{search}} \gets T_{\text{search}} + (\textsc{CurrentTime}() - T_{\text{start\_search}})$

\STATE \textbf{\texttt{\textbackslash\textbackslash} Step 2: CRACK Decision (Where to crack, see \S\ref{subsec:where-heuristics})}
\FOR{each query \( q \in \textit{Q} \)}
    \STATE $(D_\text{local}, I_\text{local}, C_\text{local}) \gets (D[q], I[q], C_{visited}[q])$ 
    \STATE $I_\text{steal} \gets \{ p \in I_\text{local} \mid D_\text{local}(p) \mathrel{\text{op}} D_P^\text{dyn}(p) \}$ \COMMENT{op = $<$ for L2, $>$ for IP}
    \STATE $good\_crack\_candidate$ = CrackHeuristic($I_{steal},   \text{State}_{\text{dyn}}$)

    \IF{$good\_crack\_candidate$} 
        \STATE $C_{\text{buffered}} \gets C_{\text{buffered}} \cup \{q\}$
        \STATE $I_{\text{buffered}} \gets I_{\text{buffered}} \cup I_{\text{steal}}$ 
        
        \STATE \texttt{\textbackslash\textbackslash}  update dynamic state
    \ENDIF
\ENDFOR

\STATE \textbf{\texttt{\textbackslash\textbackslash} Step 3: CRACK Decision (When to Commit, see \S\ref{subsec:when-to-operate})}
\STATE $\hat{T}_{\text{CRACK}} \gets \textsc{EstimateCrackCost}()$ 
\STATE $can\_afford\_crack \gets (T_{\text{build}} + \hat{T}_{\text{CRACK}} \leq \alpha \cdot (T_{\text{build}} + T_{\text{search}} + \hat{T}_{\text{CRACK}}))$
\STATE $enough\_buffered \gets (|C_{\text{buffered}}| > C_{\min})$

\IF{$can\_afford\_crack$ \textbf{and} $enough\_buffered$}
    \STATE \texttt{\textbackslash\textbackslash} UNDO if \textit{REFINE} invalidated previously good cracks
    \STATE $C_{\text{buffered}} \gets \{ c \in C_{\text{buffered}} \mid \textsc{CrackHeuristic}(c) \}$
    \STATE $\forall p \text{ where } A_P^\text{dyn}(p) \notin C_{\text{buffered}}, A_P^\text{dyn}(p) \gets A_P^\text{true}(p)$
    \STATE $I_{\text{buffered}} \gets \{ p \mid A_P^\text{dyn}(p) \neq A_P^\text{true}(p) \}$
    \STATE $T_{\text{start\_crack}} \gets \textsc{CurrentTime}()$ 
    \STATE \textsc{CRACK}$(C_{\text{buffered}},I_{\text{buffered}})$ \COMMENT{Algorithm~\ref{alg:crack}}
    \STATE $T_{\text{build}} \gets T_{\text{build}} + (\textsc{CurrentTime}() - T_{\text{start\_crack}})$ 
    \STATE \texttt{\textbackslash\textbackslash} dynamic index state now matches true state
\ELSE
    \STATE \textbf{\texttt{\textbackslash\textbackslash} Step 4: REFINE Decision (Where \& When, see \S\ref{subsec:where-heuristics} and \S\ref{subsec:when-to-operate})}

    \FOR{each query \( q \in \textit{Q} \)}
        \STATE $\hat{T}_{\text{REFINE}} \gets \textsc{EstimateRefineCost}()$
        \STATE $can\_afford\_refine \gets (T_{\text{build}} + \hat{T}_{\text{REFINE}} \leq \alpha \cdot (T_{\text{build}} + T_{\text{search}} + \hat{T}_{\text{REFINE}}))$

        \IF{$can\_afford\_refine$}
            \STATE $good\_refine\_candidate \gets \textsc{RefineHeuristic}(C_{\text{local}}, \text{State}_{\text{true}})$
            \IF{$good\_refine\_candidate$}
                \STATE $T_{\text{start\_refine}} \gets \textsc{CurrentTime}()$  \COMMENT{Start timing REFINE}
                \STATE \textsc{REFINE}$(\dots)$
                \STATE $T_{\text{build}} \gets T_{\text{build}} + (\textsc{CurrentTime}() - T_{\text{start\_refine}})$ 
                \STATE \texttt{\textbackslash\textbackslash} update dynamic state and true state
            \ENDIF
        \ENDIF
    \ENDFOR
\ENDIF

\RETURN \( D[:, :k], I[:, :k] \)
\end{algorithmic}
\end{algorithm}

\textbf{\textit{CRACK}}: Physical reorganization of points between inverted lists, and the mutation of the index does not happen after each query. Each incoming query represents a crack candidate, which is evaluated by rules that define the decision boundaries which classify it as good or bad (Section~\ref{subsec:where-heuristics}) (Line 9 Algorithm~\ref{alg:search_and_crack}). Only good cracks are further buffered and potentially committed at a later time. Each crack is essentially a new partition, where the points assigned to the partition have been ``stolen'' based on their proximity to the query. We reuse the results of the \textit{SEARCH} operations to compare if the distance of a point to the query is closer than the distance of the point to its current assignment. If it is, then it's beneficial to reassign the point to the new crack, and this assignment is buffered, to be committed when $CRACK$ is executed, as explained in Section~\ref{subsec:when-to-operate}. $CRACK$ is the procedure that performs the physical reorganization, synchronizing the buffered state of the index with the currently working state. The core individual kernels of the procedure are shown in Algorithm \ref{alg:crack}. It adds a new inverted list for each buffered crack, moves stolen points to their new assignments, and updates the partition representatives to be the centroid of the points based on their new assignments. This procedure is only executed when enough budget for build operations has been accumulated throughout the usage of the index (Line 18 Algorithm~\ref{alg:search_and_crack}). This budget control mechanism ensures that CrackIVF does not spend a disproportionate amount of time on CRACK operations. The parameter $\alpha$, the ratio of total build operations time to the total time, is what controls the available budget. Since $CRACK$ is applied lazily, the heuristics governing where a crack is added and the cost-based control mechanism for when it is added are independent.

\begin{algorithm}[t]
\footnotesize
\caption{\textsc{CRACK}}
\label{alg:crack}
\begin{algorithmic}[1]
\REQUIRE $C_{\text{buffered}}$, $I_{\text{buffered}}, State_{dyn}, State_{true}$
\ENSURE New cracks committed; points reassigned; index state updated.
\STATE \textbf{Get Local Region:} Retrieve all points indexed by $I_{\text{buffered}}$, i.e., all data points that cracks in $C_{\text{buffered}}$ visited.
\STATE \textbf{Commit Reorg (w/ $\mathbf{A_P^\text{dyn}}$) :} Reorganize storage. Commit $C_{\text{buffered}}$ to the index. Reassign points to inverted lists using tracked assignments $A_P^\text{dyn}$
\STATE \textbf{Update Centroids:} Recompute each crack’s representative to be the centroid of the points assigned to it.
\STATE \textbf{Update Index State:} Synchronize $State_{dyn}$ and $State_{true}$ to reflect the changes.
\end{algorithmic}
\end{algorithm}

\textbf{\textit{REFINE}}: Refinement is the second operation that improves the index. It is performed eagerly, focusing on the local region visited by the query. A local region is defined by the $nprobe$ nearest partitions that the query visits during search. Algorithm~\ref{alg:search_and_crack} Line 7, is where the local region $C_{local}$ for a query is extracted from the \textit{SEARCH} result $C_{visited}$ of a query batch. The goal of REFINE is to immediately correct suboptimal assignments in the regions of the current index. Thus, any buffered cracks and assignments within the local region are not considered, as they are not part of the current index state. The decision to refine a region is based on heuristics evaluating the imbalance of point assignments in local cracks. Executing a refinement operation depends on an estimate of the runtime cost, determined using a cost model specific to $REFINE$ (Line 33, Algorithm~\ref{alg:search_and_crack}). Unlike cracking, the opportunity to refine a local region takes place when both the currently visited local region is a good candidate for refinement and there is sufficient computational budget to execute the operation. If the ``where'' and ``when'' heuristics for $REFINE$ do not simultaneously agree, the next opportunity is when another query accesses the same region. $REFINE$ is shown in Algorithm~\ref{alg:refine}). It executes a sequence of kernels implementing a local K-means variation and performs physical reorganization. This process improves the placement of crack representatives within the local region and reassigns points to their nearest representative. $REFINE$ has higher computational costs as it does not rely on precomputed buffered assignments, and assignments are determined dynamically upon completion of the local K-means process. $CRACK$ has a higher data-movement cost, since typically a larger region of the vector space is affected when cracks are added throughout.

\begin{algorithm}[t]
\footnotesize
\caption{\textsc{REFINE} }
\label{alg:refine}
\begin{algorithmic}[1]
\REQUIRE $C_{local}, State_{dyn}, State_{true}$
\ENSURE Refinement local region; points reassigned; index state updated.
\STATE \textbf{Get Local Region:} Retrieve all points $P_{local}$ in local region.
\STATE \textbf{Local K-Means:} Refine region of $C_{\text{local}}$, producing $C'_{\text{local}}$.
\STATE \textbf{Commit Reorg (w/o $\mathbf{A_P^\text{dyn}}$):} Replacing $C_{\text{local}}$ with refined $C'_{\text{local}}$. Compute new local point assignments. Reorganize storage, with local reassignment of $P_{local}$ to inverted lists.
\STATE \textbf{Update Index State:} Synchronize $State_{dyn}$ and $State_{true}$ to reflect the changes.

\end{algorithmic}
\end{algorithm}

\textbf{Index State:} Since \textit{CRACK} is a lazy operation, the index has two concurrent states, the true current state and a dynamic buffered state. The true state reflects the current structure of the index, while the dynamic one tracks its expected future structure for when $CRACK$ is executed, which synchronizes the two states. $REFINE$ operations directly alter the current index state, while cracking decisions are made based on the dynamic state. This can lead to scenarios that require us to undo buffering decisions. This is done to avoid bad cracks from being committed and having a situation where we commit decisions to the index that we know to be suboptimal (Line 21, Algorithm~\ref{alg:search_and_crack}). To track the index state, we define:

\begin{equation}
    \text{State}_\ast = (A_P^\ast, D_P^\ast, H_C^\ast)
    \label{eq:state_generic}
\end{equation}

where $\ast \in \{\text{dyn}, \text{true}\}$ denotes either the \textit{dynamic} uncommitted or \textit{true} committed state. The \textit{dynamic state} consists of \textit{dynamic assignments} $A_P^\text{dyn}$, \textit{dynamic point distances} $D_P^\text{dyn}$, and a \textit{histogram} $H_C^\text{dyn}$ of both the true and buffered crack sizes. Similarly, the \textit{true state} represents the true, committed state of assignments, distances, and crack sizes in the index. We extend the FAISS-IVF ``\texttt{.add()}'' operation to return the assignments and distances computed, which are an intermediate result, in order to initialize the state tracking. The state held in  $A_P^\text{dyn}$,$D_P^\text{dyn}$ and $H_C^\text{dyn}$, is only consistent with the actual state of the index after $CRACK$ operations and before buffering any cracks. If no cracks are buffered and only $REFINE$ operations are executed, the dynamic and true state are always consistent.
All of the above state metadata is used after the SEARCH operation, along with the search results $I,D,C_{visited}$ to classify each query as a good or bad crack candidate (Section \ref{subsec:where-heuristics}).

\textbf{Rollback of buffered state:} Consider the following scenario. A $CRACK$ operation was just executed at time $t_0$, so $State_{true} = State_{dyn}$. After a few queries, at $t_1$, assume that 2 good crack candidates have been buffered, both in the same $C_{local}$ region of the vector space and thus now $State_{true} \ne State_{dyn}$. At a later point, $t_2$, a $REFINE$ is executed, also in $C_{local}$ changing the assignments to $C'_{local}$, affecting $S_{true}$. Since $REFINE$ improves the current state eagerly, the decision that the buffered cracks are good was taken at a prior time $t_1$, on a prior state $C_{local}$. Thus, at $t_2$, during $REFINE$ the buffered cracks may now have had their points stolen back, and assigned to other refined cracks in $C'_{local}$. Now, while still $|C_{buffered}|=2$, it may now contain bad cracks. Finally, when at $t_3$, it is a good time to execute a $CRACK$ operation again, all buffered cracks are re-evaluated, as shown in lines 22-24 Algorithm~\ref{alg:search_and_crack}, and only the good cracks that remain are committed. 

\textbf{Memory Overhead and Optimizations:} The additional memory cost of maintaining \(State_{\text{dyn}}\) and \(State_{\text{true}}\) is \(O(4|P| + 2|C|)\). It is negligible since \(d \gg 4\) and the in-memory data alone are $O(|P|\times d)$. Additionally, these array structures are freed once the index converges or incoming queries stop. In our case, it enables an efficient compute-storage trade-off, deferring cracking costs instead of paying them per query. There are many optimizations that a full-fledged implementation could utilize. For embedding lakes, where vectors are stored on disk, state tracking arrays can be stored alongside data points instead of in the index (irrelevant for our in-memory FAISS implementation). Since queries for an individual index are typically infrequent, \(State_{\text{dyn}}\) can be committed or flushed between queries, while \(State_{\text{true}}\) recomputed as needed. Full memory overhead is only necessary during bursty query loads with limited hardware resources, where cracking must be buffered to reduce cumulative overhead and prevent long tail response times. Finally, the cracking overhead can be hidden from query response times since results are available before any cracking operations begin. A separate thread pool can handle cracking operations while queries use the current index state, provided \(CRACK\) and \(REFINE\) remain atomic to prevent inconsistencies. Prior work on the related problem of index maintenance \cite{xu2023spfresh} demonstrated how to concurrently execute queries by dispatching such tasks to background threads. In our work, the entire thread pool is dedicated either to search or cracking operations. We also assume a continuous query load at system capacity. As a result, the memory overhead for index state tracking is unavoidable. We further cover index maintenance prior work in Section~\ref{sec:related_work}. 

\subsection{Where to apply build operations?}
\label{subsec:where-heuristics}

Each arriving query is a crack candidate, and the visited region during index traversal is a candidate for refinement. This results in a continuous stream of candidates of varying quality. As discussed, not all vector space regions are equally important, and not all operations lead to performance gains. The \textbf{goal} is \textit{to maintain a stream of high-quality candidates}, classified from incoming queries, enabling the index to grow by adding cracks in frequently accessed regions that need further partitioning, and refining poorly clustered regions with K-means. CrackIVF employs two heuristics, acting as a binary classification model for candidates. Each uses a set of decision rules that are distilled from our empirical observations of IVF index behavior. We do not claim to have defined an exhaustive list of decision rules for the model. Changing or learning the model and the parameters while still achieving the stated goal above, would not change the overall design of the index (Algorithm~\ref{alg:search_and_crack}).

\textbf{Where to CRACK (CrackHeuristic):} We observe that the optimal performance is achieved at some maximum number of partitions (Figure~\ref{fig:fig1a_response_time}). Thus, for it to grow, we treat each incoming query as a \textit{good crack candidate by default} and define a set of decision rules to classify when a query is a \textit{bad} candidate.  Query candidates classified as \textit{good}, are buffered until a future $CRACK$ operation. A \textit{bad} candidate (returning \textit{False}) is if \textit{any} of the following hold:

\textit{Don't Crack: Rule 1} rejects queries that steal too few points:
\begin{equation}
     |I_{\text{steal}}| < \text{min\_pts}
\end{equation}

\textit{Don't Crack: Rule 2} prevents excessive partitioning, constraining local partition count based on the local number of points:
\begin{equation}
    \frac{\text{total points in local region}}{\text{total cracks in local region}} > \text{pts\_crack\_thr}
\end{equation}
    
We set $\text{min\_pts} = 2$ in order to avoid adding empty cracks, and $\text{pts\_crack\_thr} = 64$ to ensure an average of at least 64 points per crack locally. Rule 2 is especially important in high skew datasets, where most crack candidates are for the same region.

\textbf{Where to REFINE (RefineHeuristic):}  We observe that refining an index improves performance at each index size (Figure~\ref{fig:fig1a_response_time}), by better distributing points across partitions and repositioning centroids to better fit the data. Optimal performance is reached at the maximum index size, so we want to avoid getting stuck refining a small index. Unlike $CRACK$, where latency is amortized by buffering, $REFINE$ is eagerly applied to immediately improve the current index, but it is also a costly operation. Thus, we treat \textit{each incoming refine candidate as bad}, leaving more of the cost budget for cracks so that the index can grow. A refine candidate is the local visited region of the query, visited during search. It is a \textit{good} candidate for refinement (returning \textit{True}) if \textit{any} of the following hold: 

\textit{Refine: Rule 1} captures local imbalance:
\begin{equation}
    \text{CV}_{\text{local}} = \frac{\sigma_{\text{local}}}{\mu_{\text{local}}} > \text{cv\_max}
\end{equation}
where $\sigma_{\text{local}}$ and $\mu_{\text{local}}$ are the standard deviation and mean of cluster sizes in the local region. A high coefficient of variation indicates substantial local imbalance, warranting refinement.

\textit{Refine: Rule 2} captures imbalance using the global distribution:

\begin{equation}
    \begin{aligned}
        \exists c_1, c_2 \quad \text{s.t.} \quad  
        S(c_1) \leq {\text{cutoff}_{\text{low}}}, \\
        S(c_2) \geq {\text{cutoff}_{\text{high}}}
    \end{aligned}
\end{equation}

where $c_1, c_2$ are cracks in the local region, and $S(c) = H_C^\text{true}[c]$ represents the size (i.e., number of assignments) of crack $c$. The $\text{cutoff}_{\text{low}}$ and $\text{cutoff}_{\text{high}}$ correspond to cluster sizes at the $size\_prctl$ and $100 - size\_prctl$ percentiles of the global partition sizes. This rule captures the unlikely scenario that a large and small crack globally are next to each other locally, indicating points can be better distributed. 

We set $\text{cv\_max} = 2$, meaning refinement is triggered if cluster sizes vary by at least twice the local mean, and $\text{size\_prctl} = 10$, defining small and large cracks based on the 10th and 90th percentiles of global cluster sizes. 

\subsection{When to apply build operations?}
\label{subsec:when-to-operate}

To mitigate the overhead of constant physical reorganization, we use a build budget to restrict the fraction of time dedicated to $CRACK$ and $REFINE$ operations. We define \( T_{\text{build}} \) as the total measured time spent on previous build operations. Similarly, \( T_{\text{search}} \) represents the cumulative time spent on all past search operations. When considering a potential build operation, we estimate its execution cost using \( \hat{T}_f \), which captures the expected time required to perform a $CRACK$ operation in the current dynamic state or a $REFINE$ operation in the true state. The build operation considered is specified by \( f \in \{\textsc{CRACK}, \textsc{REFINE}\} \). To enforce the budget constraint, the total time spent on past build operations in addition to the estimated time if the current one is executed must remain within a fraction \( \alpha \) of the system's total runtime:

\begin{equation}
\label{eq:budget_constr}
    T_{\text{build}} + \hat{T}_f \leq \alpha \cdot (T_{\text{build}} + T_{\text{search}} + \hat{T}_f).
\end{equation}

\( T_{\text{build}} \) and \( T_{\text{search}} \) are directly measured from the execution time, while \( \hat{T}_f \) is estimated using a predictive cost model. The parameter \( \alpha \) defines the proportion of total execution time that can be allocated to indexing operations. We set $\alpha=0.5$, which ensures a maximum of 50\% of the total time at any given point is spent on build operations. As more queries arrive, {CrackIVF} will progressively have a larger budget to apply for physical reorganization operations. In this way, we can remain near the optimal minimum of cumulative time across all query scales. Finally, there is a minimum size constraint on the number of buffered cracks before a $CRACK$ is executed. The idea comes from Figure~\ref{fig:fig1a_response_time}, where if a REFINE can achieve similar performance improvement at the current index size, it is better to wait until more cracks have been buffered, so that the performance improvement from index growth is substantial. Specifically, in our implementation, we set that minimum to 20\% of the current index size, $C_{min}=0.2 \times |C|$ (Line 19 Algorithm~\ref{alg:search_and_crack}).

\textbf{Predictive Cost Model:} \label{sec:cost_model} Estimating the execution time of the \textit{CRACK} and \textit{REFINE} operations happens after every incoming query. Thus, we must maintain low latency when making a prediction, while have accurate enough cost estimates to correctly use the budget constraint Eq.~\ref{eq:budget_constr}. For this, we use a simple first-order linear model fitted using multivariate regression \cite{jain1990art}. Each build function $f$ comprises a sequence of procedures, or kernels, indexed by $i$ and executed sequentially, as shown in Algorithm \ref{alg:crack} and Algorithm \ref{alg:refine}. The execution time is dominated by the following kernels:

\textsc{CRACK}:  
(i) \textit{Get Local Region},  
(ii) \textit{Commit Reorg} (w/ $A_P^{\text{dyn}}$),  
(iii) \textit{Update Centroids}.  

\textsc{REFINE}:  
(i) \textit{Get Local Region},  
(ii) \textit{Local k-Means},  
(iii) \textit{Commit Reorg} (w/o $A_P^{\text{dyn}}$).

\begin{equation}
    T_f = \sum_{i} T_{f, i}, \quad \forall f \in \{\textit{CRACK}, \textit{REFINE}\}.
\end{equation}

These kernels are executed sequentially so the total execution time $T_f$ is simply the sum of kernel execution times. Each kernel \( i \) contributes a latency \( T_{f, i} \), which we model as a linear function of its two dominant cost factors: computational complexity and data movement. For each, we learn the following predictive model:

\begin{equation}
    T_{f, i} = w_{f, i}^{(1)} \cdot C_{f, i} + w_{f, i}^{(2)} \cdot D_{f, i} + b_{f, i},
\end{equation}

where \( C_{f, i} \) represents the dominant computational complexity of kernel \( i \), and \( D_{f, i} \) denotes the dominant data movement cost. The terms \( w_{f, i}^{(1)} \) and \( w_{f, i}^{(2)} \) are learned coefficients for computation and data movement and \( b_{f, i} \) is a learned parameter that accounts for fixed overheads. We gather measurement with micro-benchmarks for each kernel, by varying input parameters and measurig execution latency to fit a regression model, learning \( w_{f, i}^{(1)}, w_{f, i}^{(2)}, b_{f, i} \).

\begin{table}[t] 
    \centering
    \caption{Features used for each kernel. \textnormal{\footnotesize{\textsuperscript{†} For REFINE, \( C_{\text{buffered}} = 1 \)}}}
    \label{tab:kernel_cost_model}
    \resizebox{\columnwidth}{!}{%
        \begin{tabular}{lcc}
            \toprule
            \textbf{Kernel} & \textbf{Compute} & \textbf{Data Movement} \\
            \midrule
            Get Local Region\textsuperscript{†} & \( |C_{\text{buffered}}| \times |C| \) & \( |P_{\text{local}}| \times d \)  \\
            Commit Reorg (w/o $A_P^\text{dyn}$) & \( |P_{\text{local}}| \times |C_{\text{local}}| \times d + |C| \) & \( |P_{\text{local}}| \times d \)  \\
            Commit Reorg (w/ $A_P^{dyn}$) & \( |C| \) & \( |P_{\text{local}}| \times d \)  \\
            Update Centroids & \( |P| \times d \) & \( (|P| + |C|) \times d \)  \\
            Local K-Means & \( n_{\text{iter}} \times |P_{train}| \times |C_{\text{local}}| \times d \) & \( |C| \times d \) \\
            \bottomrule
        \end{tabular}
    }
    \captionsetup{aboveskip=2pt, belowskip=7pt}
\end{table}

Our approach removes the need for exact analytical models coupled to specific implementations and hardware. Switching hardware requires re-running micro-benchmarks and refitting the linear model. The main limitation is that execution time is modeled as a linear function. While this captures dominant compute and memory trends, it overlooks system-level nonlinearities such as caching effects, NUMA constraints, and SIMD optimizations. Despite this, our empirical results show that the cost model performs well across all benchmarks. Importantly, we always use the true measured times \( T_{\text{build}} \) and \( T_{\text{search}} \) in the budget constraint (Equation~\ref{eq:budget_constr}), which help correct over and under estimations. In our tests, the model achieved moderate to high \( R^2 \) scores, explaining 40–95\% of the total variance, with RMSE ranging from milliseconds to a few seconds.

\section{Experiments}
\label{sec:experiments}

\begin{figure*}[t]
    \centering
    \begin{subfigure}[b]{0.22\textwidth}
         \centering
         \includegraphics[width=\textwidth]{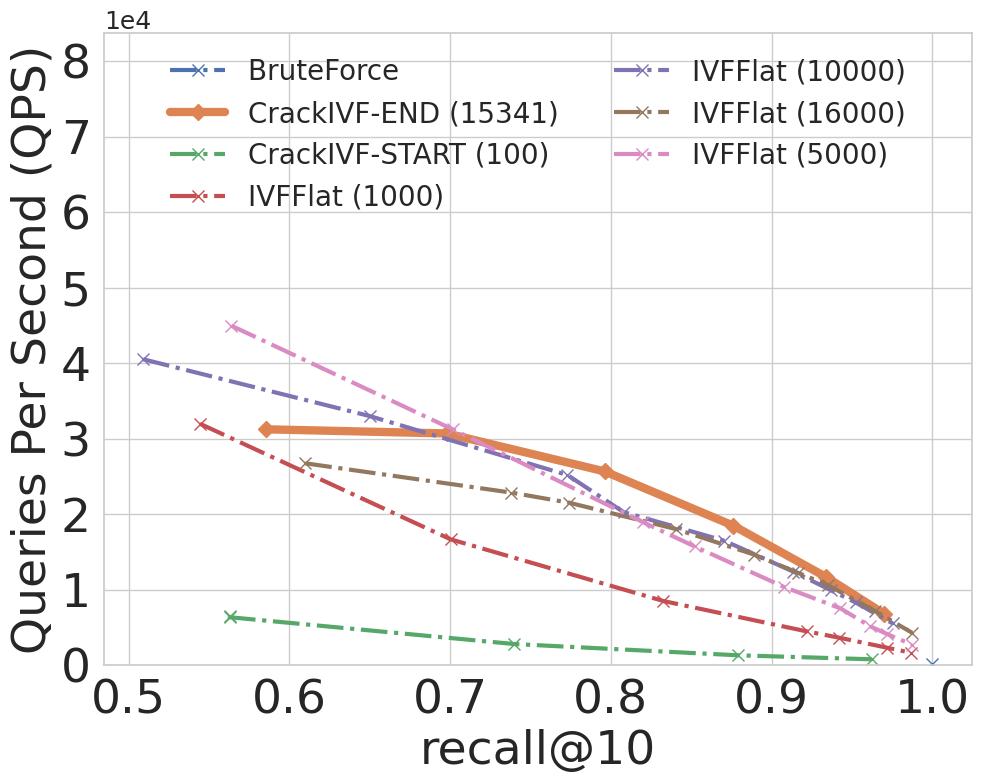}
         \caption{SIFT1M: QPS vs. Recall}
         \label{fig:sift1m-qps}
    \end{subfigure}
    \hfill
    \begin{subfigure}[b]{0.22\textwidth}
         \centering
         \includegraphics[width=\textwidth]{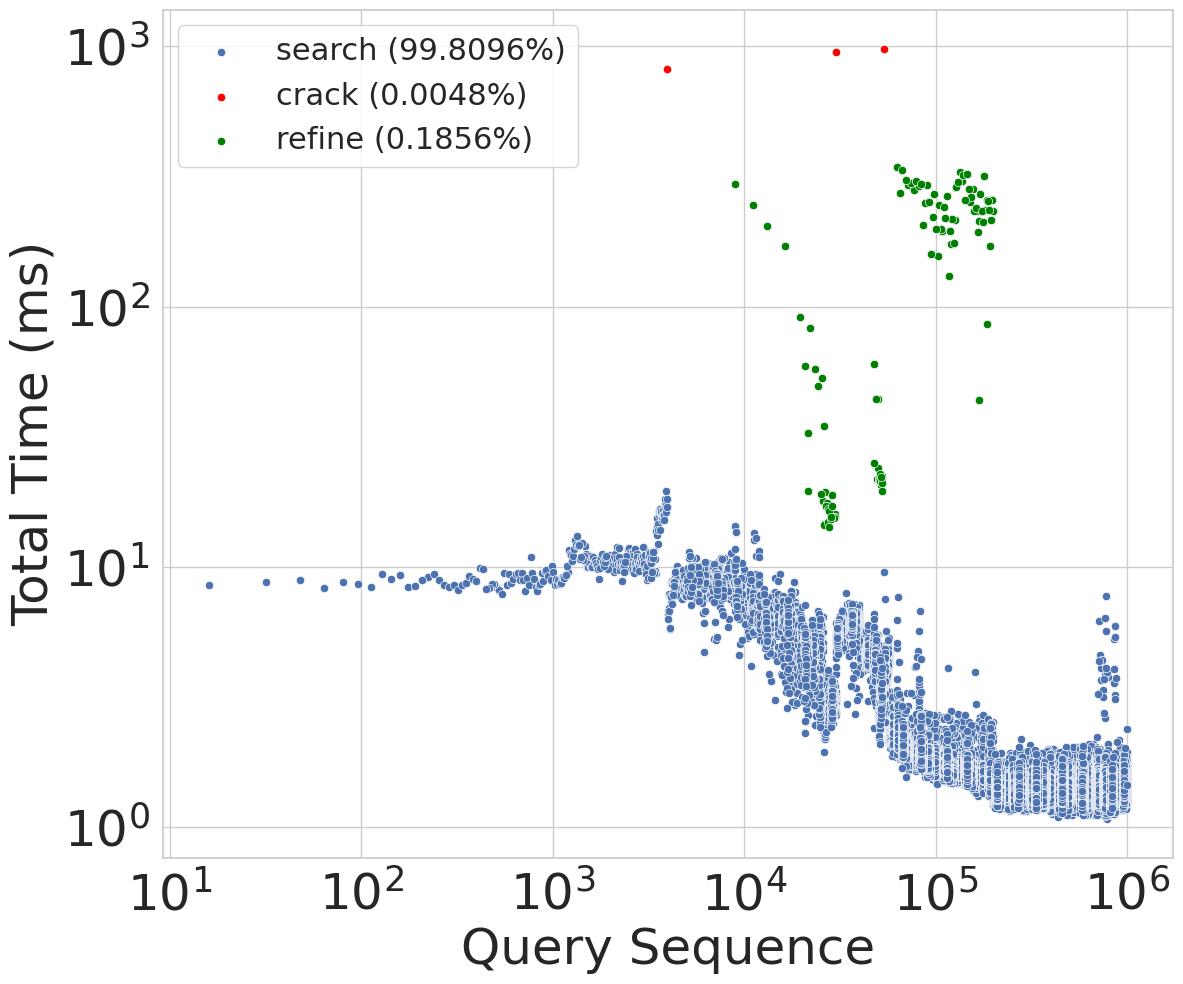}
         \caption{{SIFT1M: Time per Query}}
         \label{fig:sift1m-qt}
    \end{subfigure}
    \hfill
    \begin{subfigure}[b]{0.22\textwidth}
         \centering
         \includegraphics[width=\textwidth]{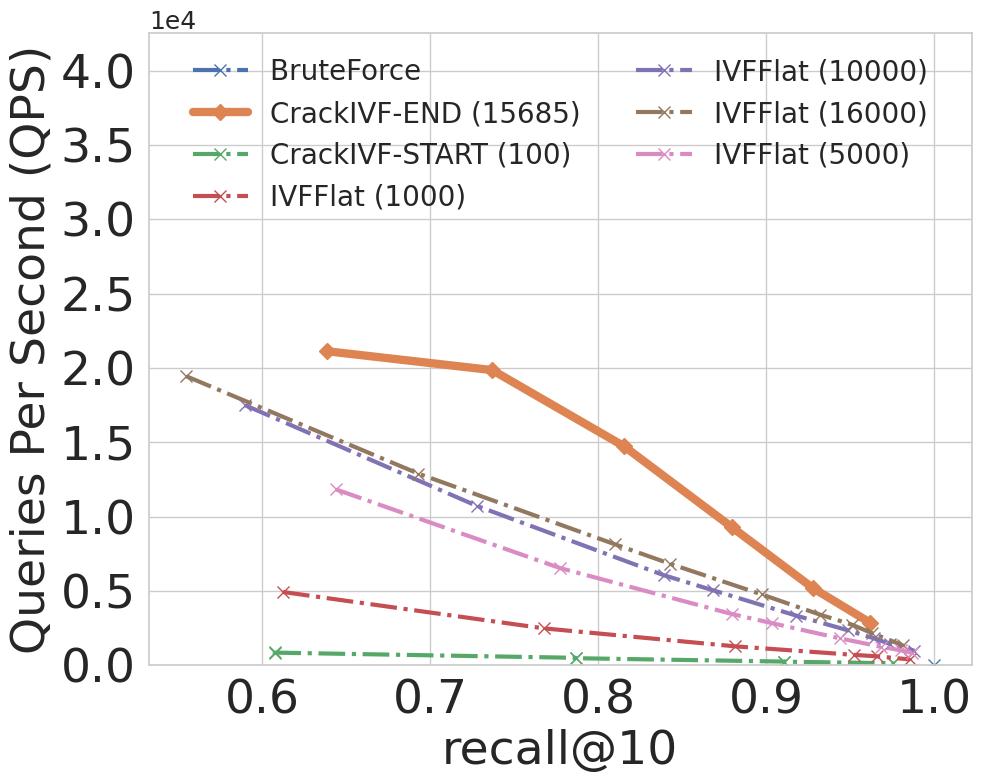}
         \caption{SIFT10M: QPS vs. Recall}
         \label{fig:sift10m-qps}
    \end{subfigure}
    \hfill
    \begin{subfigure}[b]{0.22\textwidth}
         \centering
         \includegraphics[width=\textwidth]{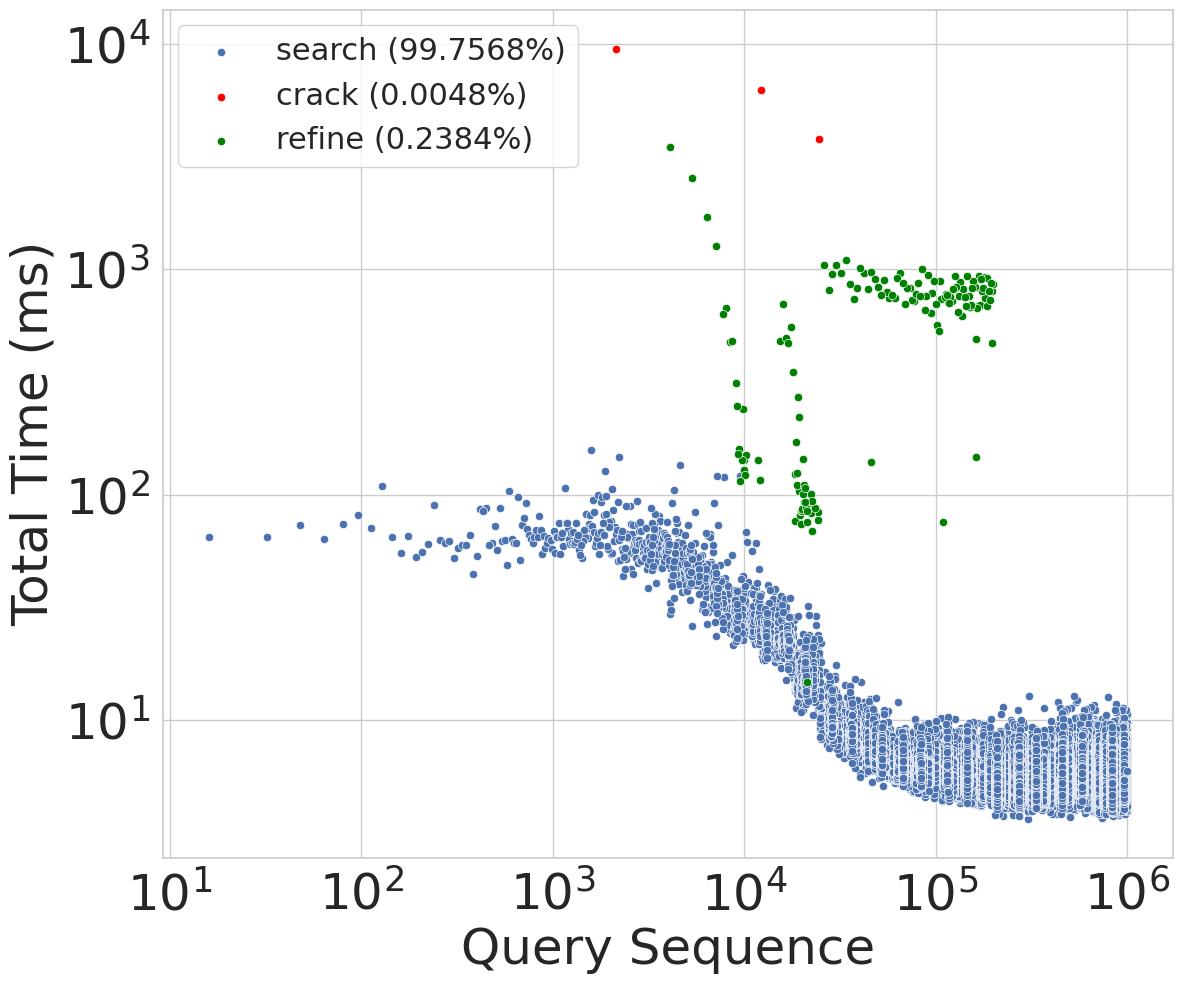}
         \caption{{SIFT10M: Time per Query}}
         \label{fig:sift10m-qt}
    \end{subfigure}
    
    \begin{subfigure}[b]{0.22\textwidth}
         \centering
         \includegraphics[width=\textwidth]{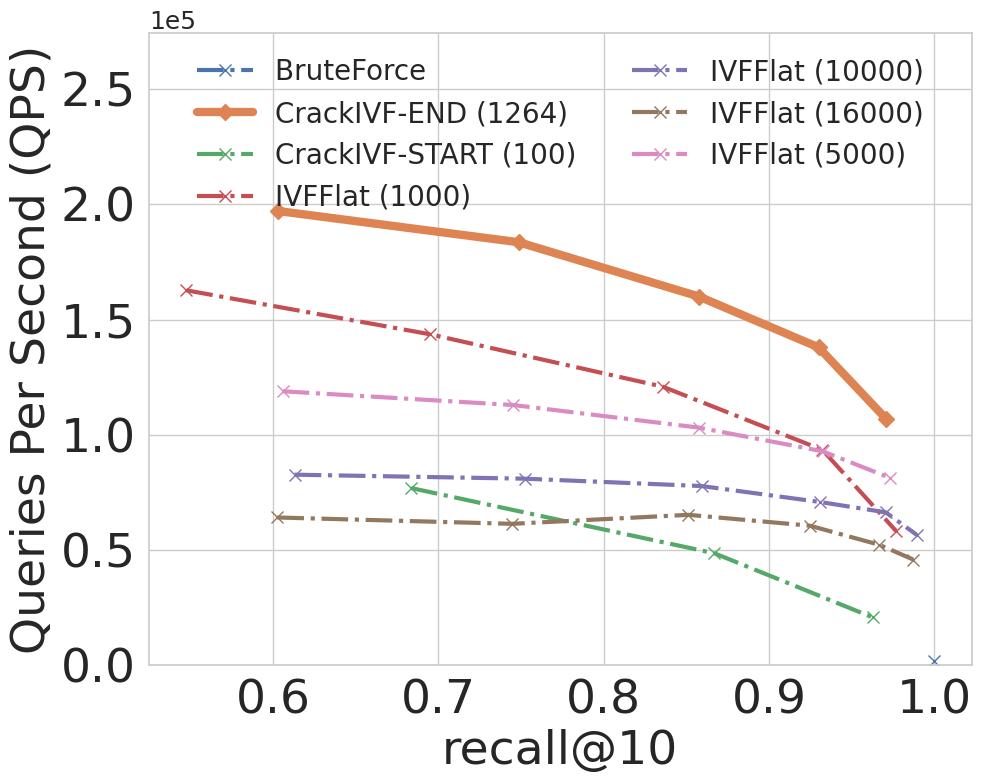}
         \caption{Last.fm: QPS vs. Recall}
         \label{fig:lastfm-qps}
    \end{subfigure}
    \hfill
    \begin{subfigure}[b]{0.22\textwidth}
         \centering
         \includegraphics[width=\textwidth]{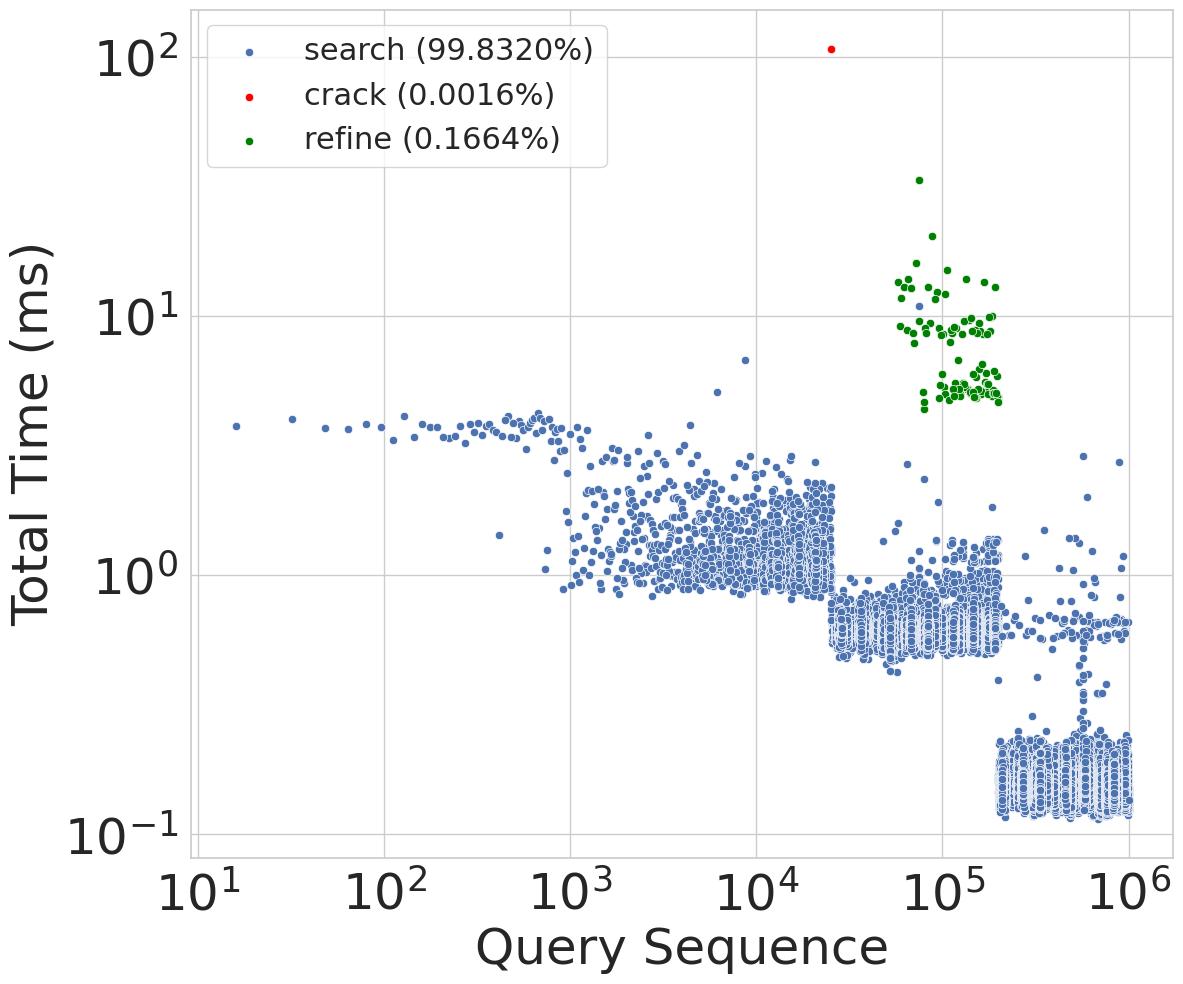}
         \caption{{Last.fm: Time per Query}}
         \label{fig:lastfm-qt}
    \end{subfigure}
    \hfill
    \begin{subfigure}[b]{0.22\textwidth}
         \centering
         \includegraphics[width=\textwidth]{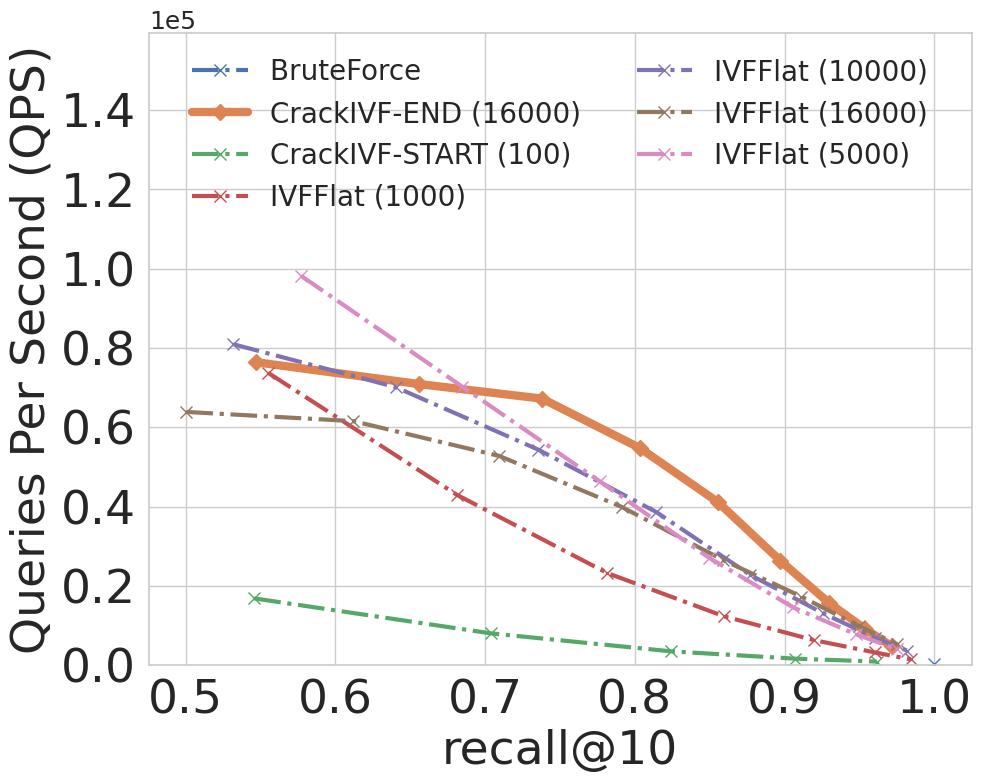}
         \caption{Glove50: QPS vs. Recall}
         \label{fig:glove50-qps}
    \end{subfigure}
    \hfill
    \begin{subfigure}[b]{0.22\textwidth}
         \centering
         \includegraphics[width=\textwidth]{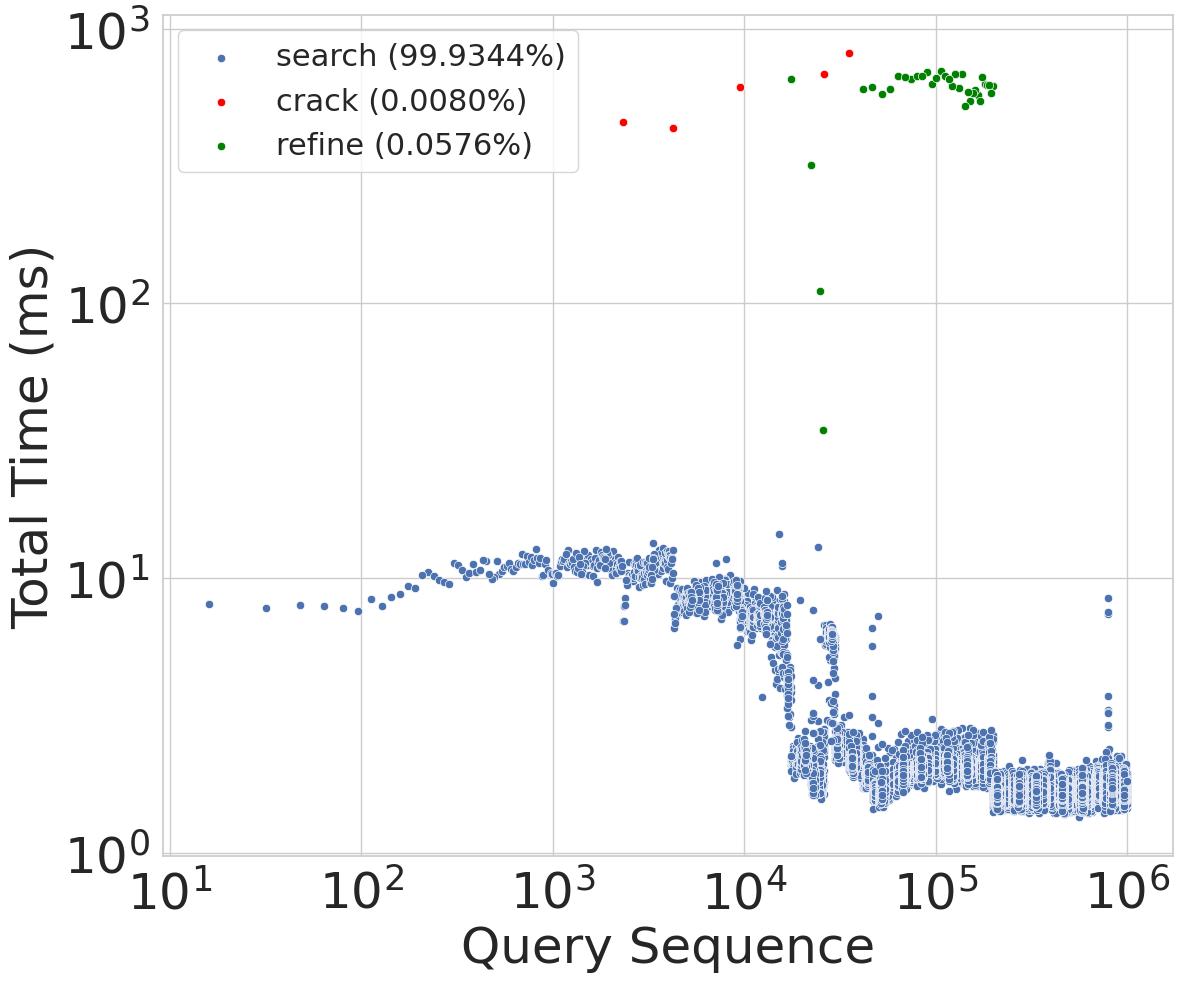}
         \caption{{Glove50: Time per Query}}
         \label{fig:glove50-qt}
    \end{subfigure}
    
    \begin{subfigure}[b]{0.22\textwidth}
         \centering
         \includegraphics[width=\textwidth]{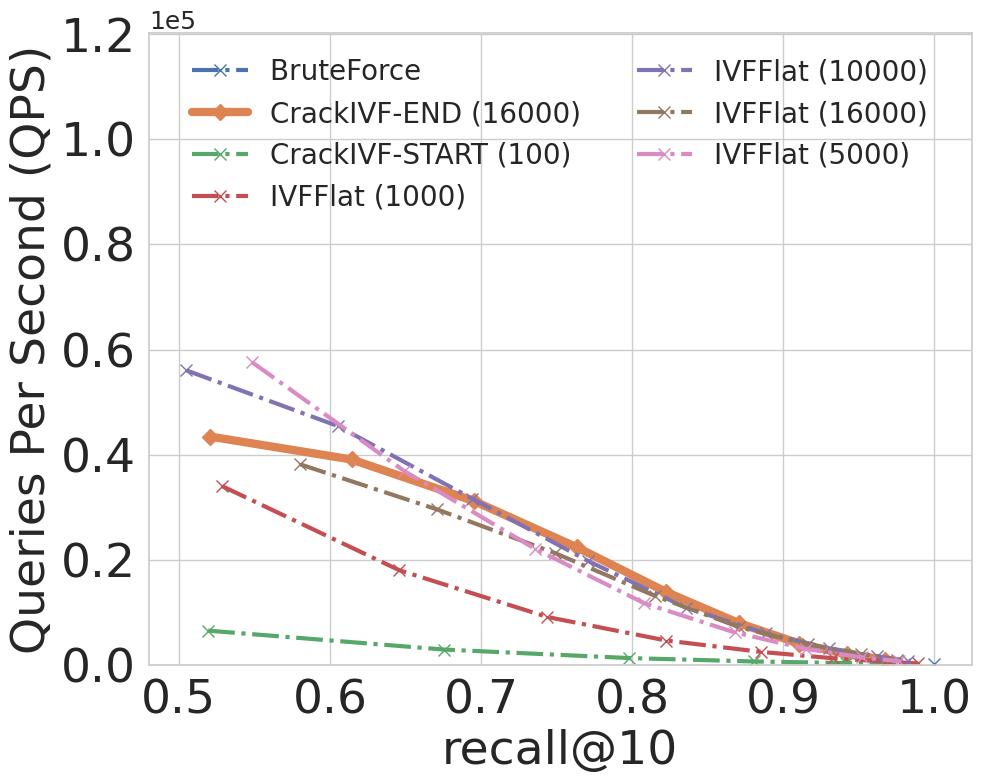}
         \caption{Glove100: QPS vs. Recall}
         \label{fig:glove100-qps}
    \end{subfigure}
    \hfill
    \begin{subfigure}[b]{0.22\textwidth}
         \centering
         \includegraphics[width=\textwidth]{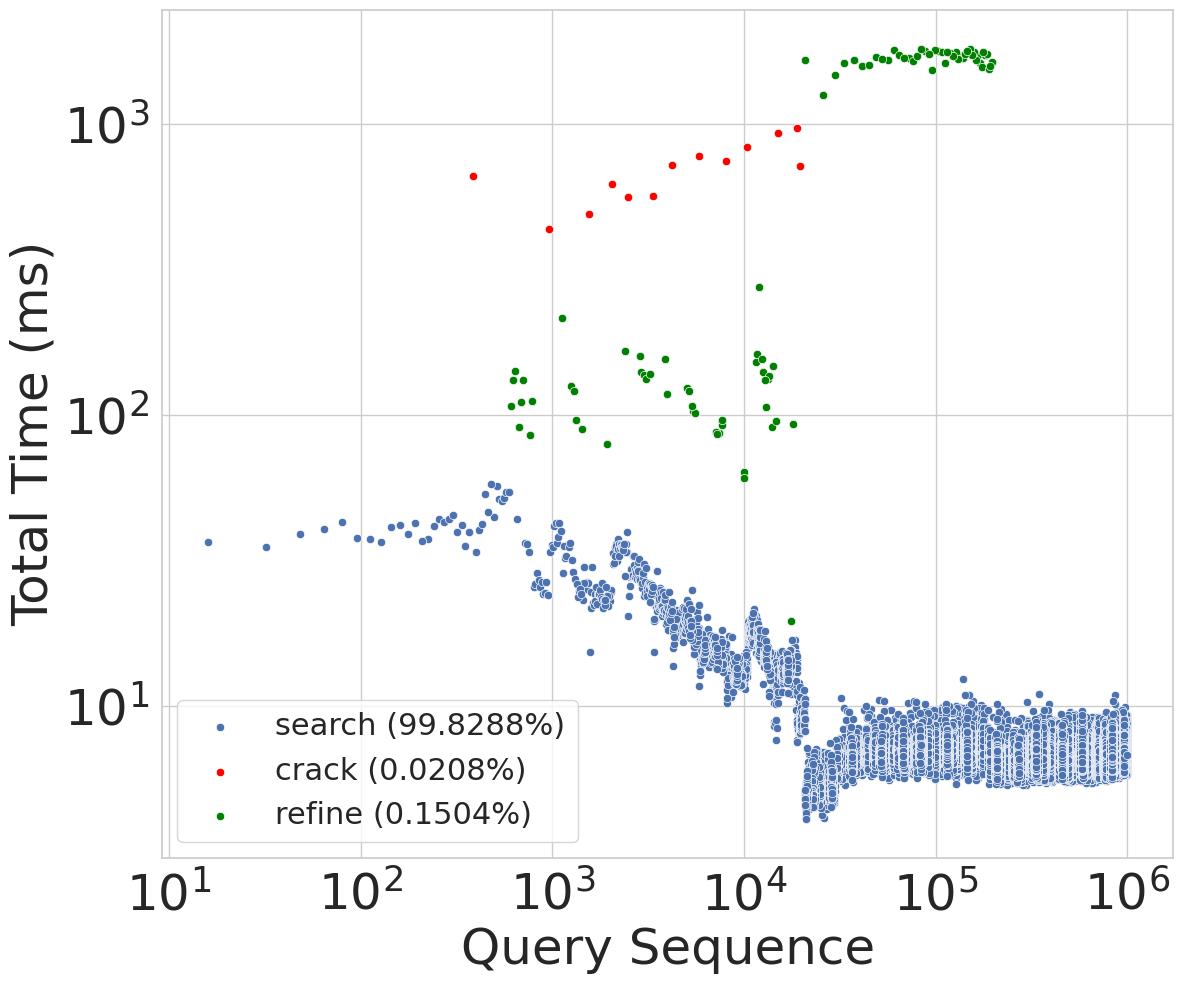}
         \caption{{Glove100: Time per Query}}
         \label{fig:glove100-qt}
    \end{subfigure}
    \hfill
    \begin{subfigure}[b]{0.22\textwidth}
         \centering
         \includegraphics[width=\textwidth]{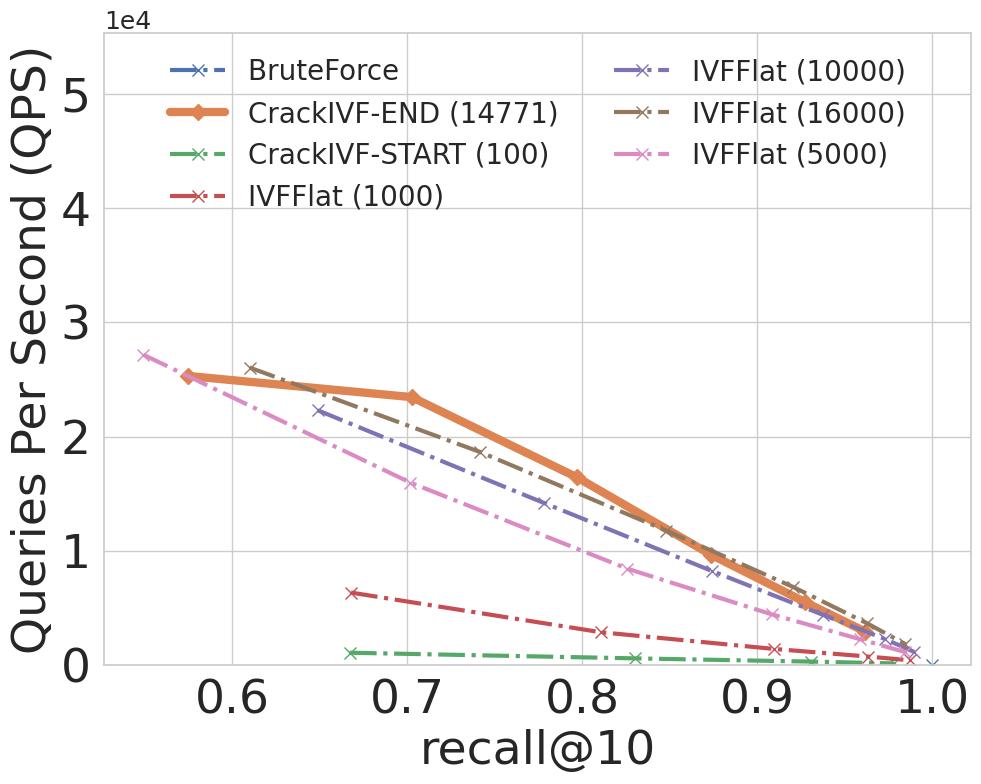}
         \caption{Deep10M: QPS vs. Recall}
         \label{fig:deep-qps}
    \end{subfigure}
    \hfill
    \begin{subfigure}[b]{0.22\textwidth}
         \centering
         \includegraphics[width=\textwidth]{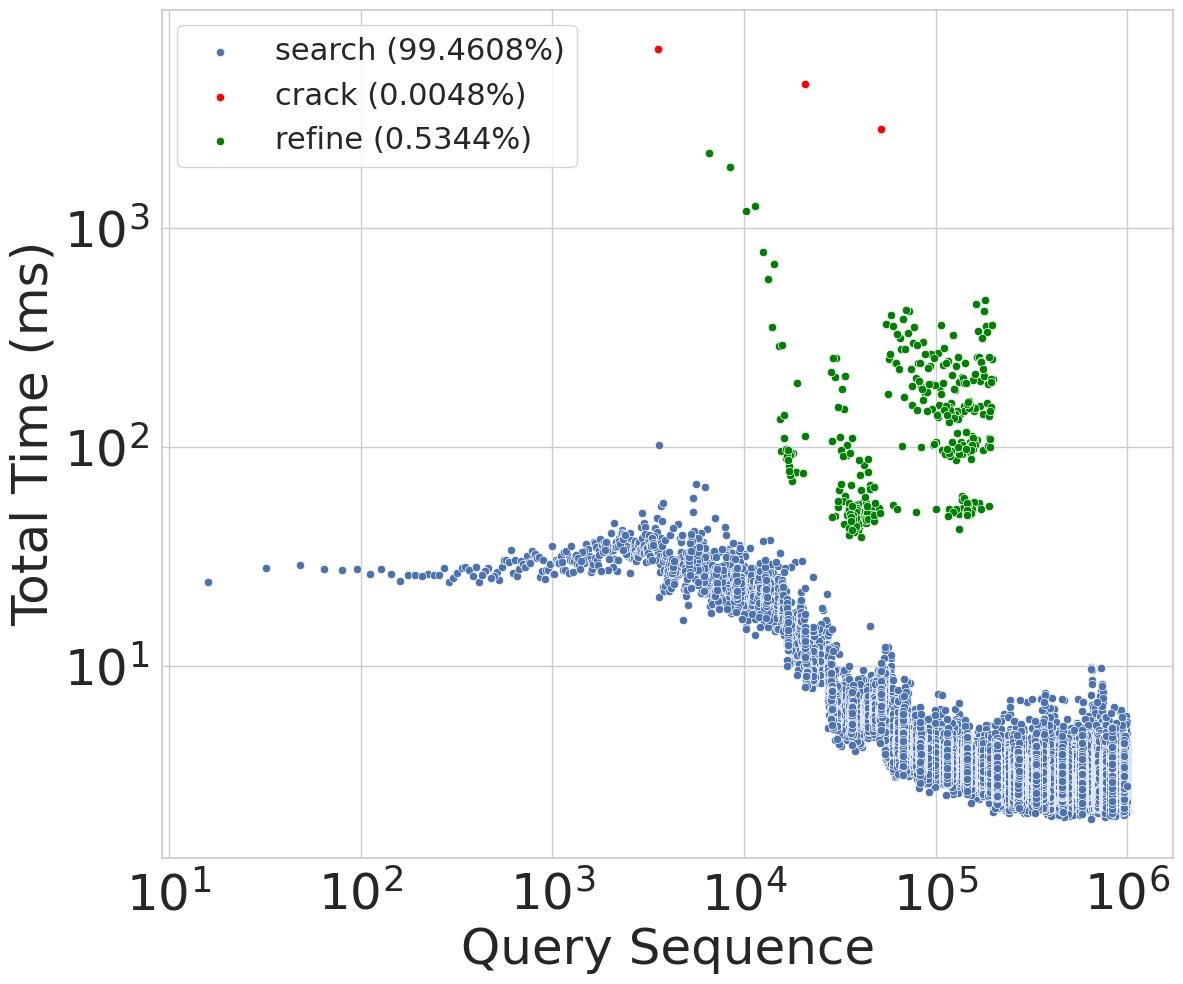}
         \caption{{Deep10M: Time per Query}}
         \label{fig:deep-qt}
    \end{subfigure}
    
    \caption{{Queries Per Second (QPS) vs. Recall and Time per each Query batch for the entire trace across different datasets.}}
    \label{fig:qps_recall_and_response_time_comparison}
\end{figure*}

We run on a dual-socket AMD EPYC 7V13 system (128 cores, 3.7 GHz, 512GB RAM). All experiments, except for Table~\ref{tab:avtree}, use 16 cores and a batch size of 16, the point at which memory bandwidth saturates. IVF and CrackIVF share the same SEARCH operation in FAISS and benefit equally from inter-query parallelism.

We evaluate on standard ANN search benchmarks, GloVe\cite{pennington2014glove}, SIFT\cite{jegou2011searching}, DEEP\cite{babenko2016efficient}, and Last.fm\cite{Bertin-Mahieux2011}.  SIFT1M and SIFT10M contain 128-dimensional SIFT descriptors; DEEP10M  consists of 96-dimensional embeddings, with 1M and 10M being the number of data points in each slice. Both use L2 distance as a metric. GloVe contains 1.18M varying dimensional embeddings and uses cosine similarity, which we implement by \textit{l2-normalizing} and computing the inner product (IP). Last.fm, contains 64-dimensional vectors and IP metric. The Last.fm query set is highly skewed, allowing us to test CrackIVF in such cases. Last.fm provides 50k unique queries, while the other datasets provide 10K unique queries. We replicate them until we match our target scales (up to 1M).

For ENN, we compare against Brute Force, and AV-Tree \cite{lampropoulos2023adaptive}, which is a cracking baseline. For ANN search, we compare against FAISS IVFFlat indexes with the number of partitions chosen shown in parentheses, e.g., {IVFFlat} (1000). We vary the number of partitions following FAISS guidelines \cite{faiss_guidelines}, for our dataset scales. For CrackIVF, we use fixed parameters values for the control mechanisms, set as explained in (Sections~\ref{subsec:where-heuristics}, \ref{subsec:when-to-operate}). We consider the index converged when the crack decision rules stop triggering and refines keep targeting the same region without effect, thus the index can no longer grow in size, nor improve at the current size. For CrackIVF and IVFFlat, $nprobe$ is set to always achieve 90–95\% Recall@10 in ANN. CrackIVF is initialized with 100 partitions on all experiments. We set this parameter by choosing the largest possible starting partitions that still give minimal startup cost. Because of FAISS parallelism, at 16 threads, the startup time for 100 partitions is identical or faster than starting with 1 partition. On a single thread for Table~\ref{tab:avtree}, the difference between them is only $\approx$120 ms.

\subsection{Comparison with AV-Tree:}

\begin{table}[b]
    \caption{{AV-Tree vs. CrackIVF. \textnormal{\footnotesize{\textsuperscript{†} Cracking-based baselines.}}}}
    \label{tab:avtree}
    \resizebox{\columnwidth}{!}{%
    \begin{tabular}{@{}llllllll@{}}
    \toprule
             & \multicolumn{7}{c}{{Cumulative Time at i-th Query (seconds)}}            \\ \midrule
             & {$10^0$} & {$10^1$} & {$10^2$} & {$10^3$} & {$10^4$}  & {$10^5$} & {$10^6$} \\ \cmidrule(l){2-8} 
    {IVF-1K}   & {12.57}  & {12.59}  & {12.76} & {14.48} & {32.10} & {205.40} & {1938.60} \\
    {IVF-5K}   & {178.17} & {178.18} & {178.35} & {179.93} & {195.94} & {354.63} & {1943.02} \\
    {AV-Tree\textsuperscript{†}}  & {0.08}  & {0.75}  & {7.48}  & {75.43} & {830.0} & {1274.85} & {5721.29}      \\
    {CrackIVF\textsuperscript{†}} & {1.10}  & {1.16}  & {1.73}  & {13.18} & {55.33}  & {233.11} & {1531.40}             \\ \bottomrule
    \end{tabular}%
    }

\end{table}

To the best of our knowledge, AV-tree \cite{lampropoulos2023adaptive} is the only similarity search index, whose underlying idea is based on cracking. It is a tree-based ENN cracking index targeting short-lived data up to one thousand queries. Unlike AV-Tree, CrackIVF is an IVF-based ANN cracking index, implemented around FAISS, offering parallelism, and targeting high dimensional RAG applications which could reach much larger query volumes. As AV-Tree is single-threaded and an ENN index, to compare fairly, we set CrackIVF and the IVF baselines to run with a single thread and make sure we probe enough partitions to get an average of more than $99\%$ recall across all query scales. We set the cracking threshold of AV-Tree to 128, which we find gives the best total runtime. We show the cumulative time for a trace of {$1$M} queries over the SIFT 1M dataset in Table~\ref{tab:avtree}. At the start, the $10^0$-th query time includes the index construction costs of CrackIVF and IVF baselines. CrackIVF is initialized as a small 100 partition index which grows as more queries arrive, whereas AV-Tree does not pre-build any index. In this experiment, AV-Tree has an advantage over CrackIVF for the first 10 queries, as it can start answering immediately. However, CrackIVF processes each query faster and outperforms AV-Tree by a large margin after the first few queries. The cumulative time of AV-Tree increases at a slower rate with more queries, showing AV-Tree is improving over time. Our CrackIVF is 3.7x faster than AV-Tree in the end of the trace. As cracking baselines, both CrackIVF and AV-Tree, vastly outperform IVF indexes on the low query scale. AV-Tree continues to outperform up to 1000 queries, matching original claims \cite{lampropoulos2023adaptive}. CrackIVF performance benefits continue even at 1 million queries. In the following experiments we focus on ANN.

\subsection{Does CrackIVF improve over time?}
We measure how CrackIVF improves over time in Figure~\ref{fig:qps_recall_and_response_time_comparison}.
QPS vs. Recall plots show the search performance trade-off in which ANN search operates under. In our experiments, by the end of the query trace, CrackIVF-END is consistently at or near the Pareto frontier across all tested datasets. On SIFT10M and Last.fm, CrackIVF is by far the best-performing index. Specifically on Last.fm, CrackIVF with 1264 final partitions, achieves $\approx$50\% higher QPS for Recall@10 = 0.9, than any other index. We attribute this to CrackIVF's approach of allocating new cracks following the query distribution. Since Last.fm is the dataset with the highest skew (Figure~\ref{fig:plot_skew_cdf}), we can fully partition the accessed space with a small number of cracks, achieving the best performance. In Figure~\ref{fig:qps_recall_and_response_time_comparison}, we also plot the response time per query batch across the entire query trace. CrackIVF search performance improves over time after incremental reorganization operations. The vast majority of outliers are due to the individual $CRACK$ and $REFINE$ reorganization operations. In our experiments, $CRACK$ operations are less than {0.03\%} and $REFINE$ less than {0.6\%} of the query trace.
    
\subsection{Does CrackIVF Minimize Cumulative Time?}

\begin{figure}[t]
    \centering
    \begin{subfigure}[b]{0.48\columnwidth}
         \centering
         \includegraphics[width=\textwidth]{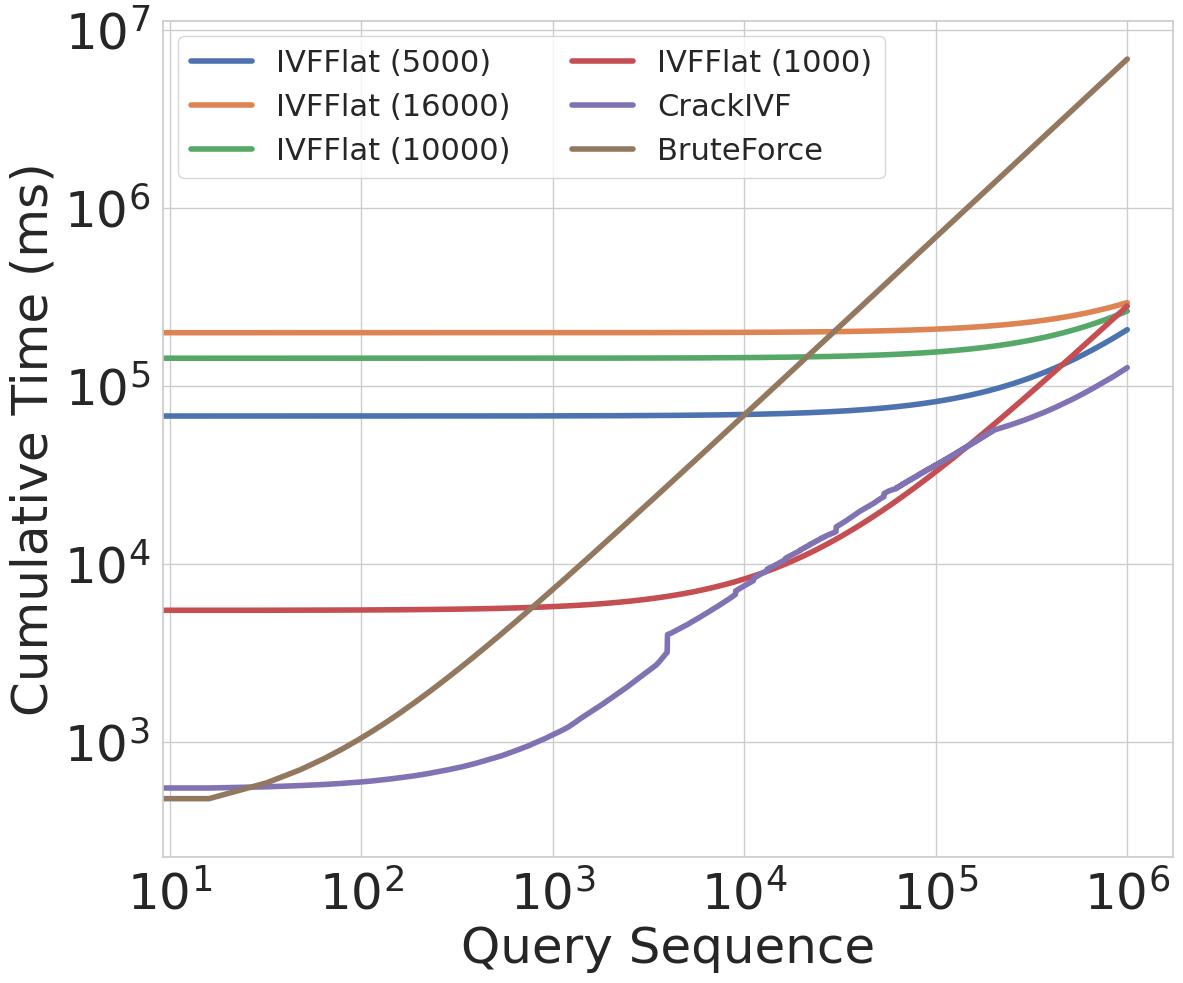}
         \caption{SIFT1M}
         \label{fig:sift1m-cu}
    \end{subfigure}
    \hfill
    \begin{subfigure}[b]{0.48\columnwidth}
         \centering
         \includegraphics[width=\textwidth]{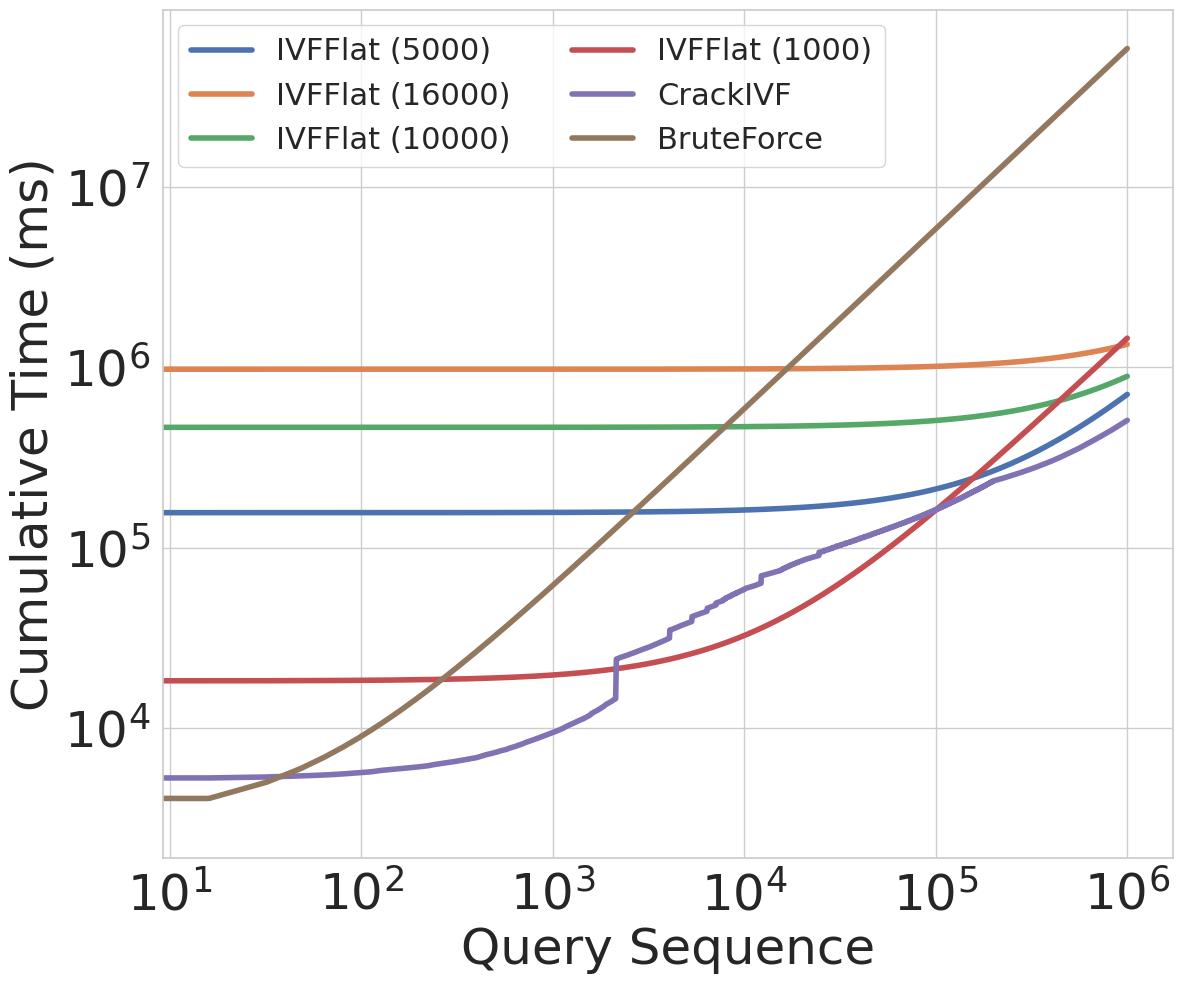}
         \caption{SIFT10M}
         \label{fig:sift10m-cu}
    \end{subfigure}
    
    \vfill
    \begin{subfigure}[b]{0.48\columnwidth}
         \centering
         \includegraphics[width=\textwidth]{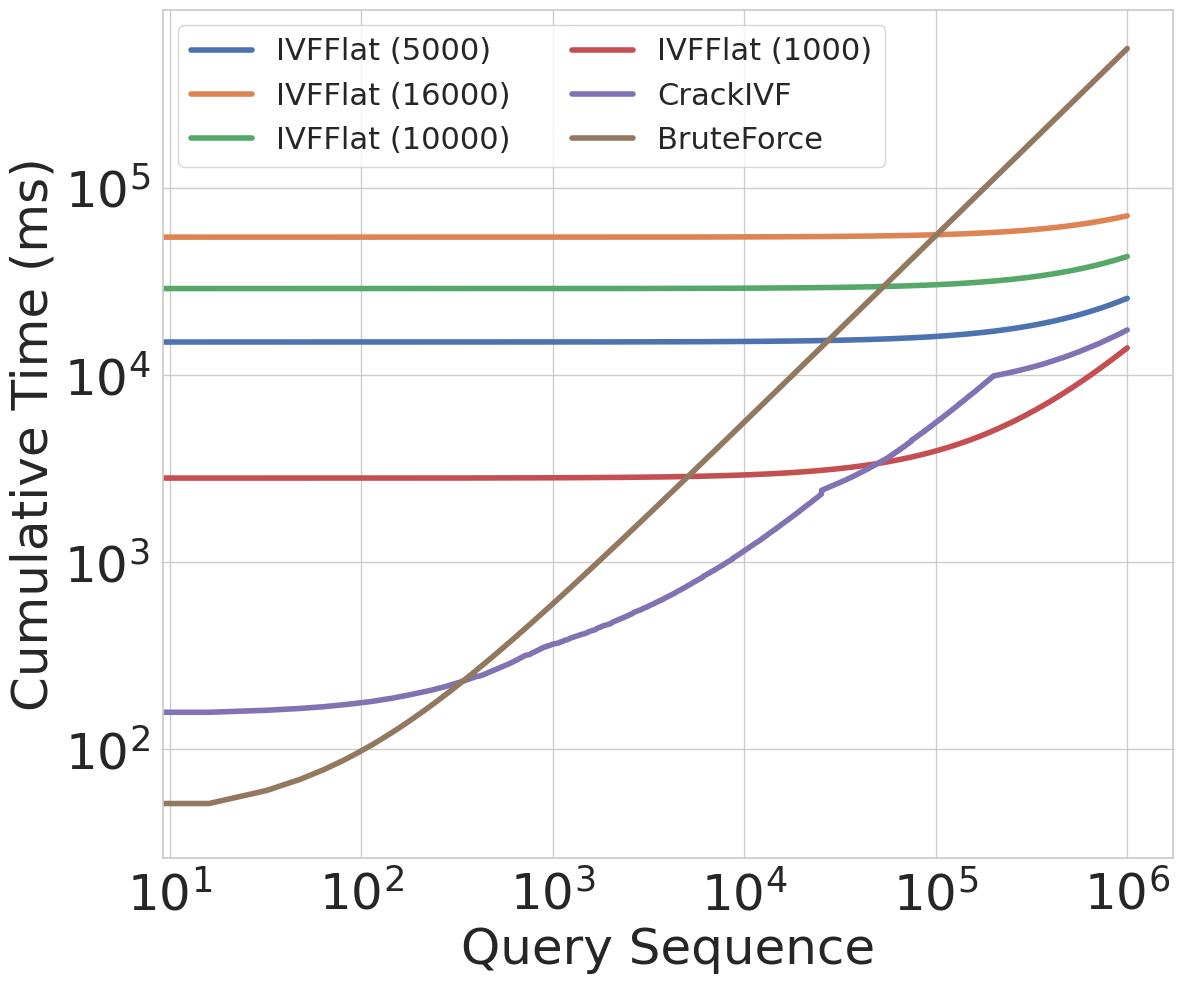}
         \caption{Last.fm}
         \label{fig:lastfm-cu}
    \end{subfigure}
    \hfill
    \begin{subfigure}[b]{0.48\columnwidth}
         \centering
         \includegraphics[width=\textwidth]{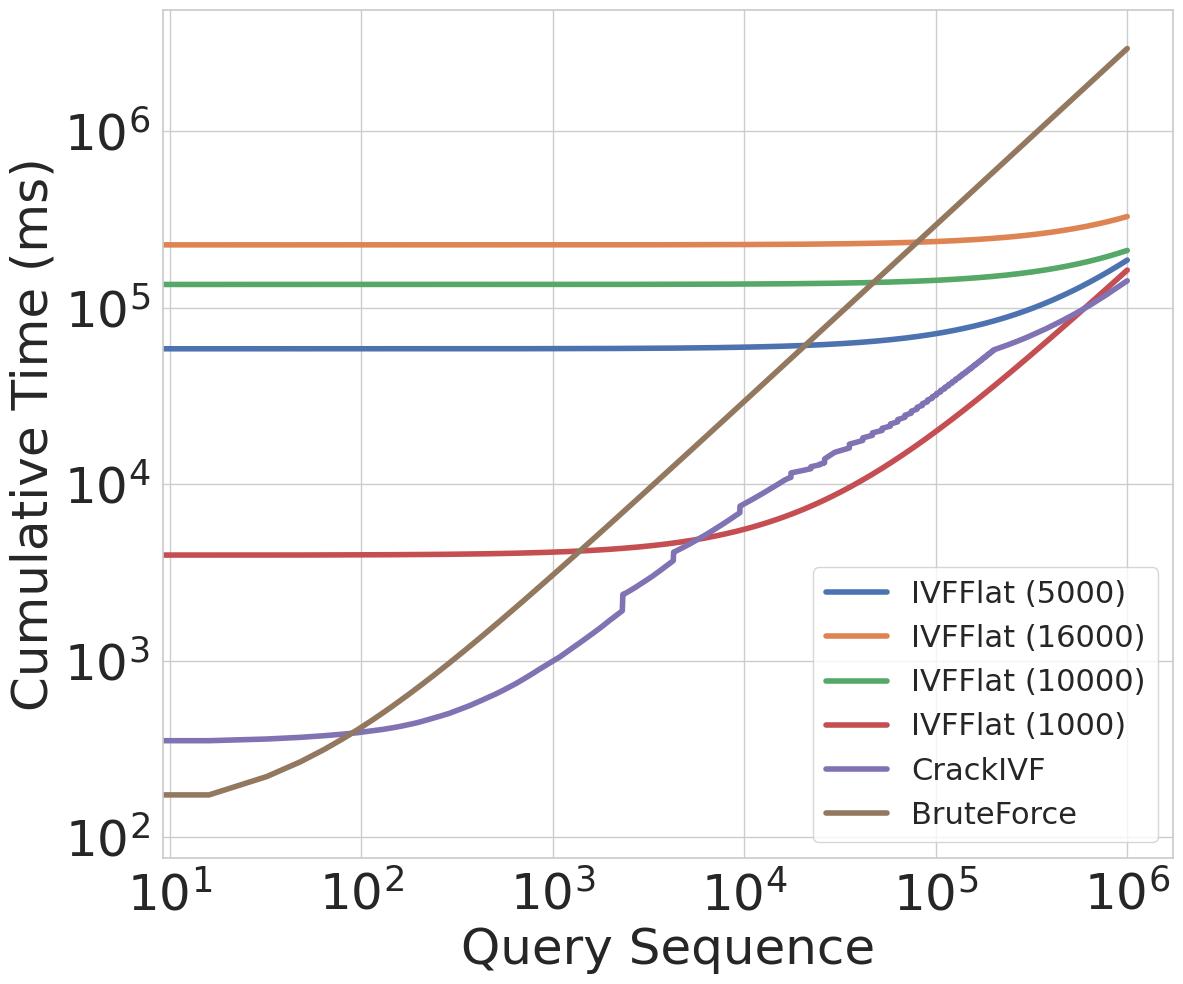}
         \caption{Glove50}
         \label{fig:glove50-cu}
    \end{subfigure}
    
    \vfill
    \begin{subfigure}[b]{0.48\columnwidth}
         \centering
         \includegraphics[width=\textwidth]{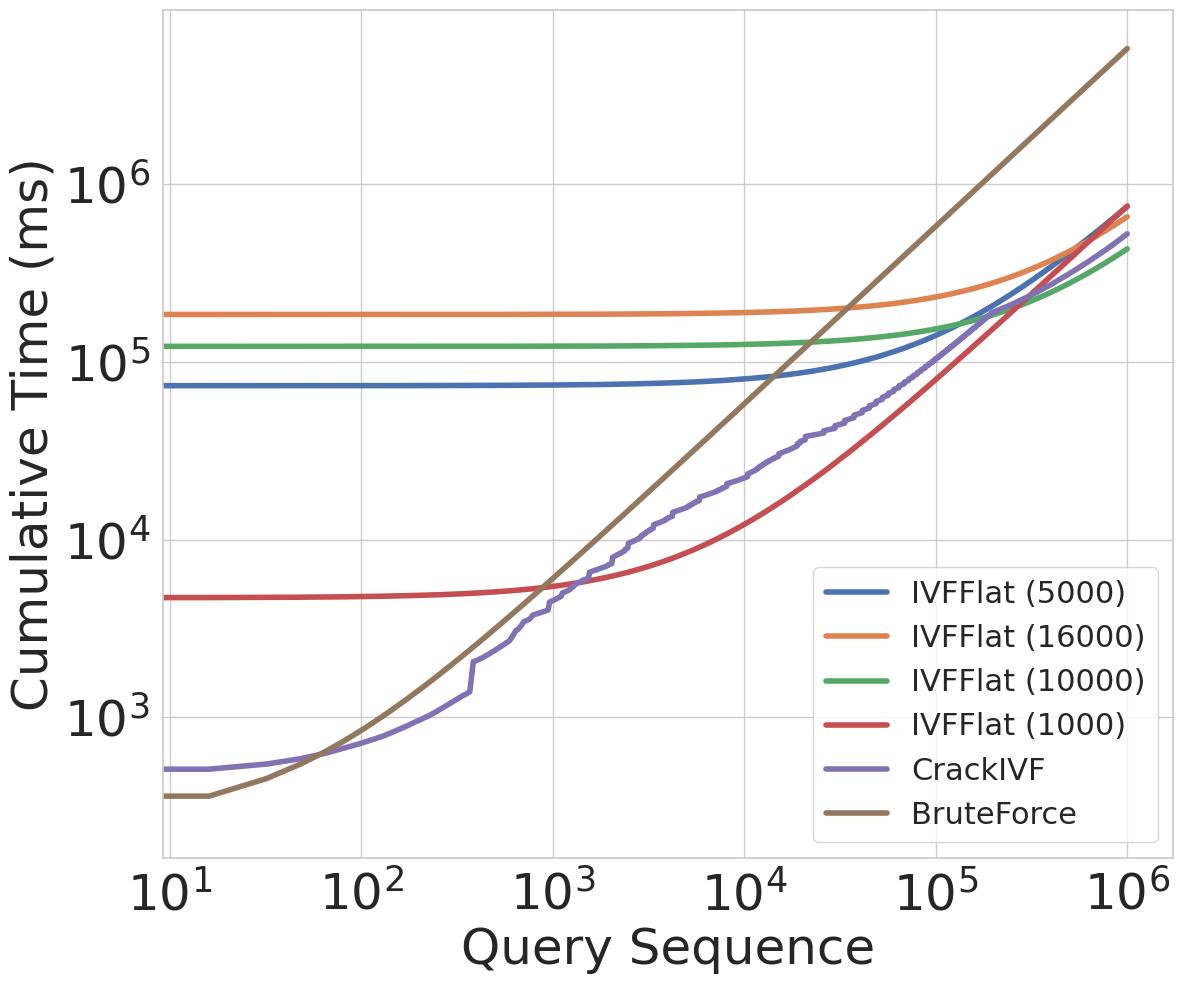}
         \caption{Glove100}
         \label{fig:glove100-cu}
    \end{subfigure}
    \hfill
    \begin{subfigure}[b]{0.48\columnwidth}
         \centering
         \includegraphics[width=\textwidth]{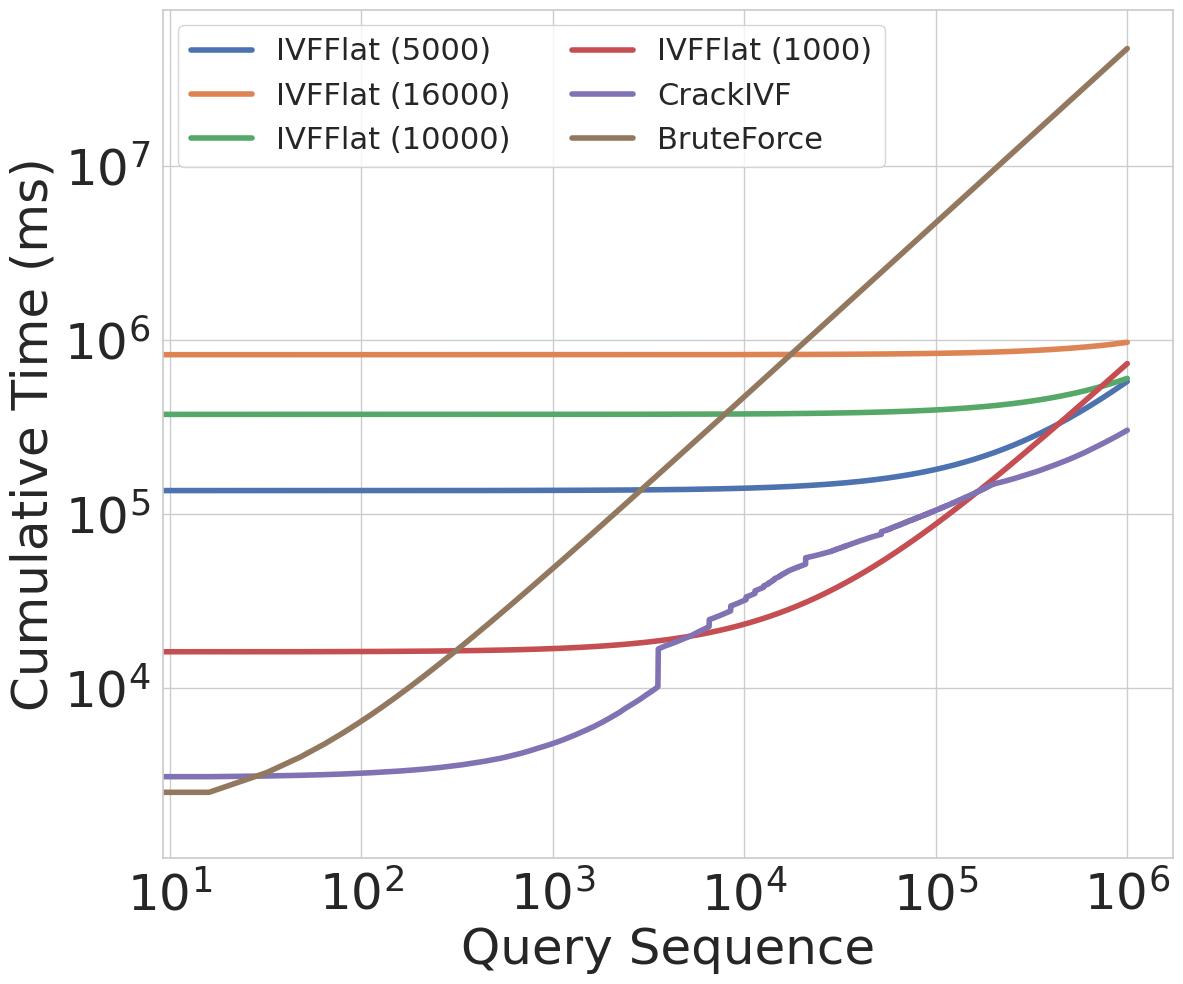}
         \caption{Deep10M}
         \label{fig:deep-cu}
    \end{subfigure}
    
    \caption{Cumulative time plots across datasets.}
    \label{fig:cumulative_times_all_datasets}
\end{figure}

We run a trace of 1 million queries for each dataset and include the upfront build cost for all baselines as the time before they answer the first query (Figure~\ref{fig:cumulative_times_all_datasets}). The smaller 1000 partition IVF indexes have relatively low startup costs but do not scale well past $\approx$10k queries. Larger indexes have a huge upfront build cost, which only pays off if a dataset receives >1 million queries. Brute-force has the minimum initialization time, making it the better choice when the number of queries is low (<100). In our experiments, CrackIVF achieves near-minimum cumulative time across all query scales. It achieves almost 1000x lower initialization time than larger indexes (Figure~\ref{fig:glove50-cu}) and can answer 100K to 1M queries before comparable indexes finish building. For example on SIFT10M (Figure~\ref{fig:sift10m-cu}), the initialization time cost is $\approx$100x lower than IndexFlat(5000) and almost 1000x lower than IndexFlat(16000), yet CrackIVF answers 100k and 1M queries respectively, before the indexes finish building.

\subsection{ Does CrackIVF Reduce Total Distance Computations for Build Operations?}

 IVF indexes incur full indexing cost upfront over the entire vector space. Indexing large datasets (100M–1B+) in reasonable time requires GPUs, but even then, K-means training can take hours due to the increasing number of training points and partitions, which have a multiplicative effect in the number of pairwise distance computations (see Table~\ref{tab:kernel_cost_model} \textit{K-means Compute}). CrackIVF amortizes this cost over time using $REFINE$ operations, which apply K-means locally on small subsets of points and partitions. As a result, total distance computations are spread across time and space and can potentially be handled on cheaper hardware (e.g., CPUs). 

\begin{figure}[t]
    \centering
    \begin{subfigure}[b]{0.48\columnwidth}
         \centering
         \includegraphics[width=\textwidth]{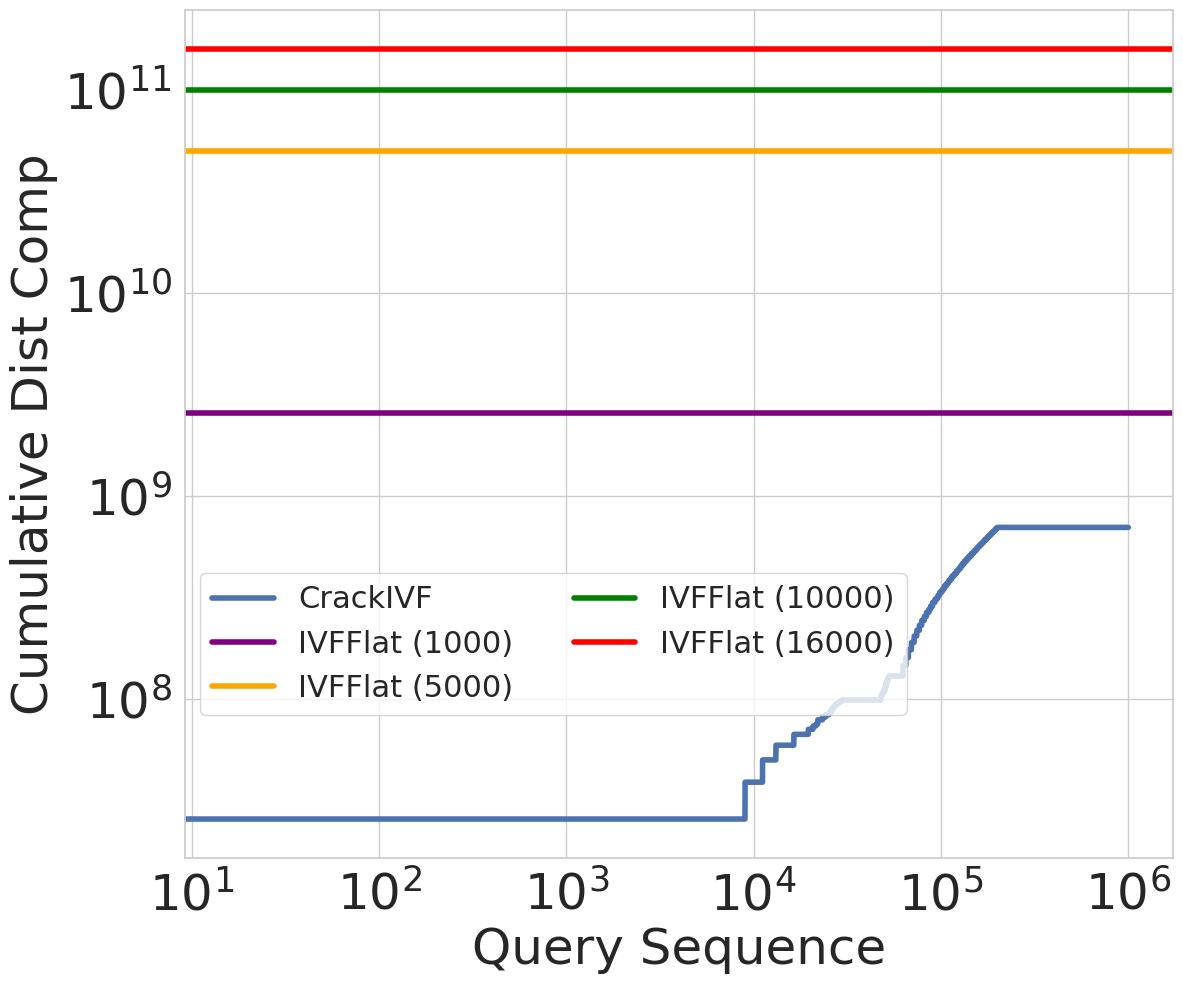}
         \caption{{SIFT1M}}
         \label{fig:sift1m-cdc}
    \end{subfigure}
    \hfill
    \begin{subfigure}[b]{0.48\columnwidth}
         \centering
         \includegraphics[width=\textwidth]{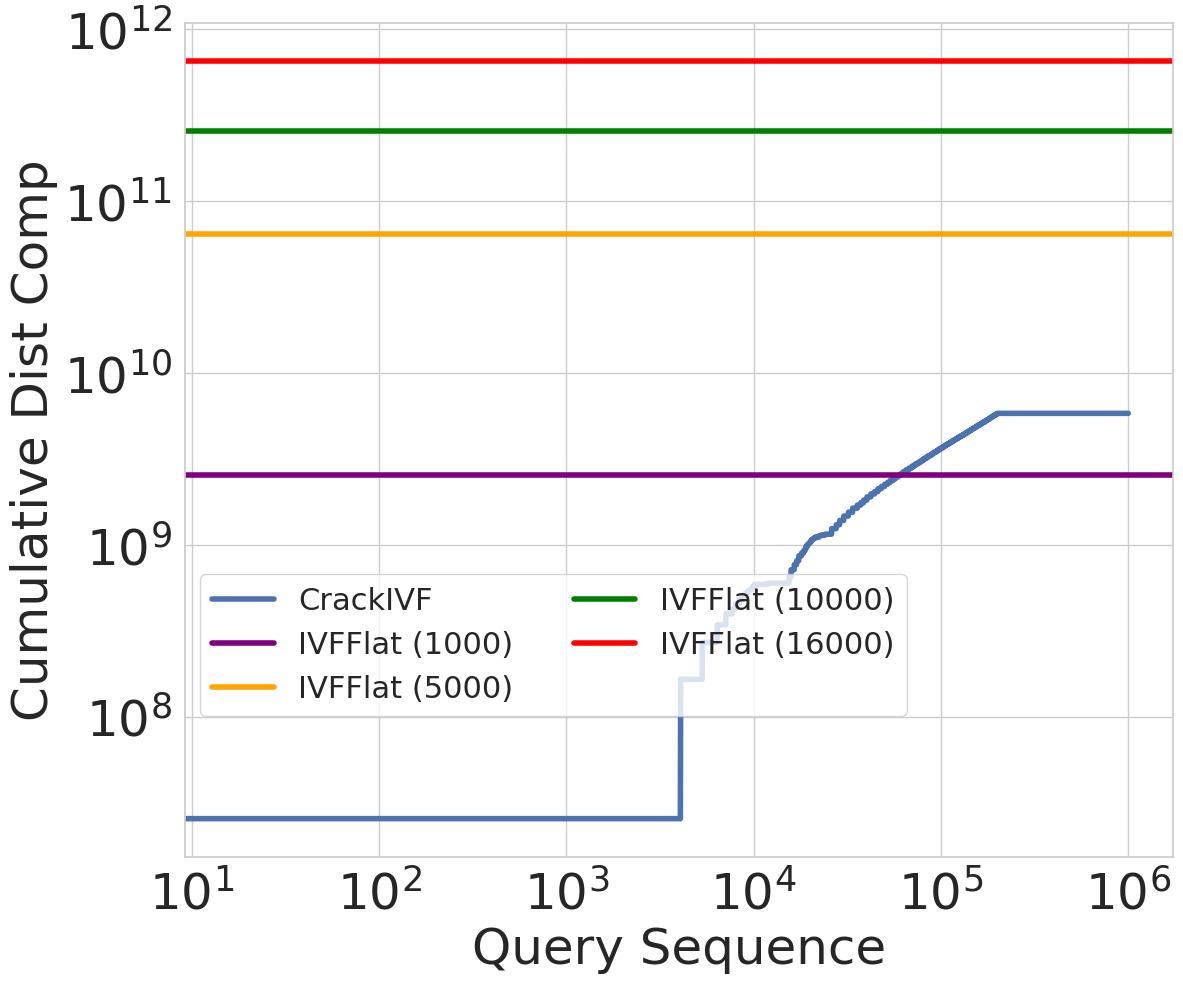}
         \caption{{SIFT10M}}
         \label{fig:sift10m-cdc}
    \end{subfigure}
    
    \vfill
    \begin{subfigure}[b]{0.48\columnwidth}
         \centering
         \includegraphics[width=\textwidth]{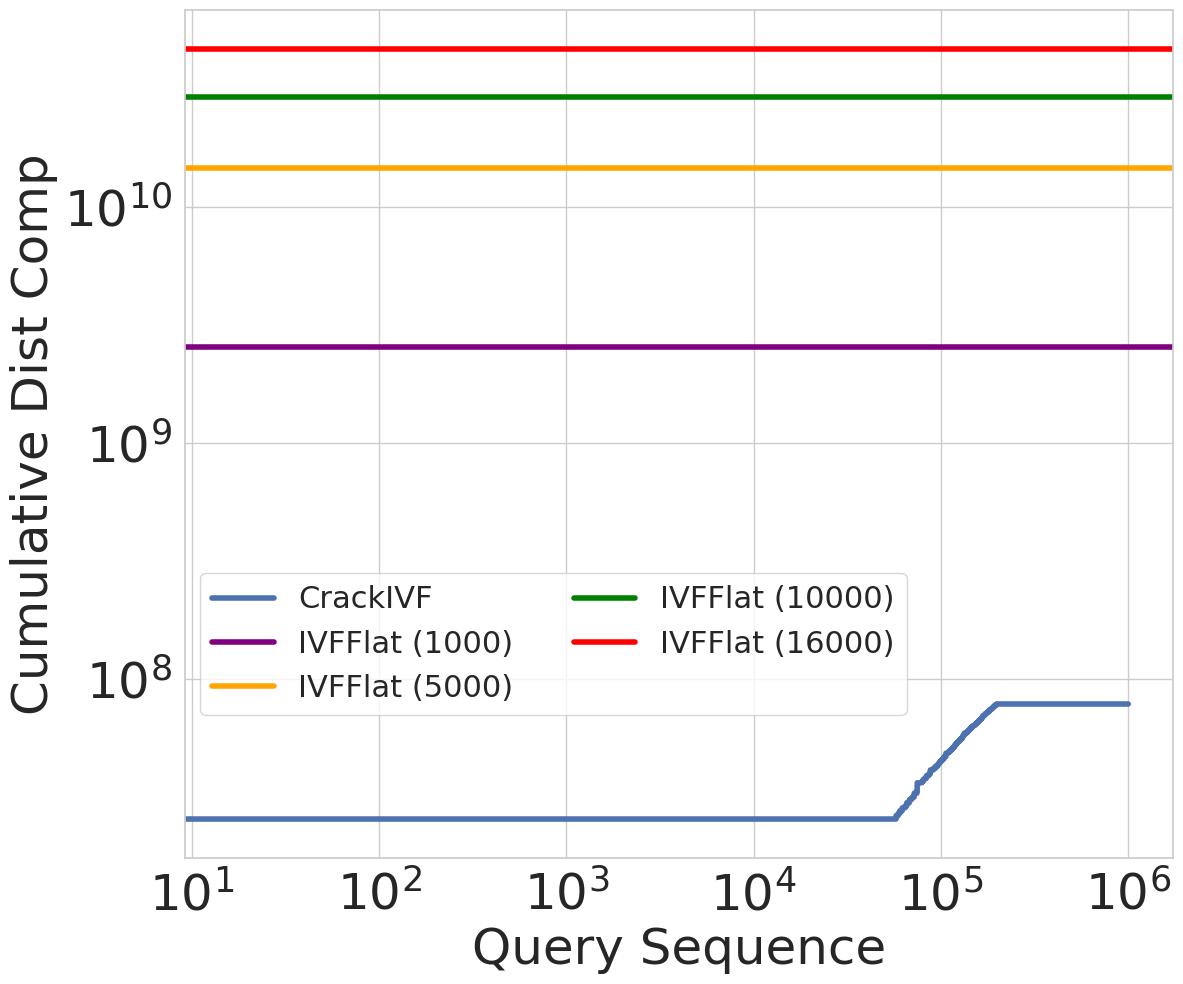}
         \caption{{Last.fm}}
         \label{fig:lastfm-cdc}
    \end{subfigure}
    \hfill
    \begin{subfigure}[b]{0.48\columnwidth}
         \centering
         \includegraphics[width=\textwidth]{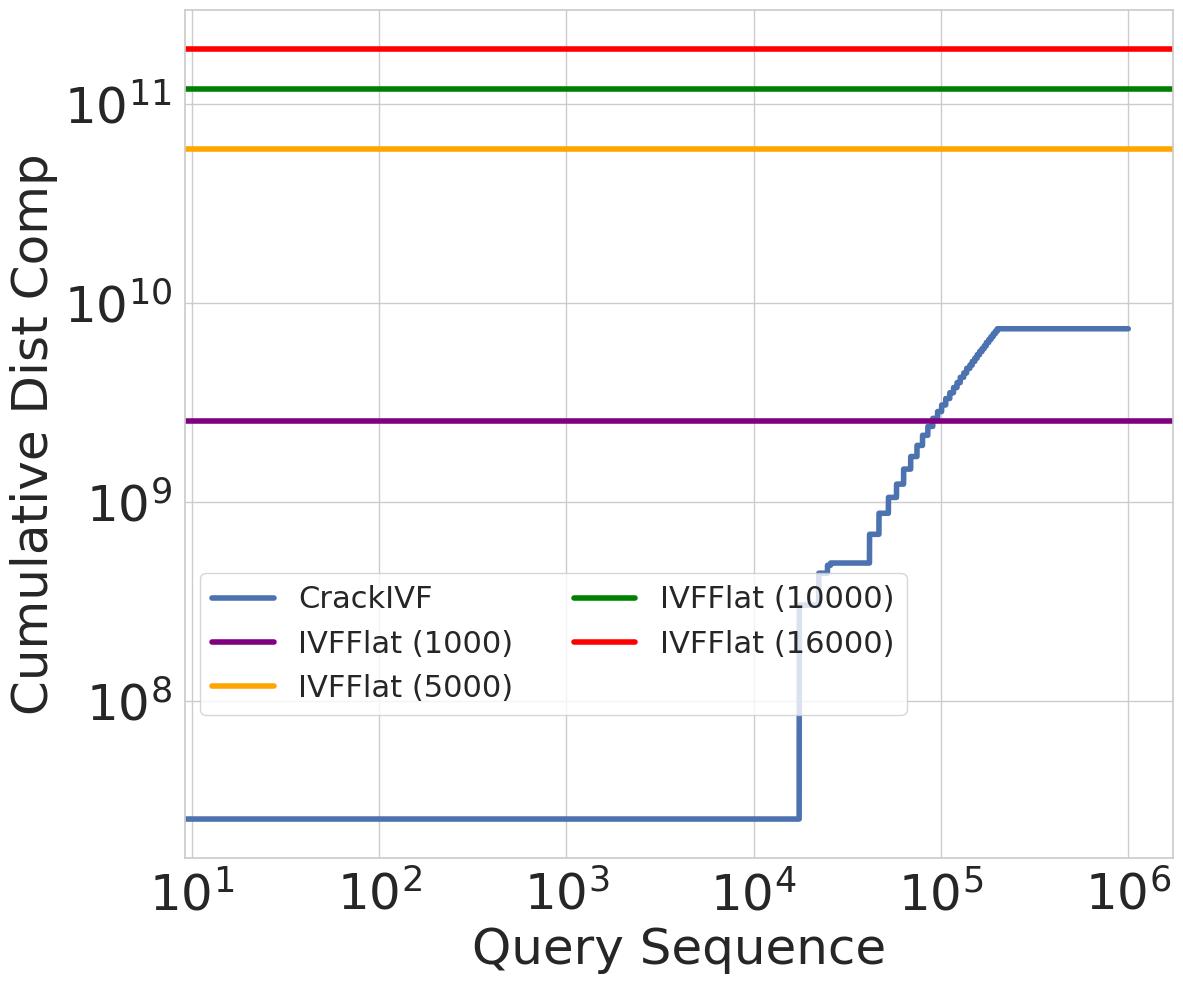}
         \caption{{Glove50}}
         \label{fig:glove50-cdc}
    \end{subfigure}
    
    \vfill
    \begin{subfigure}[b]{0.48\columnwidth}
         \centering
         \includegraphics[width=\textwidth]{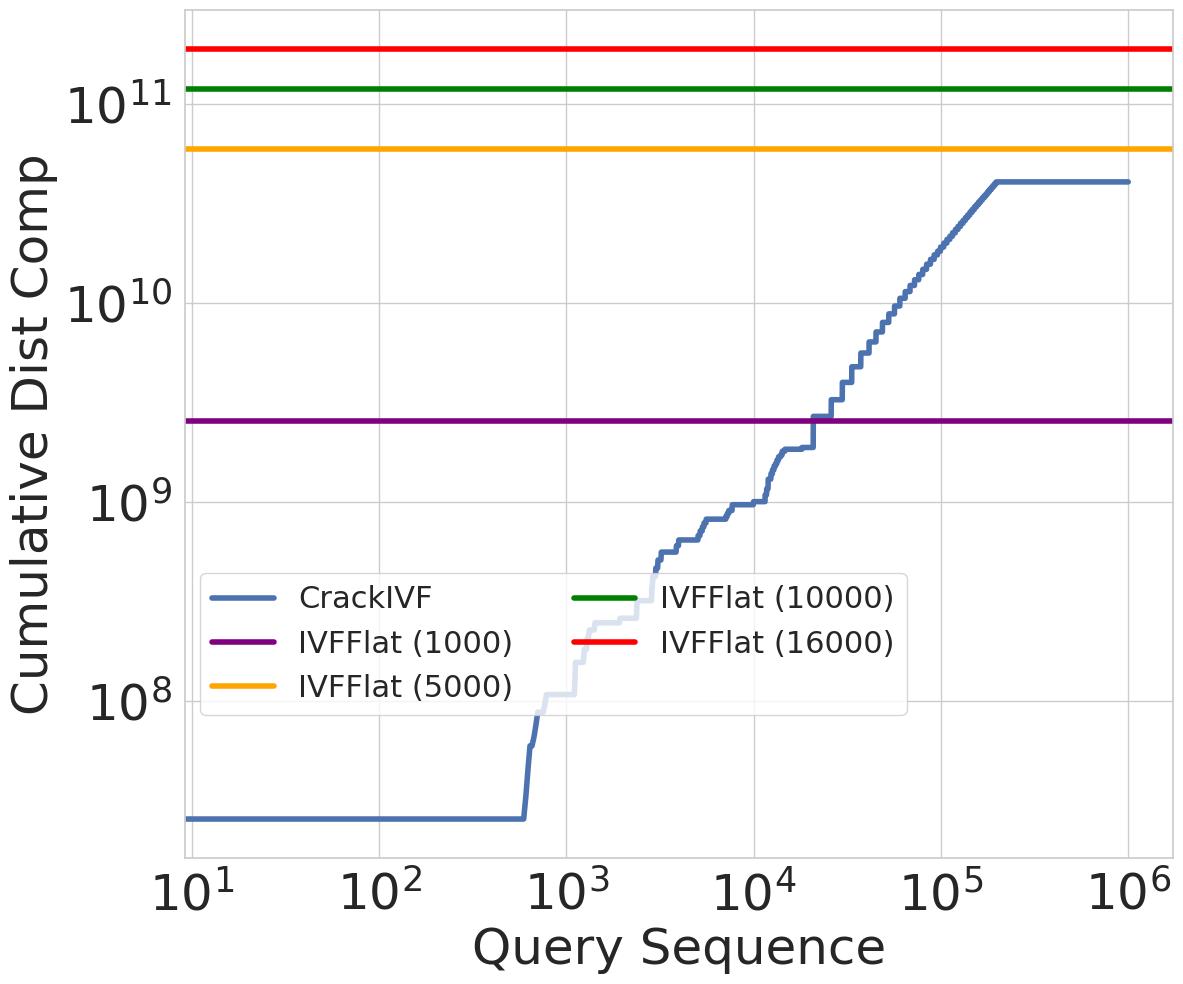}
         \caption{{Glove100}}
         \label{fig:glove100-cdc}
    \end{subfigure}
    \hfill
    \begin{subfigure}[b]{0.48\columnwidth}
         \centering
         \includegraphics[width=\textwidth]{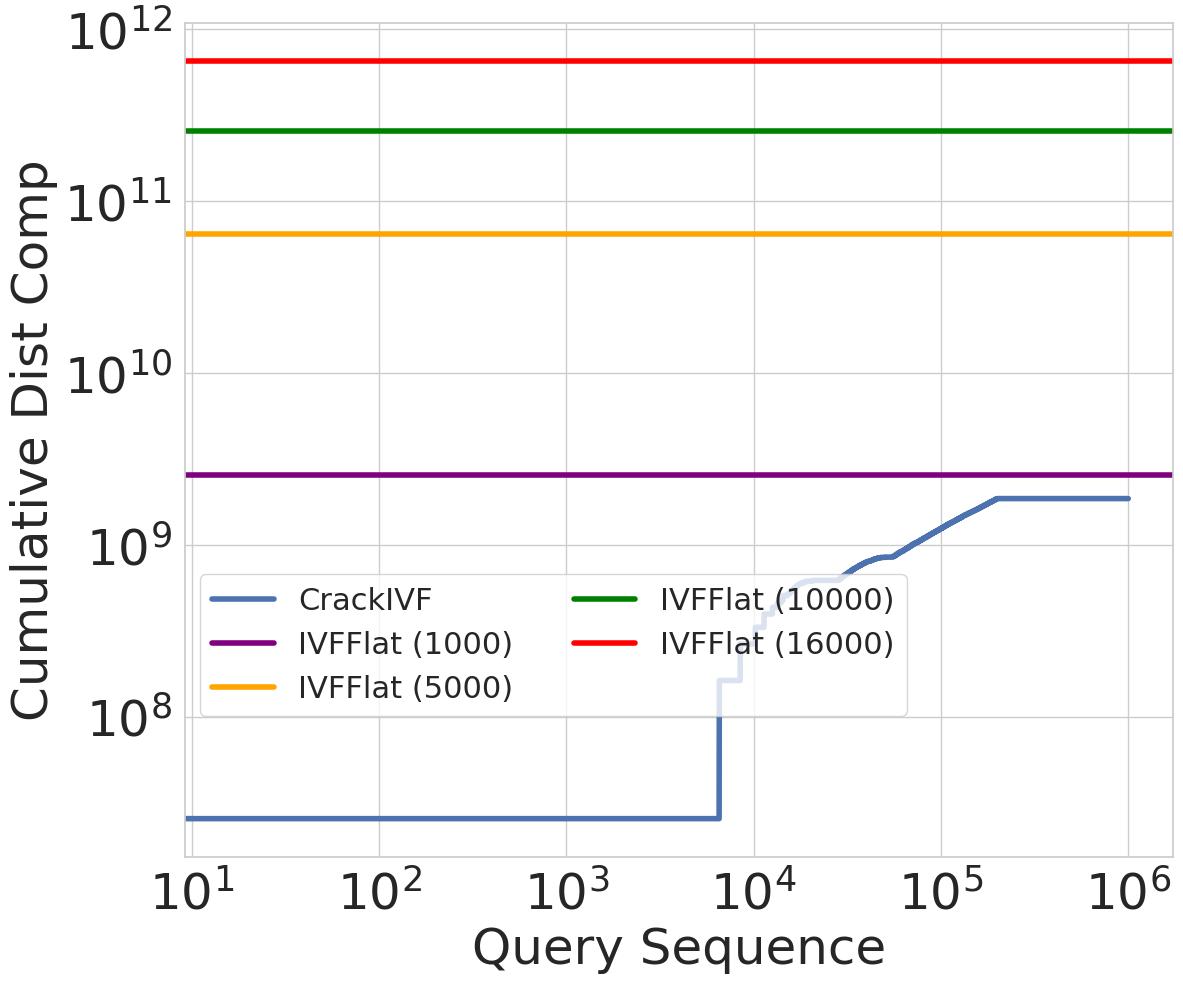}
         \caption{{Deep10M}}
         \label{fig:deep10m-cdc}
    \end{subfigure}
    
    \caption{{Cumulative distance computations for index build.}}
    \label{fig:cum_dist_comp}
\end{figure}

Figure~\ref{fig:cum_dist_comp} shows the cumulative distance computations during index build, excluding search operations. CrackIVF pays this cost gradually, and after the index converges, the final distance computations can be orders of magnitude lower than larger IVF indexes (Figure~\ref{fig:sift1m-cdc},~\ref{fig:deep10m-cdc}). Despite having comparable or better performance (Figure~\ref{fig:sift1m-qps},~\ref{fig:deep-qps}). Convergence occurs when the arrival of new crack candidates stops and refines do not trigger on new regions, example is shown in ablation. On Glove, distance computations increase faster due to $REFINE$ regions containing more points and more cracks. The median $REFINE$ in Glove100 covers $4.78$\% of the dataset with 512 cracks, while the median $REFINE$ in SIFT1M is only $0.63\%$ of the total points across 64 cracks. Nonetheless, the total cost remains amortized over time and is still almost an order of magnitude lower than that of the largest IVF index, while achieving similar final QPS-Recall performance. Finally, \textit{CRACK} is dominated by data movement, while \textit{REFINE} is mostly compute (Table~\ref{tab:kernel_cost_model}). Even though \textit{CRACK} is infrequent ($<0.03\%$ of trace Figure~\ref{fig:qps_recall_and_response_time_comparison}), its overhead is not fully captured by distance computations alone. This is why we use fixed hardware and cumulative time plots, to compare the total time across all operations, including search, in Figure~\ref{fig:cumulative_times_all_datasets}.

\begin{figure}[t]
    \centering
    \begin{subfigure}[b]{0.48\columnwidth}
         \centering
         \includegraphics[width=\textwidth]{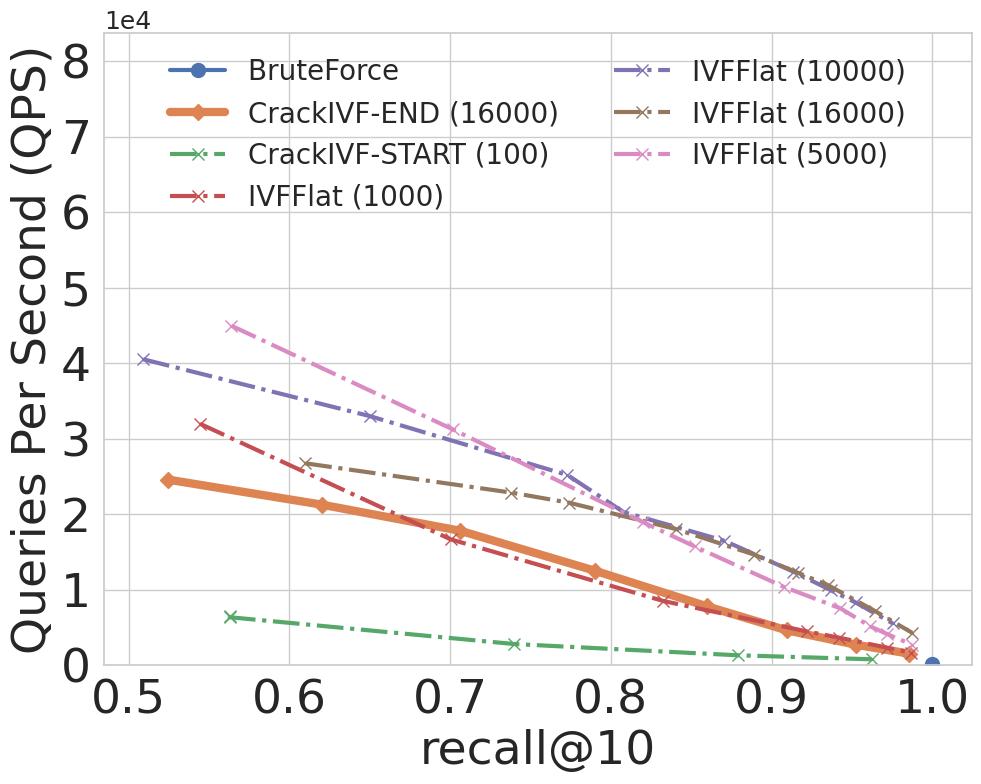}
         \caption{WHERE=OFF}
         \label{fig:where_off_qps}
    \end{subfigure}
    \hfill
    \begin{subfigure}[b]{0.48\columnwidth}
         \centering
         \includegraphics[width=\textwidth]{figures/exp/plots_SIFT1M/qps_recall.jpg}
         \caption{WHERE=ON}
         \label{fig:where_on_qps}
    \end{subfigure}
    \caption{Effect of removing the ``where'' control mechanism}    
    \label{fig:where_on_off}
\end{figure}

\begin{figure}[t]
    \centering
    
    \begin{subfigure}[b]{0.48\columnwidth}
         \centering
         \includegraphics[width=\textwidth]{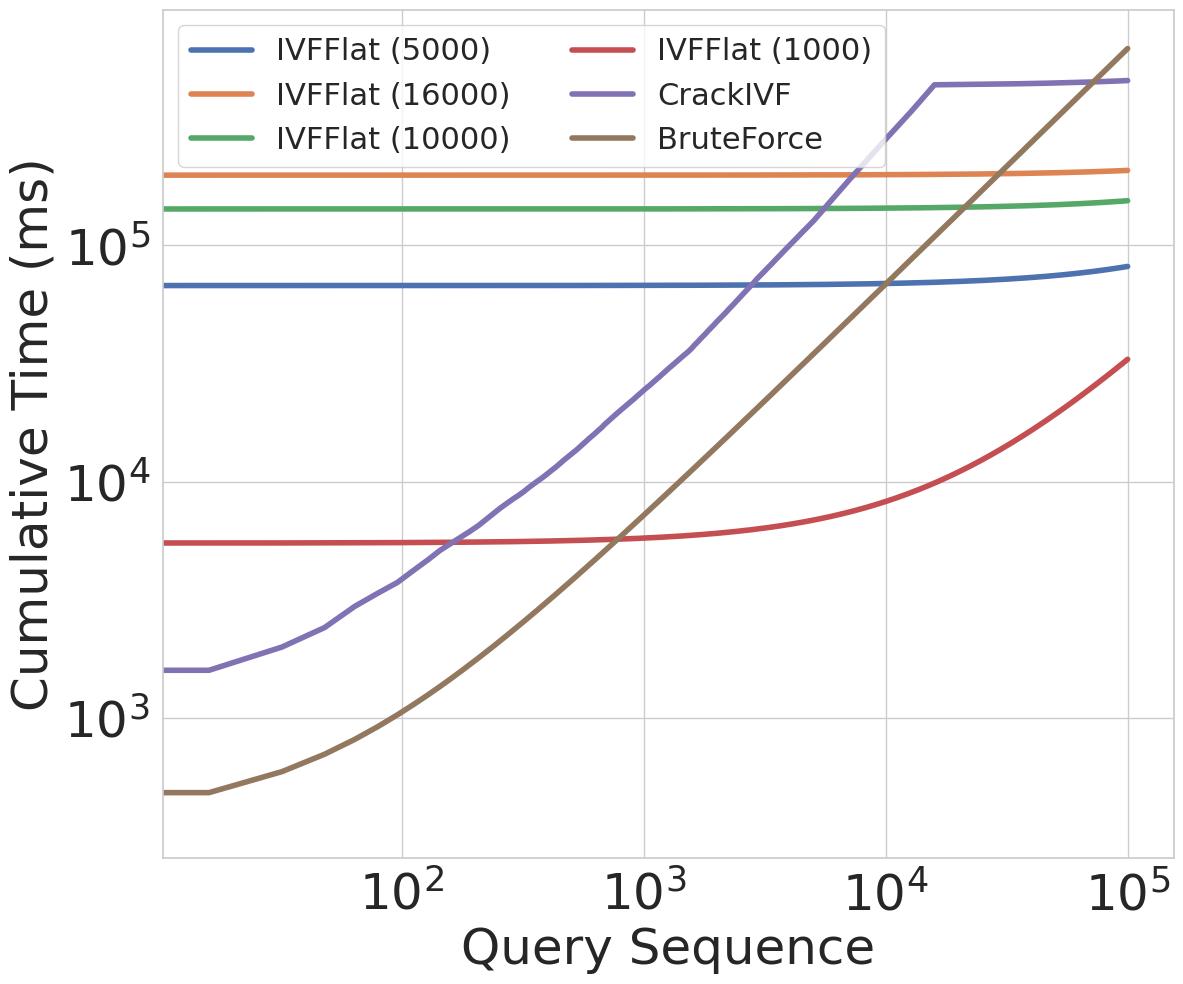}
         \caption{Cumulative Time (OFF)}
         \label{fig:when_off_cumul}
    \end{subfigure}
    \hfill
    \begin{subfigure}[b]{0.48\columnwidth}
         \centering
         \includegraphics[width=\textwidth]{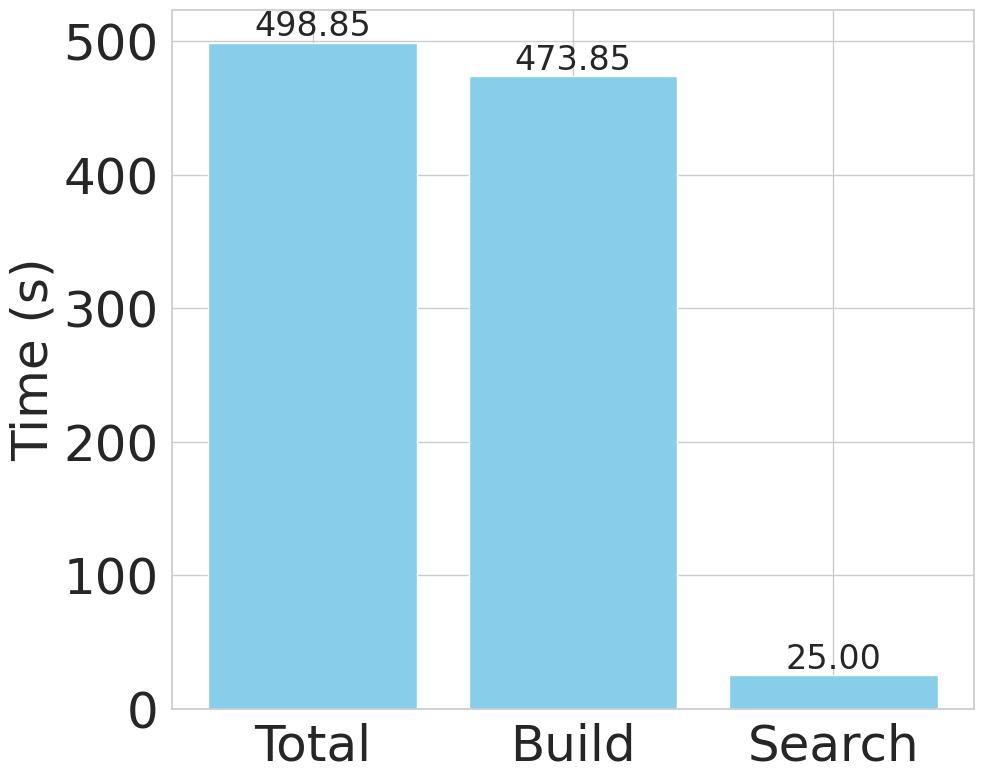}
         \caption{Time Breakdown (OFF)}
         \label{fig:when_off_total_time}
    \end{subfigure}

    \vskip\baselineskip  

    \begin{subfigure}[b]{0.48\columnwidth}
         \centering
         \includegraphics[width=\textwidth]{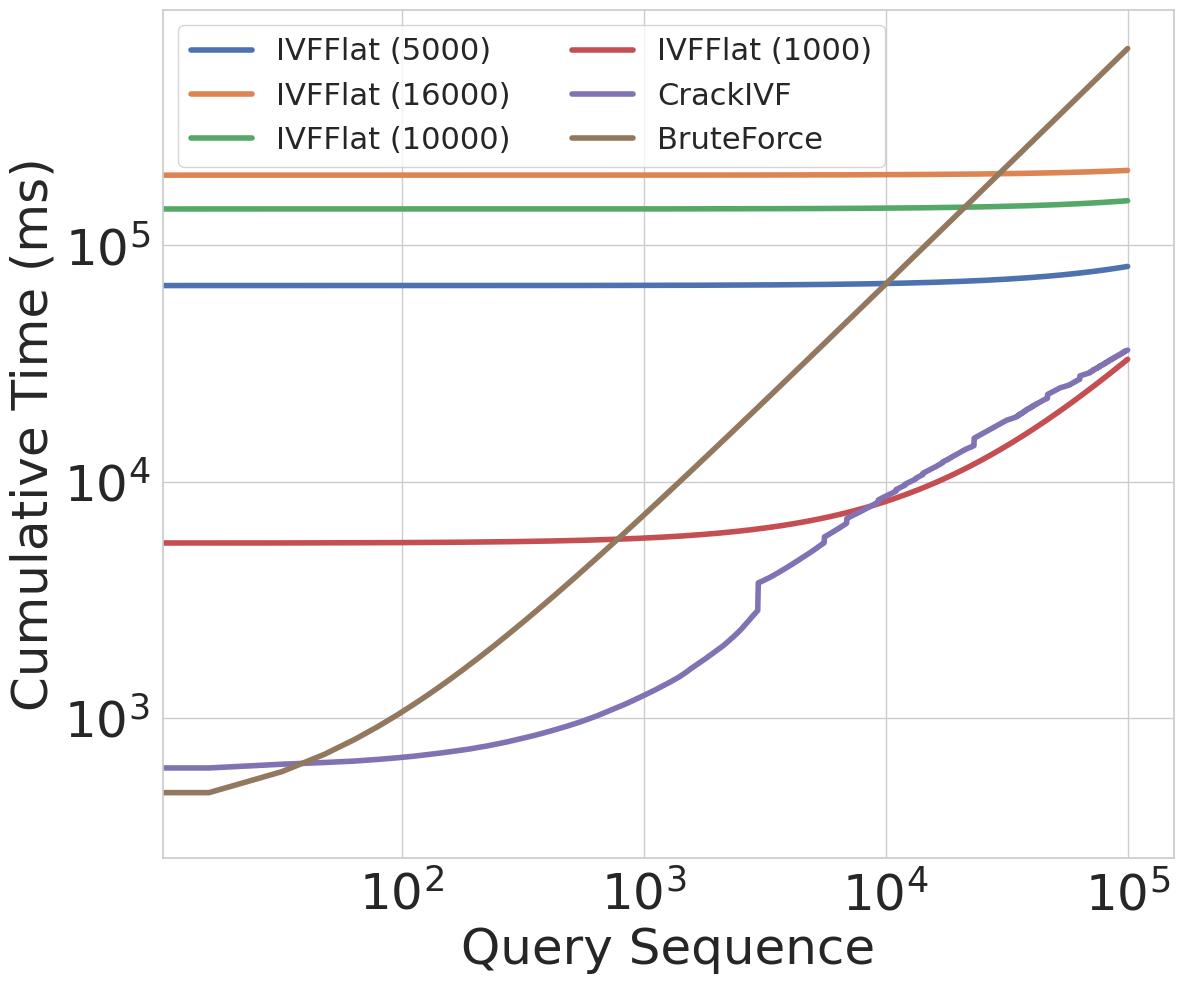}
         \caption{Cumulative Time (ON)}
         \label{fig:when_on_cumul}
    \end{subfigure}
    \hfill
    \begin{subfigure}[b]{0.48\columnwidth}
         \centering
         \includegraphics[width=\textwidth]{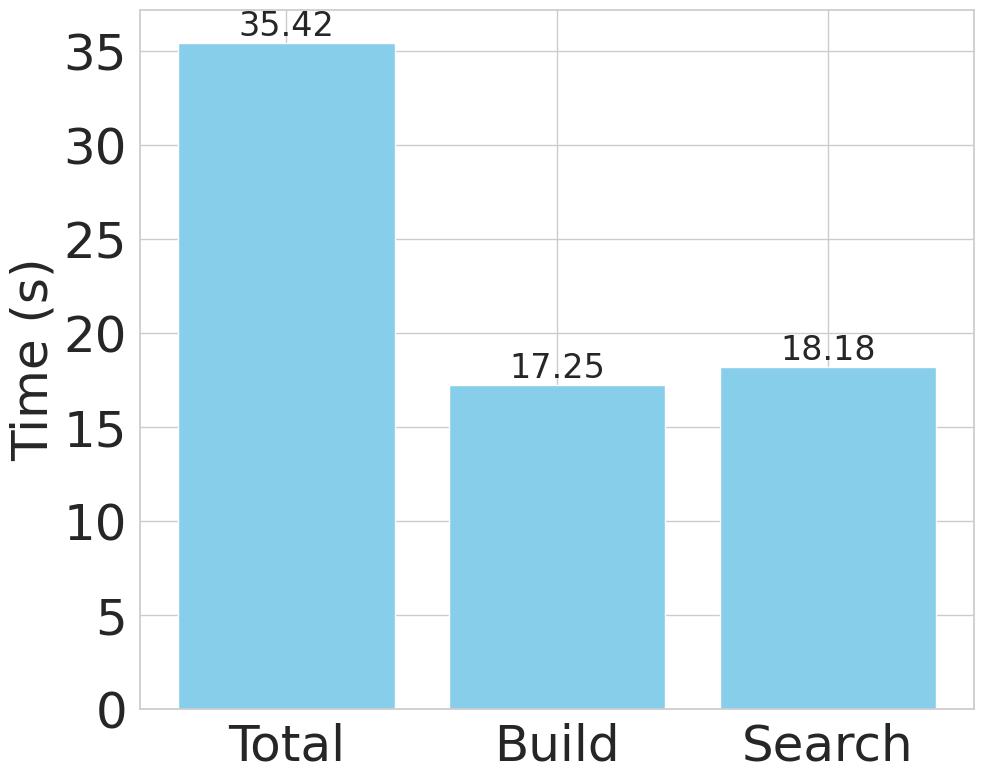}
         \caption{Time Breakdown (ON)}
         \label{fig:when_on_total_time}
    \end{subfigure}

    \caption{Effect of removing the ``when'' control mechanism}
    \label{fig:when_on_off}
\end{figure}

\subsection{Control Mechanisms Ablation Study}

This section is an ablation study exploring the control mechanisms ``where'' and ``when'' to apply $CRACK$ and $REFINE$. 

\textbf{Turning off the control mechanism for ``where'':} To illustrate the importance and accuracy of the heuristic rules for where to \textit{CRACK} and \textit{REFINE}, we switch them OFF and see the effect (Figure~\ref{fig:where_on_off}). Specifically, when ``WHERE=OFF'', and no heuristic rule is used to decide where a crack or refine should happen, the index defaults to trying to apply these operations whenever there is enough budget for them, since the ``when'' mechanism is still ON. This seemingly random addition of new cracks leads to a final performance that is only slightly better than the initial state of CrackIVF, before any query has been received (Figure~\ref{fig:where_off_qps}). On the other hand, when ``WHERE=ON'', and our heuristic rules are applied to make decisions, CrackIVF converges to the Pareto optimal QPS-Recall trade-off performance  (Figure~\ref{fig:where_on_qps}). In both examples, we test an equal number of queries and in the same order.

\begin{figure}[t]
    \centering

    \begin{subfigure}[t]{0.48\columnwidth}
        \centering
        \includegraphics[width=\textwidth]{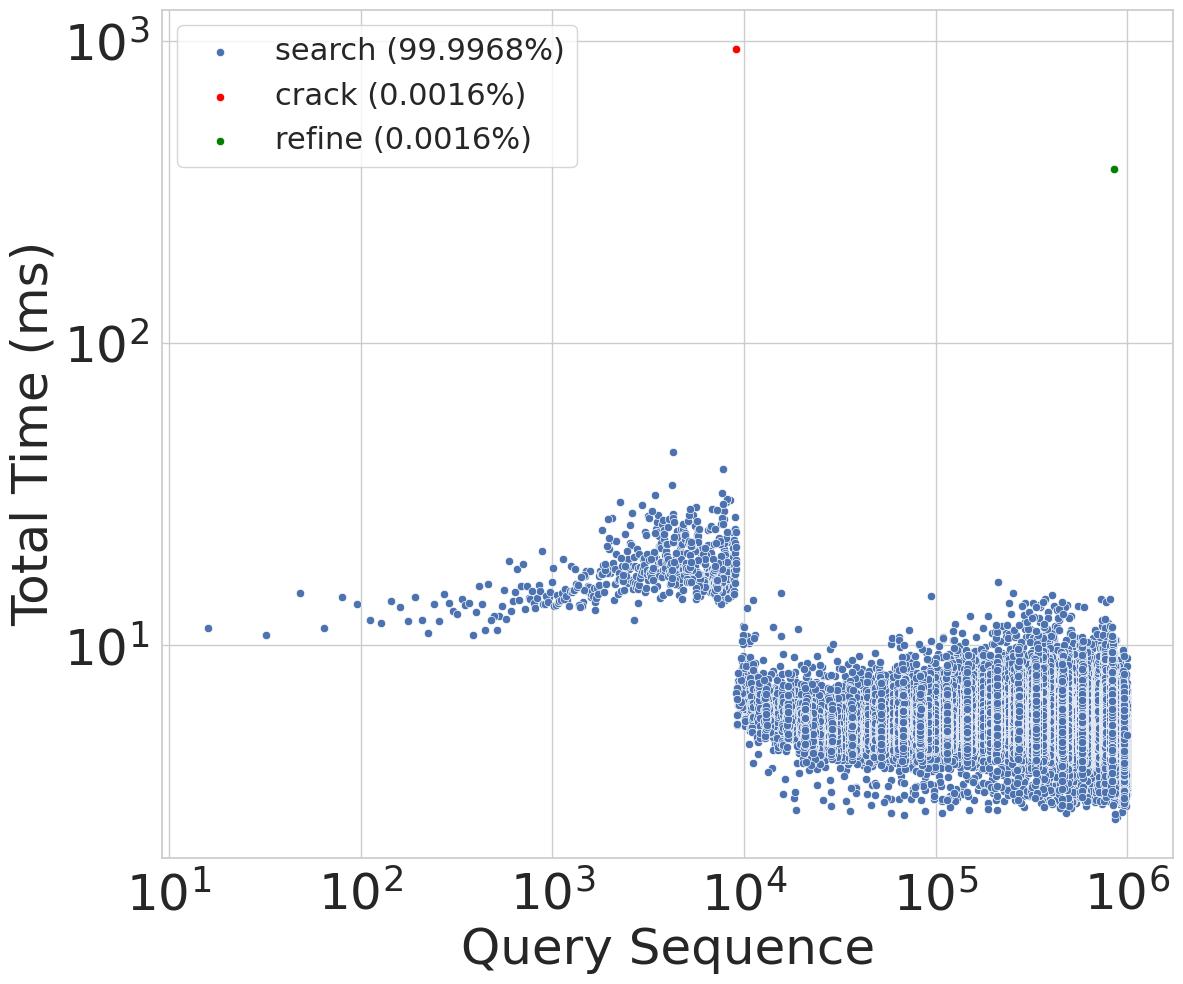}
        \caption{{SIFT1M $\alpha=0.25$}}
        \label{fig:a25}
    \end{subfigure}%
    \hfill
    \begin{subfigure}[t]{0.48\columnwidth}
        \centering
        \includegraphics[width=\textwidth]{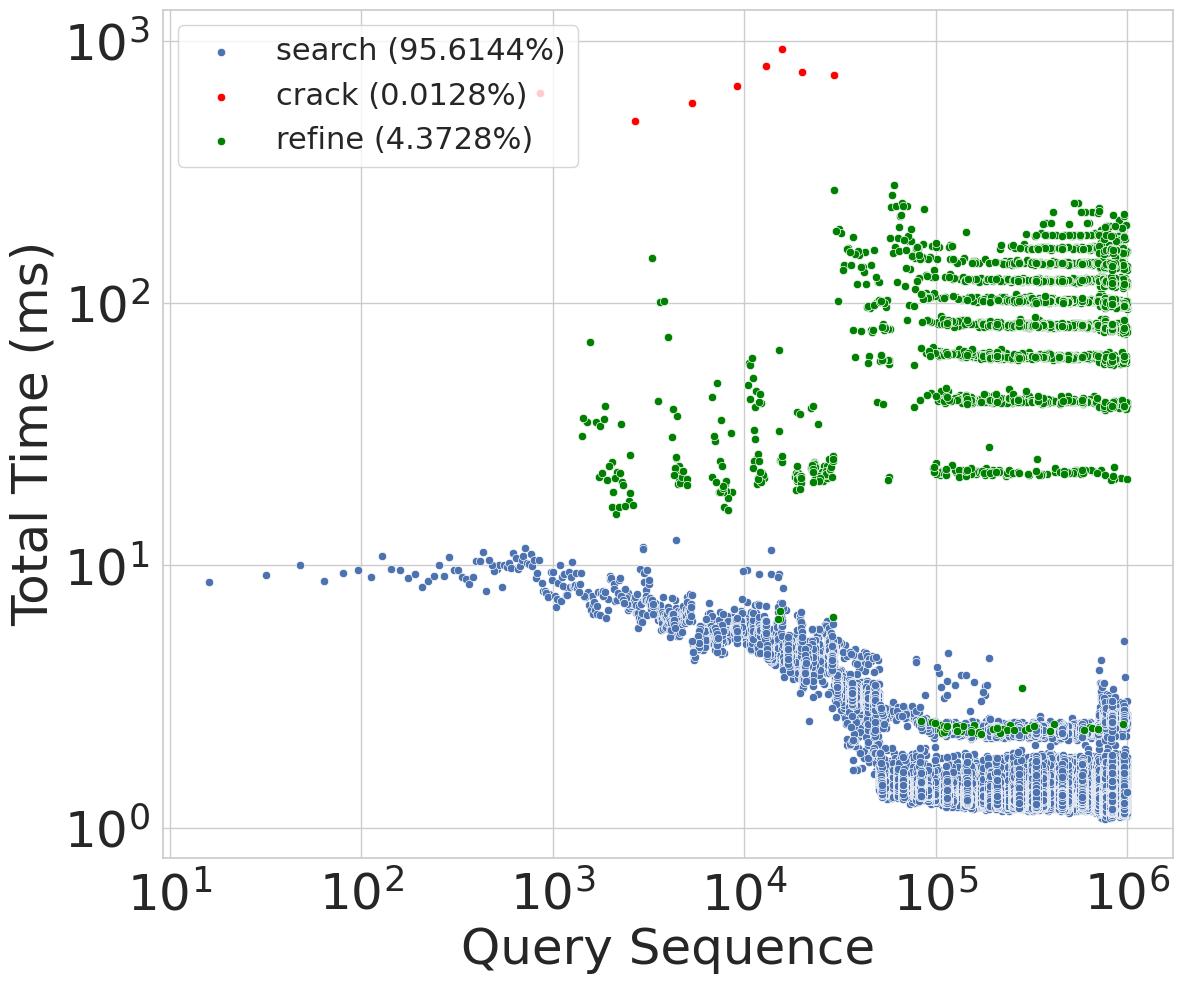}
        \caption{{SIFT1M $\alpha=0.75$}}
        \label{fig:a75}
    \end{subfigure}

    \begin{subfigure}[t]{0.48\columnwidth}
         \centering
         \includegraphics[width=\textwidth]{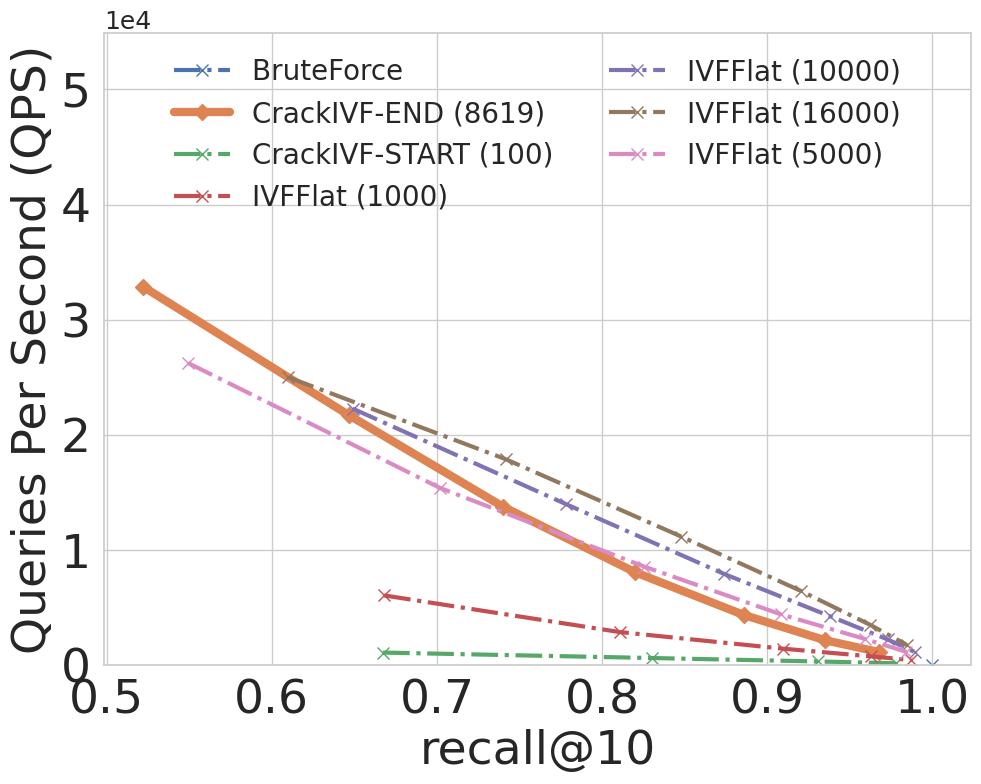}
         \caption{{Deep10M $min\_pts=32$}}
         \label{fig:deep10m-minpts-32}
    \end{subfigure}%
    \hfill
    \begin{subfigure}[t]{0.48\columnwidth}
         \centering
         \includegraphics[width=\textwidth]{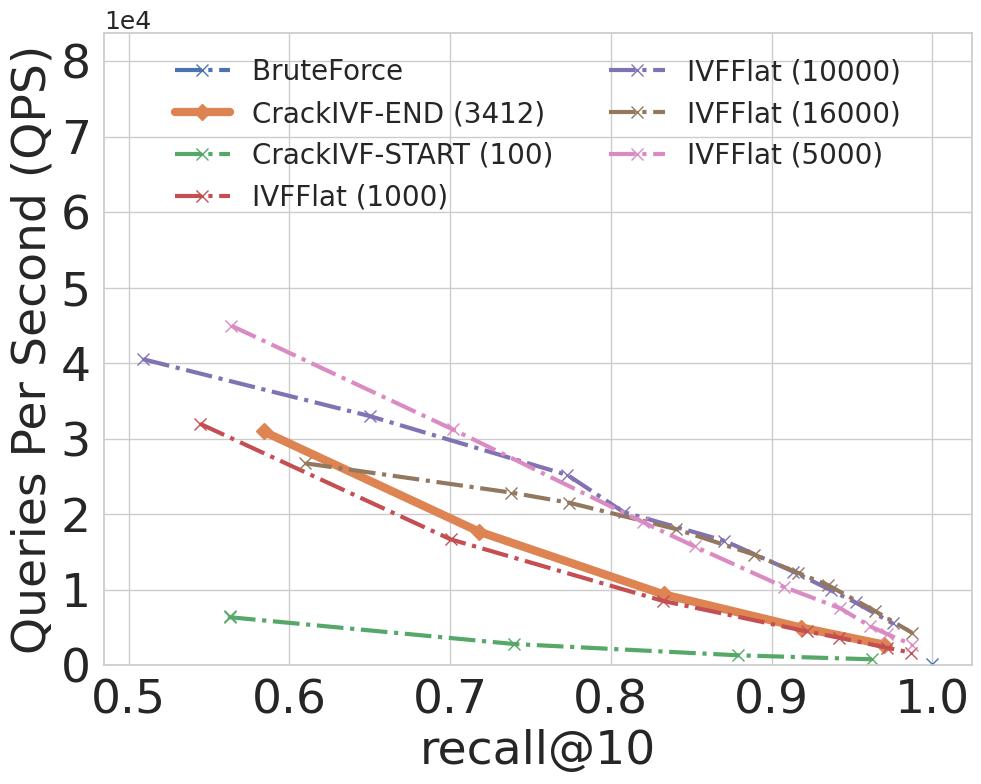}
         \caption{{SIFT1M $min\_pts=32$}}
         \label{fig:sift1m-minpts-32}
    \end{subfigure}

    \begin{subfigure}[t]{0.48\columnwidth}
         \centering
         \includegraphics[width=\textwidth]{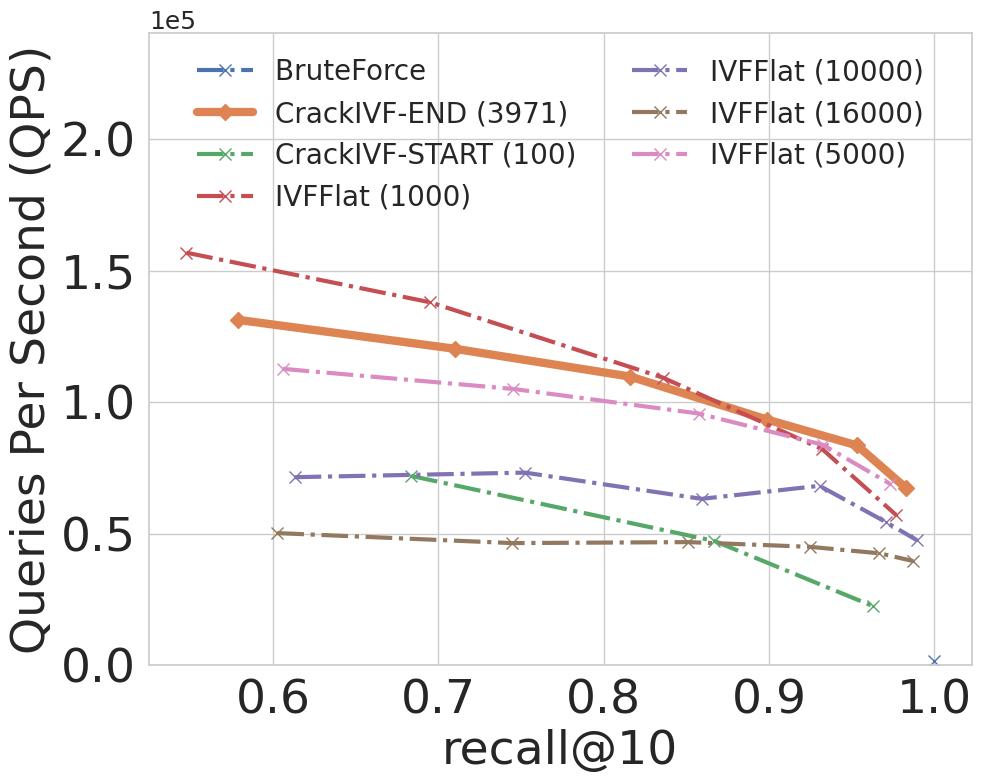}
         \caption{{lastfm $pts\_crack\_thr=16$}}
         \label{fig:lastfm-ptscrackthr-16}
    \end{subfigure}%
    \hfill
    \begin{subfigure}[t]{0.48\columnwidth}
         \centering
         \includegraphics[width=\textwidth]{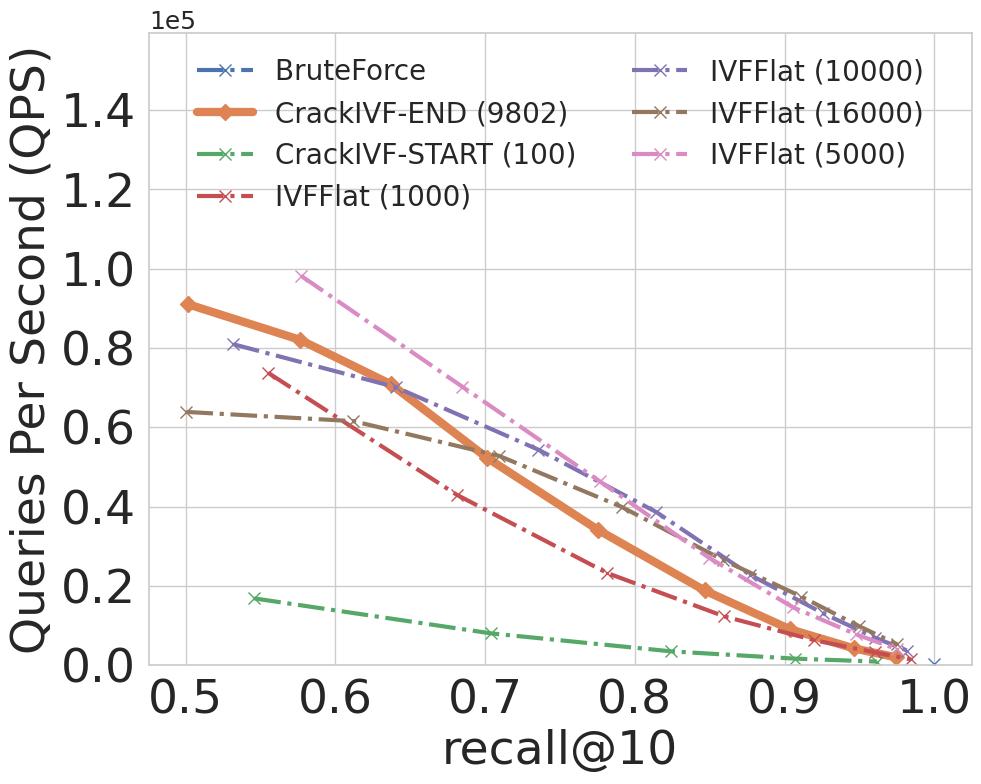}
         \caption{{Glove50 $pts\_crack\_thr=128$}}
         \label{fig:glove50-ptscrackthr-128}
    \end{subfigure}

    \begin{subfigure}[t]{0.48\columnwidth}
         \centering
         \includegraphics[width=\textwidth]{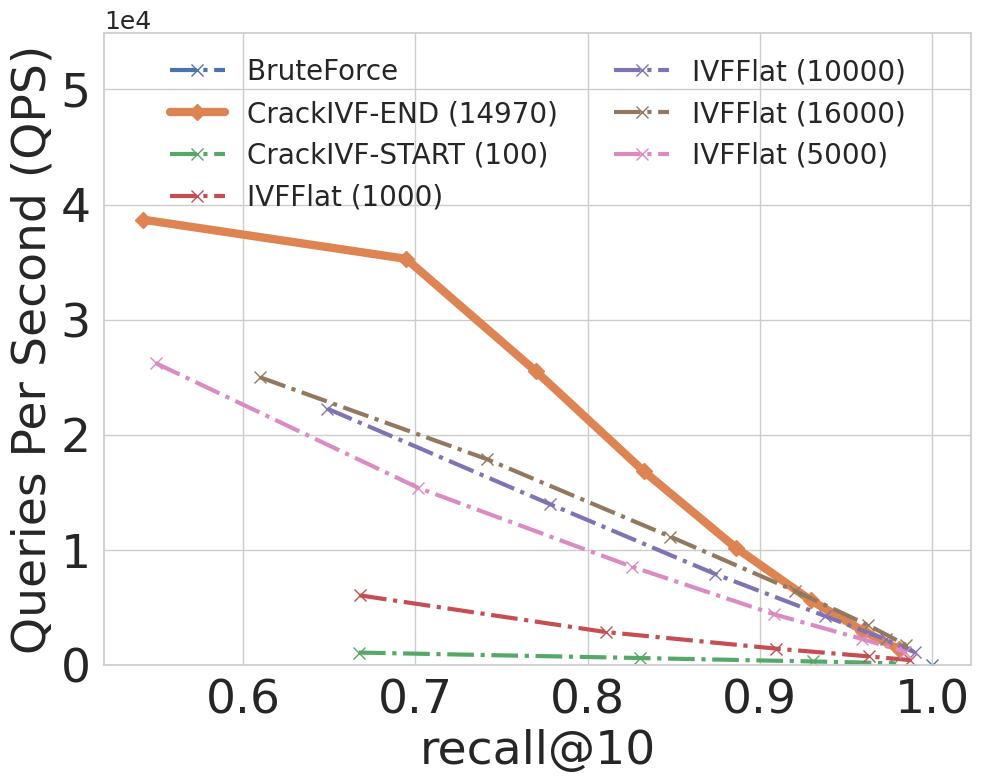}
         \caption{{Deep10M $cv\_max=8$}}
         \label{fig:deep10m-cvmax-8}
    \end{subfigure}%
    \hfill
    \begin{subfigure}[t]{0.48\columnwidth}
         \centering
         \includegraphics[width=\textwidth]{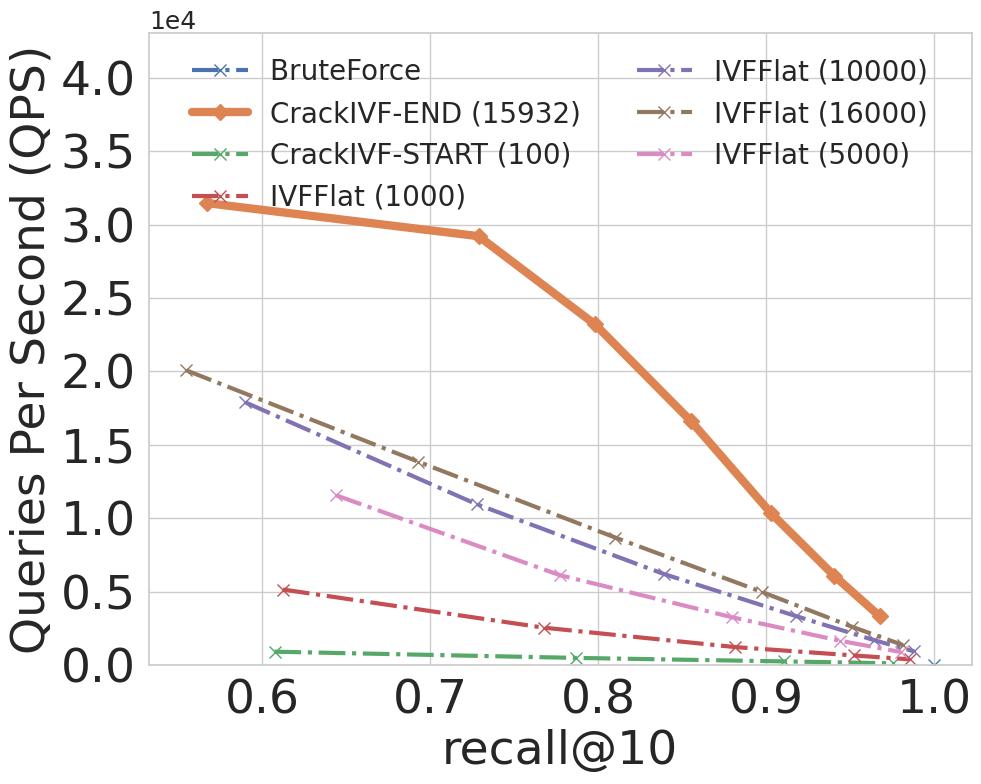}
         \caption{{SIFT10M $cv\_max=8$}}
         \label{fig:sift10m-cvmax-8}
    \end{subfigure}

    \begin{subfigure}[t]{0.48\columnwidth}
        \centering
        \includegraphics[width=\textwidth]{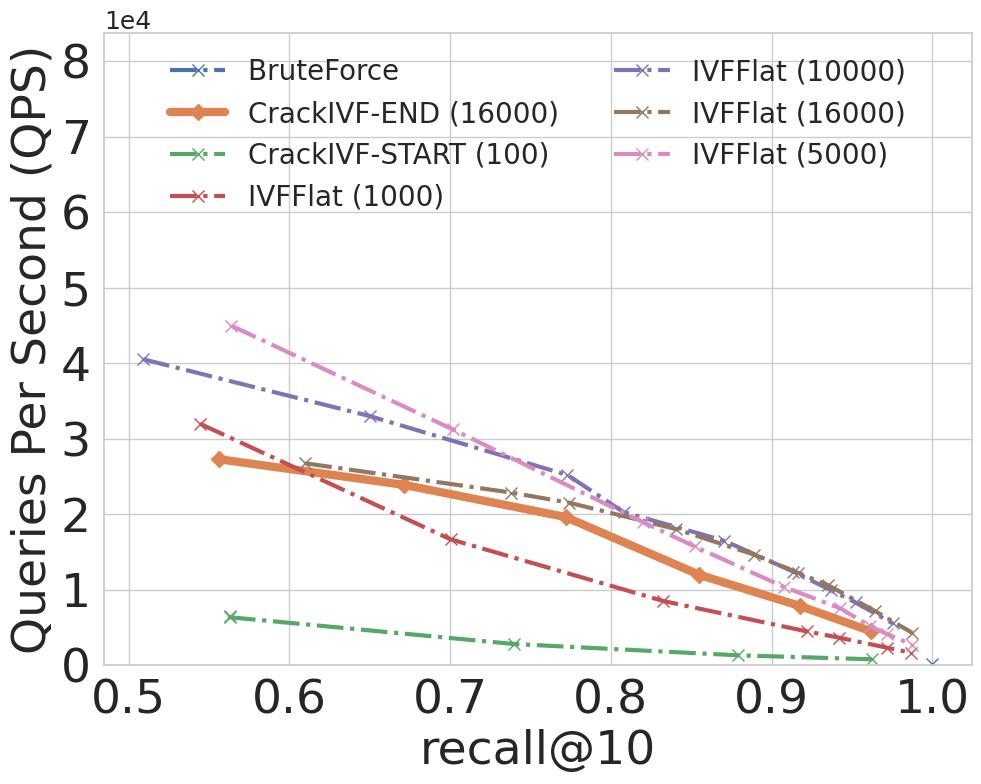}
        \caption{{SIFT1M $size\_prctl$ OFF}}
        \label{fig:rule2-qps}
    \end{subfigure}%
    \hfill
    \begin{subfigure}[t]{0.48\columnwidth}
        \centering
        \includegraphics[width=\textwidth]{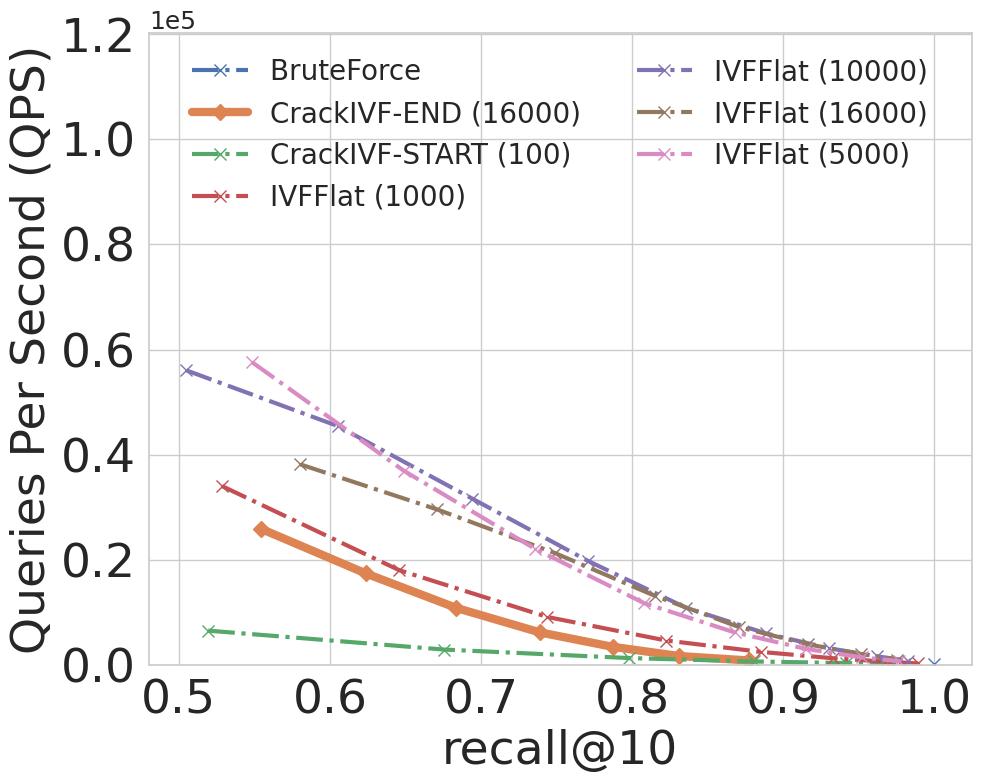}
        \caption{{Glove100 $size\_prctl$ OFF}}
        \label{fig:rule2-time}
    \end{subfigure}
    
    \caption{{Ablation study of parameters.}}
    \label{fig:ablation-study}
\end{figure}

\textbf{Turning off the control mechanism for ``when'':} The budgeting mechanism, controlling when build operations are executed, helps balance the time spent on build vs search operations. By default, we use the parameter $\alpha=0.5$, i.e., at most 50\% of the total time may be spent on $CRACK$ or $REFINE$. The effect of turning this mechanism ``WHEN=OFF'' is shown in Figure~\ref{fig:when_off_cumul} and Figure~\ref{fig:when_off_total_time}. CrackIVF with the mechanism off defaults to cracking and refining after almost every query, and it only stops when the maximum number of partitions is reached (16,000). So even though the cracks and refines happen in regions that require them, they happen disproportionally often compared to the search overhead, leading to a final time spent on build operations that completely dominates the total time. On the other hand, when the mechanism is enabled,  ``WHEN=ON'', CrackIVF is able to efficiently balance the search and build costs (Figure~\ref{fig:when_on_cumul} and Figure~\ref{fig:when_on_total_time}). These runs are for the same 100k query trace. The parameter controlling the ``when'' mechanism is $\alpha$. The effect of varying $\alpha$ is shown in Figures~\ref{fig:a25} and ~\ref{fig:a75}. If build is 25\% of the total time then it is too restrictive. On the other hand, if 75\% of the total time can go to build operations, a very large number of refines keep triggering. Figure~\ref{fig:a75} illustrates the importance of monitoring the new buffered cracks rate and the refine regions in order to converge the index. After $\approx$50k queries, no new cracks occur, preventing further index growth. Refines also repeatedly trigger in the same regions, as shown by the striated pattern, offering no additional benefit, since the median response time does not improve after the refines. This indicates the index has converged. We monitor when this state is achieved and then converge the index, avoiding further build operations.

\begin{table}[t]
    \centering
    \caption{{Parameters defining the Decision Rule Boundaries.}}
    \label{tab:ablation}
    \resizebox{\columnwidth}{!}{%
    \begin{tabular}{@{}lcl@{}}
    \toprule
    \textbf{Parameter} & \textbf{Restrict ``good candidates''} & \textbf{Tested Values}  \\ \midrule
    $min\_pts$ & ↑ & 1, \textbf{2}, 32 \\
    $pts\_crack\_thr$ & ↑  & 16, \textbf{64}, 128\\
    $cv\_max$ & ↑  & 1, \textbf{2}, 8 \\
    $size\_prctl$ & ↓  & 1, \textbf{10}, 25 \\
    \bottomrule
    \end{tabular}%
    }
\end{table}

\textbf{Ablation on decision boundary parameters:} Each parameter defines a decision boundary (Section~\ref{subsec:where-heuristics}). Changing it in one direction, makes the model more restrictive, classifying fewer candidates as good, while in the opposite direction it is more permissive. Table~\ref{tab:ablation} indicates the restrictive direction and tested values (default in \textbf{bold}). We did multiple runs for each dataset and parameter combination. For each parameter we pick the most interesting results, showcased in Figure~\ref{fig:ablation-study}. We only vary one parameter at a time, the others are fixed to their default value. We found the model to be robust to changing a single parameter. Out of the 48 edge-case parameter-dataset combinations, 15 had a noticeable effect on the final index. We attribute the robustness to each heuristic involving two decision rules (Section~\ref{subsec:where-heuristics}), thus two parameters. When one is changed, the other still provides a reasonable boundary. Setting both parameters to restrictive extremes prevents any cracking, while setting both to relaxed extremes is equivalent to \texttt{WHERE=OFF} (Figure~\ref{fig:where_off_qps}), where every query is classified as good and the operations occur when the cost budget allows. 

For $min\_pts$, we have the most consistent effect across all our observations. When we set it to a very restrictive value ($32$), the number of good cracks drops significantly, and thus the index converges to a smaller size, leading to lower performance (Figure~\ref{fig:sift1m-minpts-32}). This negatively affected all 6 datasets.

The $pts\_crack\_thr$, mostly affects the highest skew datasets. For example, on Last.fm performance worsens when the parameter is set to a permissive value, as this leads to over-partitioning of the region, as more cracks are allowed in a region (Figure~\ref{fig:lastfm-ptscrackthr-16}). But, the above value did not affect any other dataset. On the other hand, restricting it negatively affected 2 out of 6 datasets (Figure~\ref{fig:glove50-ptscrackthr-128}). This was the same effect as that of $min\_pts$. When the rule is too restrictive, not enough cracks pass it so index can not fully grow.

Restricting $cv\_max$ can have a positive effect, e.g. on SIFT10M (Figure~\ref{fig:sift10m-cvmax-8}) and on DEEP10M (Figure~\ref{fig:deep10m-cvmax-8}). This leads us to believe that in some datasets allowing a higher local imbalance, could have beneficial overfitting effects to the query distribution. The optimal value for $cv\_max$ is not consistent across datasets, two other datasets had worse results at a high value, and our default values provide the most robust results.

All the values tested for $size\_prctl$ had a negligible effect. Thus, to show the benefit of including the decision rule which $size\_prctl$ controls, we switch it off entirely in Figures~\ref{fig:rule2-qps} and~\ref{fig:rule2-time}. Disabling it greatly affects final index. While the index grows, and it does not converge to the optimal state, as fewer refines occur.

Overall, we clearly see dataset-dependent effects which can change the optimal value of each parameter per dataset. This leaves room for further tuning and ``learned'' decision boundaries instead of setting them manually. Tuning of parameters exists in any indexing technique, the more you know about your dataset the better you can tune the index for it. We did not pursue tuning in this work, as the default values empirically provide very robust performance.

\textbf{From our experiments, we make the following suggestion}: BruteForce should be used as the default method up until a dataset surpasses 10–100 queries, at which point CrackIVF becomes the optimal choice. Even though CrackIVF can start with minimal initialization cost, it is more complex and has a slight memory and runtime overhead compared to BruteForce due to the setup time and storage of internal object structures and keeping track of the first few partition representatives. At scale, in embedding lakes even a small overhead can accumulate. We recommend using CrackIVF once a dataset exceeds 100 queries and is likely to receive more.

\section{Related Work}
\label{sec:related_work}
\textbf{Vector Databases:} Vector Data Management Systems (VDBMS) \cite{wang2021milvus, weaviate, pinecone, zhang2023vbase, chen2024singlestore,guo2022manu,deng25-alayadb,kuffo25-pdx,Xu25-tribase,Cai25-filter-ann} support efficient ANN search, and have gained significant popularity. This coincides with the increase in use of Retrieval-Augmented Generation (RAG) \cite{lewis2020retrieval,asai2023retrieval}, to extend the capabilities of LLMs beyond what they have been trained on. VDBMS applications also include semantic search \cite{salton1975vector, mitra2018introduction}, recommendation systems \cite{li2017fexipro} and product search \cite{van2016learning}. Many relational engines are incorporating vector search \cite{oracle_vector_search, pgvector}. Many of these systems support extensions to dense vector ANN search, allowing for specific query types such as filtered search \cite{gollapudi2023filtered}, sparse vector search \cite{formal2021splade}, and hybrid search \cite{vespa_hybrid_search}. A recent survey covering vector database management systems can be found in \cite{pan2024survey}. 

\textbf{Vector Search Indexes:} 
Approximate Nearest Neighbor (ANN) search has been well-studied. The two dominant categories of ANN indexes are partition-based and graph-based indices. Inverted file index (IVF) is a partition-based index and includes many different implementations. ScaNN~\cite{guo2020accelerating} and SOAR~\cite{sun2024soar} identify key mathematical properties of ANNS and use them to improve index efficiency. SPANN \cite{chen2021spann} uses disk-based storage to search over billion-scale datasets and supports updates~\cite{xu2023spfresh}. FAISS-IVF \cite{douze2024faiss} provides an in-memory implementation IVF which can utilize GPUs \cite{johnson2019billion}. In this work we show how IVF indexes can be extended with cracking, and our approach is directly applicable to the above. We focus on IVF indexes as they offer the best trade-off between index build time, search performance, and memory overhead, which is critical to minimize overhead in large scale deployments on embedding data lakes, as covered in Section~\ref{sec:solution_overview}. The cracking paradigm can be generalized to other ANN indexes as well. With AV-tree\cite{lampropoulos2023adaptive} the authors show how to extend a tree-based index with cracking to supports k-NN queries targeting ENN search. Hash-based methods \cite{andoni2015practical, indyk1998approximate, datar2004locality, jain2008fast, wang2017survey} and tree-based methods \cite{muja2014scalable,liu2004investigation,wang2013trinary} also broadly fall in the partition-based category. Each index has a unique construction method and data structures. Adopting cracking ideas for them would require solving unique challenges. Finally, graph-based methods \cite{malkov2018efficient, malkov2014approximate, wang2012scalable, dong2011efficient, wang2013trinary,hajebi2011fast} currently deliver the state-of-the-art performance in terms of the Queries Per Second vs. Recall trade-off, but this comes at high memory and index construction cost. Thus, graph-based indexes are ill-suited when the goal is minimizing the index construction cost and memory overhead. Cracking is applicable here and could push search performance even further. For example, future research can show how to dynamically change the connections in the graphs based on the arriving query distribution.

\textbf{Adaptive indexes for specialized data types:} Adaptive indexing techniques, such as Dumpy \cite{wang2023dumpy, wang2024dumpyos} and ADS \cite{zoumpatianos2014indexing} are effective in data series similarity search. They are specifically designed for time-series data and include methods like Symbolic Aggregate Approximation (SAX) \cite{lin2003symbolic} and indexable SAX (iSAX) \cite{shieh2008sax}. We focus on ANN indexing methods that are the standard used by RAG systems, and due to their broad applicability across various data modalities which lack sequential relationships. Specialized, adaptive and incrementally constructed indexes are relevant candidates for embedding lake deployments, particularly for the subset of the datasets with inherent sequential dependencies such as energy, weather, financial data, audio, and video  \cite{raza2015practical,mirylenka2016characterizing,shasha1999tuning,kashino1999time}. 

\textbf{ANN index maintenance:} Work on index maintenance focuses on mitigating performance degradation caused by data distribution shifts due to data updates (insertions and deletions). DeDrift~\cite{baranchuk2023dedrift} handles data or query distribution shifts keeping the total number of partitions fixed. LIRE~\cite{xu2023spfresh} performs localized split and merge operations when partitions change after insertions/deletions. Ada-IVF~\cite{mohoney2024incremental} monitors partition size changes from updates and the partition access distribution from the queries. It combines this information to prioritize the maintenance operations. Finally, Quake \cite{mohoney2025quake} is an adaptive index for dynamic workloads with changing access patterns and heavy updates. They use a cost model to decide on maintenance operations, and dynamically set the nprobe parameter. Similar to us, these approaches demonstrate the effectiveness of localized operations for IVF indexes and the advantages of adapting to query distribution. However, they cannot be applied directly for our setting of incremental index construction to minimize cumulative time as the number of queries increases. These works trigger maintenance operations from incoming updates, or focus on distribution shits. Our focus is on the static setting, and we follow the database cracking paradigm: ``index maintenance should be a byproduct of query processing, not of updates''~\cite{idreos2007database}. Thus, CrackIVF operations are triggered solely from the queries. 

\section{Conclusion}

We introduced CrackIVF, a cracking-based IVF index for Approximate Nearest Neighbor Search (ANNS), aimed to be used with RAG systems over embedding data lakes. By dynamically adapting to increasing query workloads, it eliminates the need for costly up-front indexing, allowing immediate query processing. Our experimental results show that CrackIVF can process over 1 million queries before conventional methods even finish building an index, achieving 10-1000x faster initialization times. This makes it particularly effective for cold data, infrequently accessed datasets, and rapid access to unseen data.

\begin{acks}
We would like to thank AMD Research for the generous donation of the equipment in the ETHZ-AMD Heterogeneous Accelerated Compute Cluster (HACC) (\url{https://systems.ethz.ch/research/data-processing-on-modern-hardware/hacc.html}) that was used to run the experiments reported in this paper. 
\end{acks}

\bibliographystyle{ACM-Reference-Format}
\bibliography{main.bib}


\begin{thebibliography}{95}


\ifx \showCODEN    \undefined \def \showCODEN     #1{\unskip}     \fi
\ifx \showDOI      \undefined \def \showDOI       #1{#1}\fi
\ifx \showISBNx    \undefined \def \showISBNx     #1{\unskip}     \fi
\ifx \showISBNxiii \undefined \def \showISBNxiii  #1{\unskip}     \fi
\ifx \showISSN     \undefined \def \showISSN      #1{\unskip}     \fi
\ifx \showLCCN     \undefined \def \showLCCN      #1{\unskip}     \fi
\ifx \shownote     \undefined \def \shownote      #1{#1}          \fi
\ifx \showarticletitle \undefined \def \showarticletitle #1{#1}   \fi
\ifx \showURL      \undefined \def \showURL       {\relax}        \fi
\providecommand\bibfield[2]{#2}
\providecommand\bibinfo[2]{#2}
\providecommand\natexlab[1]{#1}
\providecommand\showeprint[2][]{arXiv:#2}

\bibitem[\protect\citeauthoryear{Anderson, Fritz, Lee, Li, Lindblad, Lindeman, Meyer, Parmar, Ranade, Shah, et~al\mbox{.}}{Anderson et~al\mbox{.}}{2024}]%
        {anderson2024design}
\bibfield{author}{\bibinfo{person}{Eric Anderson}, \bibinfo{person}{Jonathan Fritz}, \bibinfo{person}{Austin Lee}, \bibinfo{person}{Bohou Li}, \bibinfo{person}{Mark Lindblad}, \bibinfo{person}{Henry Lindeman}, \bibinfo{person}{Alex Meyer}, \bibinfo{person}{Parth Parmar}, \bibinfo{person}{Tanvi Ranade}, \bibinfo{person}{Mehul~A Shah}, {et~al\mbox{.}}} \bibinfo{year}{2024}\natexlab{}.
\newblock \showarticletitle{The Design of an LLM-powered Unstructured Analytics System}.
\newblock \bibinfo{journal}{\emph{arXiv preprint arXiv:2409.00847}} (\bibinfo{year}{2024}).
\newblock


\bibitem[\protect\citeauthoryear{Andoni, Indyk, Laarhoven, Razenshteyn, and Schmidt}{Andoni et~al\mbox{.}}{2015}]%
        {andoni2015practical}
\bibfield{author}{\bibinfo{person}{Alexandr Andoni}, \bibinfo{person}{Piotr Indyk}, \bibinfo{person}{Thijs Laarhoven}, \bibinfo{person}{Ilya Razenshteyn}, {and} \bibinfo{person}{Ludwig Schmidt}.} \bibinfo{year}{2015}\natexlab{}.
\newblock \showarticletitle{Practical and optimal LSH for angular distance}. In \bibinfo{booktitle}{\emph{Proceedings of the 29th International Conference on Neural Information Processing Systems - Volume 1}} (Montreal, Canada) \emph{(\bibinfo{series}{NIPS'15})}. \bibinfo{publisher}{MIT Press}, \bibinfo{address}{Cambridge, MA, USA}, \bibinfo{pages}{1225–1233}.
\newblock


\bibitem[\protect\citeauthoryear{Armbrust, Xin, Lian, Huai, Li, Mukherjee, Nijkamp, Ousterhout, Paranjpye, Ramasamy, Soh, Tumanov, Yavuz, and Zaharia}{Armbrust et~al\mbox{.}}{2021}]%
        {Armbrust2021Lakehouse}
\bibfield{author}{\bibinfo{person}{Michael Armbrust}, \bibinfo{person}{Reynold~S. Xin}, \bibinfo{person}{Shixiong Lian}, \bibinfo{person}{Yin Huai}, \bibinfo{person}{Cheng Li}, \bibinfo{person}{Tathagata Mukherjee}, \bibinfo{person}{Erik Nijkamp}, \bibinfo{person}{Kay Ousterhout}, \bibinfo{person}{Ruchi Paranjpye}, \bibinfo{person}{Kartik Ramasamy}, \bibinfo{person}{Dalton Soh}, \bibinfo{person}{Alexey Tumanov}, \bibinfo{person}{Burak Yavuz}, {and} \bibinfo{person}{Matei Zaharia}.} \bibinfo{year}{2021}\natexlab{}.
\newblock \showarticletitle{Lakehouse: A New Generation of Open Platforms that Unify Data Warehousing and Advanced Analytics}.
\newblock \bibinfo{journal}{\emph{CIDR}} (\bibinfo{year}{2021}).
\newblock
\urldef\tempurl%
\url{https://www.cidrdb.org/cidr2021/papers/cidr2021_paper17.pdf}
\showURL{%
\tempurl}


\bibitem[\protect\citeauthoryear{Asai, Min, Zhong, and Chen}{Asai et~al\mbox{.}}{2023}]%
        {asai2023retrieval}
\bibfield{author}{\bibinfo{person}{Akari Asai}, \bibinfo{person}{Sewon Min}, \bibinfo{person}{Zexuan Zhong}, {and} \bibinfo{person}{Danqi Chen}.} \bibinfo{year}{2023}\natexlab{}.
\newblock \showarticletitle{Retrieval-based language models and applications}. In \bibinfo{booktitle}{\emph{Proceedings of the 61st Annual Meeting of the Association for Computational Linguistics (Volume 6: Tutorial Abstracts)}}. \bibinfo{pages}{41--46}.
\newblock


\bibitem[\protect\citeauthoryear{Aumüller, Bernhardsson, and Faithfull}{Aumüller et~al\mbox{.}}{2020}]%
        {aumuller2020ann}
\bibfield{author}{\bibinfo{person}{Martin Aumüller}, \bibinfo{person}{Erik Bernhardsson}, {and} \bibinfo{person}{Alexander Faithfull}.} \bibinfo{year}{2020}\natexlab{}.
\newblock \showarticletitle{ANN-Benchmarks: A benchmarking tool for approximate nearest neighbor algorithms}.
\newblock \bibinfo{journal}{\emph{Information Systems}}  \bibinfo{volume}{87} (\bibinfo{year}{2020}), \bibinfo{pages}{101374}.
\newblock
\showISSN{0306-4379}
\urldef\tempurl%
\url{https://doi.org/10.1016/j.is.2019.02.006}
\showDOI{\tempurl}


\bibitem[\protect\citeauthoryear{Babenko and Lempitsky}{Babenko and Lempitsky}{2016}]%
        {babenko2016efficient}
\bibfield{author}{\bibinfo{person}{Artem Babenko} {and} \bibinfo{person}{Victor Lempitsky}.} \bibinfo{year}{2016}\natexlab{}.
\newblock \showarticletitle{Efficient indexing of billion-scale datasets of deep descriptors}. In \bibinfo{booktitle}{\emph{Proceedings of the IEEE Conference on Computer Vision and Pattern Recognition}}. \bibinfo{pages}{2055--2063}.
\newblock


\bibitem[\protect\citeauthoryear{Baranchuk, Douze, Upadhyay, and Yalniz}{Baranchuk et~al\mbox{.}}{2023}]%
        {baranchuk2023dedrift}
\bibfield{author}{\bibinfo{person}{Dmitry Baranchuk}, \bibinfo{person}{Matthijs Douze}, \bibinfo{person}{Yash Upadhyay}, {and} \bibinfo{person}{I~Zeki Yalniz}.} \bibinfo{year}{2023}\natexlab{}.
\newblock \showarticletitle{Dedrift: Robust similarity search under content drift}. In \bibinfo{booktitle}{\emph{Proceedings of the IEEE/CVF International Conference on Computer Vision}}. \bibinfo{pages}{11026--11035}.
\newblock


\bibitem[\protect\citeauthoryear{Benjelloun, Chen, and Noy}{Benjelloun et~al\mbox{.}}{2020}]%
        {benjelloun2020google}
\bibfield{author}{\bibinfo{person}{Omar Benjelloun}, \bibinfo{person}{Shiyu Chen}, {and} \bibinfo{person}{Natasha Noy}.} \bibinfo{year}{2020}\natexlab{}.
\newblock \showarticletitle{Google dataset search by the numbers}. In \bibinfo{booktitle}{\emph{International Semantic Web Conference}}. Springer, \bibinfo{pages}{667--682}.
\newblock


\bibitem[\protect\citeauthoryear{Bentley}{Bentley}{1975}]%
        {bentley1975multidimensional}
\bibfield{author}{\bibinfo{person}{Jon~Louis Bentley}.} \bibinfo{year}{1975}\natexlab{}.
\newblock \showarticletitle{Multidimensional binary search trees used for associative searching}.
\newblock \bibinfo{journal}{\emph{Commun. ACM}} \bibinfo{volume}{18}, \bibinfo{number}{9} (\bibinfo{date}{Sept.} \bibinfo{year}{1975}), \bibinfo{pages}{509–517}.
\newblock
\showISSN{0001-0782}
\urldef\tempurl%
\url{https://doi.org/10.1145/361002.361007}
\showDOI{\tempurl}


\bibitem[\protect\citeauthoryear{Bertin-Mahieux, Ellis, Whitman, and Lamere}{Bertin-Mahieux et~al\mbox{.}}{2011}]%
        {Bertin-Mahieux2011}
\bibfield{author}{\bibinfo{person}{Thierry Bertin-Mahieux}, \bibinfo{person}{Daniel~P.W. Ellis}, \bibinfo{person}{Brian Whitman}, {and} \bibinfo{person}{Paul Lamere}.} \bibinfo{year}{2011}\natexlab{}.
\newblock \showarticletitle{The Million Song Dataset}. In \bibinfo{booktitle}{\emph{{Proceedings of the 12th International Conference on Music Information Retrieval ({ISMIR} 2011)}}}.
\newblock


\bibitem[\protect\citeauthoryear{Brickley, Burgess, and Noy}{Brickley et~al\mbox{.}}{2019}]%
        {brickley2019google}
\bibfield{author}{\bibinfo{person}{Dan Brickley}, \bibinfo{person}{Matthew Burgess}, {and} \bibinfo{person}{Natasha Noy}.} \bibinfo{year}{2019}\natexlab{}.
\newblock \showarticletitle{Google Dataset Search: Building a search engine for datasets in an open Web ecosystem}. In \bibinfo{booktitle}{\emph{The world wide web conference}}. \bibinfo{pages}{1365--1375}.
\newblock


\bibitem[\protect\citeauthoryear{Brown, Mann, Ryder, Subbiah, Kaplan, Dhariwal, Neelakantan, Shyam, Sastry, Askell, et~al\mbox{.}}{Brown et~al\mbox{.}}{2020}]%
        {brown2020language}
\bibfield{author}{\bibinfo{person}{Tom Brown}, \bibinfo{person}{Benjamin Mann}, \bibinfo{person}{Nick Ryder}, \bibinfo{person}{Melanie Subbiah}, \bibinfo{person}{Jared~D Kaplan}, \bibinfo{person}{Prafulla Dhariwal}, \bibinfo{person}{Arvind Neelakantan}, \bibinfo{person}{Pranav Shyam}, \bibinfo{person}{Girish Sastry}, \bibinfo{person}{Amanda Askell}, {et~al\mbox{.}}} \bibinfo{year}{2020}\natexlab{}.
\newblock \showarticletitle{Language models are few-shot learners}.
\newblock \bibinfo{journal}{\emph{Advances in neural information processing systems}}  \bibinfo{volume}{33} (\bibinfo{year}{2020}), \bibinfo{pages}{1877--1901}.
\newblock


\bibitem[\protect\citeauthoryear{Cai, Shi, Chen, and Zheng}{Cai et~al\mbox{.}}{2024}]%
        {Cai25-filter-ann}
\bibfield{author}{\bibinfo{person}{Yuzheng Cai}, \bibinfo{person}{Jiayang Shi}, \bibinfo{person}{Yizhuo Chen}, {and} \bibinfo{person}{Weiguo Zheng}.} \bibinfo{year}{2024}\natexlab{}.
\newblock \showarticletitle{Navigating Labels and Vectors: A Unified Approach to Filtered Approximate Nearest Neighbor Search}.
\newblock \bibinfo{journal}{\emph{Proc. ACM Manag. Data}} \bibinfo{volume}{2}, \bibinfo{number}{6}, Article \bibinfo{articleno}{246} (\bibinfo{date}{Dec.} \bibinfo{year}{2024}), \bibinfo{numpages}{27}~pages.
\newblock
\urldef\tempurl%
\url{https://doi.org/10.1145/3698822}
\showDOI{\tempurl}


\bibitem[\protect\citeauthoryear{Chapman, Simperl, Koesten, Konstantinidis, Ib{\'a}{\~n}ez, Kacprzak, and Groth}{Chapman et~al\mbox{.}}{2020}]%
        {chapman2020dataset}
\bibfield{author}{\bibinfo{person}{Adriane Chapman}, \bibinfo{person}{Elena Simperl}, \bibinfo{person}{Laura Koesten}, \bibinfo{person}{George Konstantinidis}, \bibinfo{person}{Luis-Daniel Ib{\'a}{\~n}ez}, \bibinfo{person}{Emilia Kacprzak}, {and} \bibinfo{person}{Paul Groth}.} \bibinfo{year}{2020}\natexlab{}.
\newblock \showarticletitle{Dataset search: a survey}.
\newblock \bibinfo{journal}{\emph{The VLDB Journal}} \bibinfo{volume}{29}, \bibinfo{number}{1} (\bibinfo{year}{2020}), \bibinfo{pages}{251--272}.
\newblock


\bibitem[\protect\citeauthoryear{Chen, Jin, Zhang, Podolsky, Wu, Wang, Hanson, Sun, Walzer, and Wang}{Chen et~al\mbox{.}}{2024}]%
        {chen2024singlestore}
\bibfield{author}{\bibinfo{person}{Cheng Chen}, \bibinfo{person}{Chenzhe Jin}, \bibinfo{person}{Yunan Zhang}, \bibinfo{person}{Sasha Podolsky}, \bibinfo{person}{Chun Wu}, \bibinfo{person}{Szu-Po Wang}, \bibinfo{person}{Eric Hanson}, \bibinfo{person}{Zhou Sun}, \bibinfo{person}{Robert Walzer}, {and} \bibinfo{person}{Jianguo Wang}.} \bibinfo{year}{2024}\natexlab{}.
\newblock \showarticletitle{SingleStore-V: An Integrated Vector Database System in SingleStore}.
\newblock \bibinfo{journal}{\emph{Proceedings of the VLDB Endowment}} \bibinfo{volume}{17}, \bibinfo{number}{12} (\bibinfo{year}{2024}), \bibinfo{pages}{3772--3785}.
\newblock


\bibitem[\protect\citeauthoryear{Chen, Zhao, Wang, Li, Liu, Li, Yang, and Wang}{Chen et~al\mbox{.}}{2021}]%
        {chen2021spann}
\bibfield{author}{\bibinfo{person}{Qi Chen}, \bibinfo{person}{Bing Zhao}, \bibinfo{person}{Haidong Wang}, \bibinfo{person}{Mingqin Li}, \bibinfo{person}{Chuanjie Liu}, \bibinfo{person}{Zengzhong Li}, \bibinfo{person}{Mao Yang}, {and} \bibinfo{person}{Jingdong Wang}.} \bibinfo{year}{2021}\natexlab{}.
\newblock \showarticletitle{Spann: Highly-efficient billion-scale approximate nearest neighborhood search}.
\newblock \bibinfo{journal}{\emph{Advances in Neural Information Processing Systems}}  \bibinfo{volume}{34} (\bibinfo{year}{2021}), \bibinfo{pages}{5199--5212}.
\newblock


\bibitem[\protect\citeauthoryear{Chen, Gu, Cao, Fan, Madden, and Tang}{Chen et~al\mbox{.}}{2023}]%
        {chen2023symphony}
\bibfield{author}{\bibinfo{person}{Zui Chen}, \bibinfo{person}{Zihui Gu}, \bibinfo{person}{Lei Cao}, \bibinfo{person}{Ju Fan}, \bibinfo{person}{Samuel Madden}, {and} \bibinfo{person}{Nan Tang}.} \bibinfo{year}{2023}\natexlab{}.
\newblock \showarticletitle{Symphony: Towards Natural Language Query Answering over Multi-modal Data Lakes.}. In \bibinfo{booktitle}{\emph{CIDR}}.
\newblock


\bibitem[\protect\citeauthoryear{Corporation}{Corporation}{2025}]%
        {oracle_vector_search}
\bibfield{author}{\bibinfo{person}{Oracle Corporation}.} \bibinfo{year}{2025}\natexlab{}.
\newblock \bibinfo{title}{Oracle AI Vector Search}.
\newblock \bibinfo{howpublished}{\url{https://www.oracle.com/database/ai-vector-search/}}.
\newblock
\newblock
\shownote{Accessed: 2025-03-01.}


\bibitem[\protect\citeauthoryear{{Databricks}}{{Databricks}}{2024}]%
        {Databricks2024}
\bibfield{author}{\bibinfo{person}{{Databricks}}.} \bibinfo{year}{2024}\natexlab{}.
\newblock \bibinfo{title}{Vector Search in Databricks}.
\newblock
\newblock
\urldef\tempurl%
\url{https://docs.databricks.com/en/generative-ai/vector-search.html}
\showURL{%
\tempurl}
\newblock
\shownote{Accessed: 2025-02-16.}


\bibitem[\protect\citeauthoryear{Datar, Immorlica, Indyk, and Mirrokni}{Datar et~al\mbox{.}}{2004}]%
        {datar2004locality}
\bibfield{author}{\bibinfo{person}{Mayur Datar}, \bibinfo{person}{Nicole Immorlica}, \bibinfo{person}{Piotr Indyk}, {and} \bibinfo{person}{Vahab~S Mirrokni}.} \bibinfo{year}{2004}\natexlab{}.
\newblock \showarticletitle{Locality-sensitive hashing scheme based on p-stable distributions}. In \bibinfo{booktitle}{\emph{Proceedings of the twentieth annual symposium on Computational geometry}}. \bibinfo{pages}{253--262}.
\newblock


\bibitem[\protect\citeauthoryear{Deng, You, Xiang, Li, Yuan, Hong, Zheng, Li, Li, Liu, Mouratidis, Yiu, Li, Shen, Mao, and Tang}{Deng et~al\mbox{.}}{2025}]%
        {deng25-alayadb}
\bibfield{author}{\bibinfo{person}{Yangshen Deng}, \bibinfo{person}{Zhengxin You}, \bibinfo{person}{Long Xiang}, \bibinfo{person}{Qilong Li}, \bibinfo{person}{Peiqi Yuan}, \bibinfo{person}{Zhaoyang Hong}, \bibinfo{person}{Yitao Zheng}, \bibinfo{person}{Wanting Li}, \bibinfo{person}{Runzhong Li}, \bibinfo{person}{Haotian Liu}, \bibinfo{person}{Kyriakos Mouratidis}, \bibinfo{person}{Man~Lung Yiu}, \bibinfo{person}{Huan Li}, \bibinfo{person}{Qiaomu Shen}, \bibinfo{person}{Rui Mao}, {and} \bibinfo{person}{Bo Tang}.} \bibinfo{year}{2025}\natexlab{}.
\newblock \showarticletitle{AlayaDB: The Data Foundation for Efficient and Effective Long-context LLM Inference}. In \bibinfo{booktitle}{\emph{Companion of the 2025 International Conference on Management of Data}} (Berlin, Germany) \emph{(\bibinfo{series}{SIGMOD/PODS '25})}. \bibinfo{publisher}{Association for Computing Machinery}, \bibinfo{address}{New York, NY, USA}, \bibinfo{pages}{364–377}.
\newblock
\showISBNx{9798400715648}
\urldef\tempurl%
\url{https://doi.org/10.1145/3722212.3724428}
\showDOI{\tempurl}


\bibitem[\protect\citeauthoryear{Dong, Moses, and Li}{Dong et~al\mbox{.}}{2011}]%
        {dong2011efficient}
\bibfield{author}{\bibinfo{person}{Wei Dong}, \bibinfo{person}{Charikar Moses}, {and} \bibinfo{person}{Kai Li}.} \bibinfo{year}{2011}\natexlab{}.
\newblock \showarticletitle{Efficient k-nearest neighbor graph construction for generic similarity measures}. In \bibinfo{booktitle}{\emph{Proceedings of the 20th international conference on World wide web}}. \bibinfo{pages}{577--586}.
\newblock


\bibitem[\protect\citeauthoryear{Douze, Guzhva, Deng, Johnson, Szilvasy, Mazar{\'e}, Lomeli, Hosseini, and J{\'e}gou}{Douze et~al\mbox{.}}{2024}]%
        {douze2024faiss}
\bibfield{author}{\bibinfo{person}{Matthijs Douze}, \bibinfo{person}{Alexandr Guzhva}, \bibinfo{person}{Chengqi Deng}, \bibinfo{person}{Jeff Johnson}, \bibinfo{person}{Gergely Szilvasy}, \bibinfo{person}{Pierre-Emmanuel Mazar{\'e}}, \bibinfo{person}{Maria Lomeli}, \bibinfo{person}{Lucas Hosseini}, {and} \bibinfo{person}{Herv{\'e} J{\'e}gou}.} \bibinfo{year}{2024}\natexlab{}.
\newblock \showarticletitle{The faiss library}.
\newblock \bibinfo{journal}{\emph{arXiv preprint arXiv:2401.08281}} (\bibinfo{year}{2024}).
\newblock


\bibitem[\protect\citeauthoryear{{Facebook Research}}{{Facebook Research}}{2025}]%
        {faiss_guidelines}
\bibfield{author}{\bibinfo{person}{{Facebook Research}}.} \bibinfo{year}{2025}\natexlab{}.
\newblock \bibinfo{title}{{Guidelines to Choose an Index}}.
\newblock \bibinfo{howpublished}{\url{https://github.com/facebookresearch/faiss/wiki/Guidelines-to-choose-an-index}}.
\newblock
\newblock
\shownote{Accessed: 2025-02-17.}


\bibitem[\protect\citeauthoryear{Formal, Piwowarski, and Clinchant}{Formal et~al\mbox{.}}{2021}]%
        {formal2021splade}
\bibfield{author}{\bibinfo{person}{Thibault Formal}, \bibinfo{person}{Benjamin Piwowarski}, {and} \bibinfo{person}{St{\'e}phane Clinchant}.} \bibinfo{year}{2021}\natexlab{}.
\newblock \showarticletitle{SPLADE: Sparse lexical and expansion model for first stage ranking}. In \bibinfo{booktitle}{\emph{Proceedings of the 44th International ACM SIGIR Conference on Research and Development in Information Retrieval}}. \bibinfo{pages}{2288--2292}.
\newblock


\bibitem[\protect\citeauthoryear{Gollapudi, Karia, Sivashankar, Krishnaswamy, Begwani, Raz, Lin, Zhang, Mahapatro, Srinivasan, et~al\mbox{.}}{Gollapudi et~al\mbox{.}}{2023}]%
        {gollapudi2023filtered}
\bibfield{author}{\bibinfo{person}{Siddharth Gollapudi}, \bibinfo{person}{Neel Karia}, \bibinfo{person}{Varun Sivashankar}, \bibinfo{person}{Ravishankar Krishnaswamy}, \bibinfo{person}{Nikit Begwani}, \bibinfo{person}{Swapnil Raz}, \bibinfo{person}{Yiyong Lin}, \bibinfo{person}{Yin Zhang}, \bibinfo{person}{Neelam Mahapatro}, \bibinfo{person}{Premkumar Srinivasan}, {et~al\mbox{.}}} \bibinfo{year}{2023}\natexlab{}.
\newblock \showarticletitle{Filtered-diskann: Graph algorithms for approximate nearest neighbor search with filters}. In \bibinfo{booktitle}{\emph{Proceedings of the ACM Web Conference 2023}}. \bibinfo{pages}{3406--3416}.
\newblock


\bibitem[\protect\citeauthoryear{Gualtieri}{Gualtieri}{2016}]%
        {Gualtieri2016Hadoop}
\bibfield{author}{\bibinfo{person}{Mike Gualtieri}.} \bibinfo{year}{2016}\natexlab{}.
\newblock \bibinfo{booktitle}{\emph{Hadoop Is Data’s Darling For A Reason}}.
\newblock Forrester Research.
\newblock
\urldef\tempurl%
\url{https://www.forrester.com/blogs/hadoop-is-datas-darling-for-a-reason/}
\showURL{%
\tempurl}


\bibitem[\protect\citeauthoryear{Guo, Luan, Xiang, Yan, Yi, Luo, Cheng, Xu, Luo, Liu, et~al\mbox{.}}{Guo et~al\mbox{.}}{2022}]%
        {guo2022manu}
\bibfield{author}{\bibinfo{person}{Rentong Guo}, \bibinfo{person}{Xiaofan Luan}, \bibinfo{person}{Long Xiang}, \bibinfo{person}{Xiao Yan}, \bibinfo{person}{Xiaomeng Yi}, \bibinfo{person}{Jigao Luo}, \bibinfo{person}{Qianya Cheng}, \bibinfo{person}{Weizhi Xu}, \bibinfo{person}{Jiarui Luo}, \bibinfo{person}{Frank Liu}, {et~al\mbox{.}}} \bibinfo{year}{2022}\natexlab{}.
\newblock \showarticletitle{Manu: a cloud native vector database management system}.
\newblock \bibinfo{journal}{\emph{arXiv preprint arXiv:2206.13843}} (\bibinfo{year}{2022}).
\newblock


\bibitem[\protect\citeauthoryear{Guo, Sun, Lindgren, Geng, Simcha, Chern, and Kumar}{Guo et~al\mbox{.}}{2020}]%
        {guo2020accelerating}
\bibfield{author}{\bibinfo{person}{Ruiqi Guo}, \bibinfo{person}{Philip Sun}, \bibinfo{person}{Erik Lindgren}, \bibinfo{person}{Quan Geng}, \bibinfo{person}{David Simcha}, \bibinfo{person}{Felix Chern}, {and} \bibinfo{person}{Sanjiv Kumar}.} \bibinfo{year}{2020}\natexlab{}.
\newblock \showarticletitle{Accelerating large-scale inference with anisotropic vector quantization}. In \bibinfo{booktitle}{\emph{International Conference on Machine Learning}}. PMLR, \bibinfo{pages}{3887--3896}.
\newblock


\bibitem[\protect\citeauthoryear{Guttman}{Guttman}{1984}]%
        {guttman1984r}
\bibfield{author}{\bibinfo{person}{Antonin Guttman}.} \bibinfo{year}{1984}\natexlab{}.
\newblock \showarticletitle{R-trees: a dynamic index structure for spatial searching}. In \bibinfo{booktitle}{\emph{Proceedings of the 1984 ACM SIGMOD International Conference on Management of Data}} (Boston, Massachusetts) \emph{(\bibinfo{series}{SIGMOD '84})}. \bibinfo{publisher}{Association for Computing Machinery}, \bibinfo{address}{New York, NY, USA}, \bibinfo{pages}{47–57}.
\newblock
\showISBNx{0897911288}
\urldef\tempurl%
\url{https://doi.org/10.1145/602259.602266}
\showDOI{\tempurl}


\bibitem[\protect\citeauthoryear{Hajebi, Abbasi-Yadkori, Shahbazi, and Zhang}{Hajebi et~al\mbox{.}}{2011}]%
        {hajebi2011fast}
\bibfield{author}{\bibinfo{person}{Kiana Hajebi}, \bibinfo{person}{Yasin Abbasi-Yadkori}, \bibinfo{person}{Hossein Shahbazi}, {and} \bibinfo{person}{Hong Zhang}.} \bibinfo{year}{2011}\natexlab{}.
\newblock \showarticletitle{Fast approximate nearest-neighbor search with k-nearest neighbor graph}. In \bibinfo{booktitle}{\emph{IJCAI Proceedings-International Joint Conference on Artificial Intelligence}}, Vol.~\bibinfo{volume}{22}. \bibinfo{pages}{1312}.
\newblock


\bibitem[\protect\citeauthoryear{Halim, Idreos, Karras, and Yap}{Halim et~al\mbox{.}}{2012}]%
        {halim2012stochastic}
\bibfield{author}{\bibinfo{person}{Felix Halim}, \bibinfo{person}{Stratos Idreos}, \bibinfo{person}{Panagiotis Karras}, {and} \bibinfo{person}{Roland~HC Yap}.} \bibinfo{year}{2012}\natexlab{}.
\newblock \showarticletitle{Stochastic database cracking: Towards robust adaptive indexing in main-memory column-stores}.
\newblock \bibinfo{journal}{\emph{arXiv preprint arXiv:1203.0055}} (\bibinfo{year}{2012}).
\newblock


\bibitem[\protect\citeauthoryear{Harbert}{Harbert}{2021}]%
        {Harbert2021Tapping}
\bibfield{author}{\bibinfo{person}{Tam Harbert}.} \bibinfo{year}{2021}\natexlab{}.
\newblock \bibinfo{booktitle}{\emph{Tapping the Power of Unstructured Data}}.
\newblock MIT Sloan School of Management.
\newblock
\urldef\tempurl%
\url{https://mitsloan.mit.edu/ideas-made-to-matter/tapping-power-unstructured-data}
\showURL{%
\tempurl}


\bibitem[\protect\citeauthoryear{Heidorn}{Heidorn}{2008}]%
        {heidorn2008shedding}
\bibfield{author}{\bibinfo{person}{P~Bryan Heidorn}.} \bibinfo{year}{2008}\natexlab{}.
\newblock \showarticletitle{Shedding light on the dark data in the long tail of science}.
\newblock \bibinfo{journal}{\emph{Library trends}} \bibinfo{volume}{57}, \bibinfo{number}{2} (\bibinfo{year}{2008}), \bibinfo{pages}{280--299}.
\newblock


\bibitem[\protect\citeauthoryear{Holanda, Nerone, de~Almeida, and Manegold}{Holanda et~al\mbox{.}}{2018}]%
        {holanda2018cracking}
\bibfield{author}{\bibinfo{person}{Pedro Holanda}, \bibinfo{person}{Matheus Nerone}, \bibinfo{person}{Eduardo~C de Almeida}, {and} \bibinfo{person}{Stefan Manegold}.} \bibinfo{year}{2018}\natexlab{}.
\newblock \showarticletitle{Cracking KD-Tree: The First Multidimensional Adaptive Indexing (Position Paper).}. In \bibinfo{booktitle}{\emph{DATA}}. \bibinfo{pages}{393--399}.
\newblock


\bibitem[\protect\citeauthoryear{Idreos, Kersten, Manegold, et~al\mbox{.}}{Idreos et~al\mbox{.}}{2007}]%
        {idreos2007database}
\bibfield{author}{\bibinfo{person}{Stratos Idreos}, \bibinfo{person}{Martin~L Kersten}, \bibinfo{person}{Stefan Manegold}, {et~al\mbox{.}}} \bibinfo{year}{2007}\natexlab{}.
\newblock \showarticletitle{Database Cracking.}. In \bibinfo{booktitle}{\emph{CIDR}}, Vol.~\bibinfo{volume}{7}. \bibinfo{pages}{68--78}.
\newblock


\bibitem[\protect\citeauthoryear{Indyk and Motwani}{Indyk and Motwani}{1998}]%
        {indyk1998approximate}
\bibfield{author}{\bibinfo{person}{Piotr Indyk} {and} \bibinfo{person}{Rajeev Motwani}.} \bibinfo{year}{1998}\natexlab{}.
\newblock \showarticletitle{Approximate nearest neighbors: towards removing the curse of dimensionality}. In \bibinfo{booktitle}{\emph{Proceedings of the thirtieth annual ACM symposium on Theory of computing}}. \bibinfo{pages}{604--613}.
\newblock


\bibitem[\protect\citeauthoryear{Jain, Kulis, and Grauman}{Jain et~al\mbox{.}}{2008}]%
        {jain2008fast}
\bibfield{author}{\bibinfo{person}{Prateek Jain}, \bibinfo{person}{Brian Kulis}, {and} \bibinfo{person}{Kristen Grauman}.} \bibinfo{year}{2008}\natexlab{}.
\newblock \showarticletitle{Fast image search for learned metrics}. In \bibinfo{booktitle}{\emph{2008 IEEE Conference on computer vision and pattern recognition}}. IEEE, \bibinfo{pages}{1--8}.
\newblock


\bibitem[\protect\citeauthoryear{Jain}{Jain}{1991}]%
        {jain1990art}
\bibfield{author}{\bibinfo{person}{R. Jain}.} \bibinfo{year}{1991}\natexlab{}.
\newblock \bibinfo{booktitle}{\emph{{The art of computer systems performance analysis: techniques for experimental design, measurement, simulation, and modeling}}}.
\newblock \bibinfo{publisher}{Wiley New York}.
\newblock


\bibitem[\protect\citeauthoryear{J{\'e}gou, Tavenard, Douze, and Amsaleg}{J{\'e}gou et~al\mbox{.}}{2011}]%
        {jegou2011searching}
\bibfield{author}{\bibinfo{person}{Herv{\'e} J{\'e}gou}, \bibinfo{person}{Romain Tavenard}, \bibinfo{person}{Matthijs Douze}, {and} \bibinfo{person}{Laurent Amsaleg}.} \bibinfo{year}{2011}\natexlab{}.
\newblock \showarticletitle{Searching in one billion vectors: re-rank with source coding}. In \bibinfo{booktitle}{\emph{2011 IEEE International Conference on Acoustics, Speech and Signal Processing (ICASSP)}}. IEEE, \bibinfo{pages}{861--864}.
\newblock


\bibitem[\protect\citeauthoryear{Johnson, Douze, and J{\'e}gou}{Johnson et~al\mbox{.}}{2019}]%
        {johnson2019billion}
\bibfield{author}{\bibinfo{person}{Jeff Johnson}, \bibinfo{person}{Matthijs Douze}, {and} \bibinfo{person}{Herv{\'e} J{\'e}gou}.} \bibinfo{year}{2019}\natexlab{}.
\newblock \showarticletitle{Billion-scale similarity search with GPUs}.
\newblock \bibinfo{journal}{\emph{IEEE Transactions on Big Data}} \bibinfo{volume}{7}, \bibinfo{number}{3} (\bibinfo{year}{2019}), \bibinfo{pages}{535--547}.
\newblock


\bibitem[\protect\citeauthoryear{Karp, Motwani, and Raghavan}{Karp et~al\mbox{.}}{1988}]%
        {karp1988deferred}
\bibfield{author}{\bibinfo{person}{Richard~M Karp}, \bibinfo{person}{Rajeev Motwani}, {and} \bibinfo{person}{Prabhakar Raghavan}.} \bibinfo{year}{1988}\natexlab{}.
\newblock \showarticletitle{Deferred data structuring}.
\newblock \bibinfo{journal}{\emph{SIAM J. Comput.}} \bibinfo{volume}{17}, \bibinfo{number}{5} (\bibinfo{year}{1988}), \bibinfo{pages}{883--902}.
\newblock


\bibitem[\protect\citeauthoryear{Kashino, Smith, and Murase}{Kashino et~al\mbox{.}}{1999}]%
        {kashino1999time}
\bibfield{author}{\bibinfo{person}{Kunio Kashino}, \bibinfo{person}{Gavin Smith}, {and} \bibinfo{person}{Hiroshi Murase}.} \bibinfo{year}{1999}\natexlab{}.
\newblock \showarticletitle{Time-series active search for quick retrieval of audio and video}. In \bibinfo{booktitle}{\emph{1999 IEEE International Conference on Acoustics, Speech, and Signal Processing. Proceedings. ICASSP99 (Cat. No. 99CH36258)}}, Vol.~\bibinfo{volume}{6}. IEEE, \bibinfo{pages}{2993--2996}.
\newblock


\bibitem[\protect\citeauthoryear{Kuffo, Krippner, and Boncz}{Kuffo et~al\mbox{.}}{2025}]%
        {kuffo25-pdx}
\bibfield{author}{\bibinfo{person}{Leonardo Kuffo}, \bibinfo{person}{Elena Krippner}, {and} \bibinfo{person}{Peter Boncz}.} \bibinfo{year}{2025}\natexlab{}.
\newblock \showarticletitle{PDX: A Data Layout for Vector Similarity Search}.
\newblock \bibinfo{journal}{\emph{Proc. ACM Manag. Data}} \bibinfo{volume}{3}, \bibinfo{number}{3}, Article \bibinfo{articleno}{196} (\bibinfo{date}{June} \bibinfo{year}{2025}), \bibinfo{numpages}{26}~pages.
\newblock
\urldef\tempurl%
\url{https://doi.org/10.1145/3725333}
\showDOI{\tempurl}


\bibitem[\protect\citeauthoryear{Lampropoulos, Zardbani, Mamoulis, and Karras}{Lampropoulos et~al\mbox{.}}{2023}]%
        {lampropoulos2023adaptive}
\bibfield{author}{\bibinfo{person}{Konstantinos Lampropoulos}, \bibinfo{person}{Fatemeh Zardbani}, \bibinfo{person}{Nikos Mamoulis}, {and} \bibinfo{person}{Panagiotis Karras}.} \bibinfo{year}{2023}\natexlab{}.
\newblock \showarticletitle{Adaptive indexing in high-dimensional metric spaces}.
\newblock \bibinfo{journal}{\emph{Proceedings of the VLDB Endowment}} \bibinfo{volume}{16}, \bibinfo{number}{10} (\bibinfo{year}{2023}), \bibinfo{pages}{2525--2537}.
\newblock


\bibitem[\protect\citeauthoryear{{LanceDB Developers}}{{LanceDB Developers}}{2024}]%
        {lance_format}
\bibfield{author}{\bibinfo{person}{{LanceDB Developers}}.} \bibinfo{year}{2024}\natexlab{}.
\newblock \bibinfo{title}{{Lance: Modern Columnar Data Format for ML}}.
\newblock
\newblock
\urldef\tempurl%
\url{https://github.com/lancedb/lance}
\showURL{%
\tempurl}
\newblock
\shownote{Accessed: February 16, 2025.}


\bibitem[\protect\citeauthoryear{Lewis, Perez, Piktus, Petroni, Karpukhin, Goyal, K{\"u}ttler, Lewis, Yih, Rockt{\"a}schel, et~al\mbox{.}}{Lewis et~al\mbox{.}}{2020}]%
        {lewis2020retrieval}
\bibfield{author}{\bibinfo{person}{Patrick Lewis}, \bibinfo{person}{Ethan Perez}, \bibinfo{person}{Aleksandra Piktus}, \bibinfo{person}{Fabio Petroni}, \bibinfo{person}{Vladimir Karpukhin}, \bibinfo{person}{Naman Goyal}, \bibinfo{person}{Heinrich K{\"u}ttler}, \bibinfo{person}{Mike Lewis}, \bibinfo{person}{Wen-tau Yih}, \bibinfo{person}{Tim Rockt{\"a}schel}, {et~al\mbox{.}}} \bibinfo{year}{2020}\natexlab{}.
\newblock \showarticletitle{Retrieval-augmented generation for knowledge-intensive nlp tasks}.
\newblock \bibinfo{journal}{\emph{Advances in Neural Information Processing Systems}}  \bibinfo{volume}{33} (\bibinfo{year}{2020}), \bibinfo{pages}{9459--9474}.
\newblock


\bibitem[\protect\citeauthoryear{Li, Chan, Yiu, and Mamoulis}{Li et~al\mbox{.}}{2017}]%
        {li2017fexipro}
\bibfield{author}{\bibinfo{person}{Hui Li}, \bibinfo{person}{Tsz~Nam Chan}, \bibinfo{person}{Man~Lung Yiu}, {and} \bibinfo{person}{Nikos Mamoulis}.} \bibinfo{year}{2017}\natexlab{}.
\newblock \showarticletitle{FEXIPRO: fast and exact inner product retrieval in recommender systems}. In \bibinfo{booktitle}{\emph{Proceedings of the 2017 ACM International Conference on Management of Data}}. \bibinfo{pages}{835--850}.
\newblock


\bibitem[\protect\citeauthoryear{Lin, Keogh, Lonardi, and Chiu}{Lin et~al\mbox{.}}{2003}]%
        {lin2003symbolic}
\bibfield{author}{\bibinfo{person}{Jessica Lin}, \bibinfo{person}{Eamonn Keogh}, \bibinfo{person}{Stefano Lonardi}, {and} \bibinfo{person}{Bill Chiu}.} \bibinfo{year}{2003}\natexlab{}.
\newblock \showarticletitle{A symbolic representation of time series, with implications for streaming algorithms}. In \bibinfo{booktitle}{\emph{Proceedings of the 8th ACM SIGMOD workshop on Research issues in data mining and knowledge discovery}}. \bibinfo{pages}{2--11}.
\newblock


\bibitem[\protect\citeauthoryear{Liu, Russo, Cafarella, Cao, Chen, Chen, Franklin, Kraska, Madden, Shahout, et~al\mbox{.}}{Liu et~al\mbox{.}}{2025}]%
        {liupalimpzest}
\bibfield{author}{\bibinfo{person}{Chunwei Liu}, \bibinfo{person}{Matthew Russo}, \bibinfo{person}{Michael Cafarella}, \bibinfo{person}{Lei Cao}, \bibinfo{person}{Peter~Baile Chen}, \bibinfo{person}{Zui Chen}, \bibinfo{person}{Michael Franklin}, \bibinfo{person}{Tim Kraska}, \bibinfo{person}{Samuel Madden}, \bibinfo{person}{Rana Shahout}, {et~al\mbox{.}}} \bibinfo{year}{2025}\natexlab{}.
\newblock \showarticletitle{Palimpzest: Optimizing AI-Powered Analytics with Declarative Query Processing}.
\newblock \bibinfo{journal}{\emph{Conference on Innovative Data Systems Research (CIDR)}} (\bibinfo{year}{2025}).
\newblock


\bibitem[\protect\citeauthoryear{Liu, Moore, Gray, and Yang}{Liu et~al\mbox{.}}{2004}]%
        {liu2004investigation}
\bibfield{author}{\bibinfo{person}{Ting Liu}, \bibinfo{person}{Andrew~W. Moore}, \bibinfo{person}{Alexander Gray}, {and} \bibinfo{person}{Ke Yang}.} \bibinfo{year}{2004}\natexlab{}.
\newblock \showarticletitle{An investigation of practical approximate nearest neighbor algorithms}. In \bibinfo{booktitle}{\emph{Proceedings of the 18th International Conference on Neural Information Processing Systems}} (Vancouver, British Columbia, Canada) \emph{(\bibinfo{series}{NIPS'04})}. \bibinfo{publisher}{MIT Press}, \bibinfo{address}{Cambridge, MA, USA}, \bibinfo{pages}{825–832}.
\newblock


\bibitem[\protect\citeauthoryear{Malkov, Ponomarenko, Logvinov, and Krylov}{Malkov et~al\mbox{.}}{2014}]%
        {malkov2014approximate}
\bibfield{author}{\bibinfo{person}{Yury Malkov}, \bibinfo{person}{Alexander Ponomarenko}, \bibinfo{person}{Andrey Logvinov}, {and} \bibinfo{person}{Vladimir Krylov}.} \bibinfo{year}{2014}\natexlab{}.
\newblock \showarticletitle{Approximate nearest neighbor algorithm based on navigable small world graphs}.
\newblock \bibinfo{journal}{\emph{Information Systems}}  \bibinfo{volume}{45} (\bibinfo{year}{2014}), \bibinfo{pages}{61--68}.
\newblock


\bibitem[\protect\citeauthoryear{Malkov and Yashunin}{Malkov and Yashunin}{2018}]%
        {malkov2018efficient}
\bibfield{author}{\bibinfo{person}{Yu~A Malkov} {and} \bibinfo{person}{Dmitry~A Yashunin}.} \bibinfo{year}{2018}\natexlab{}.
\newblock \showarticletitle{Efficient and robust approximate nearest neighbor search using hierarchical navigable small world graphs}.
\newblock \bibinfo{journal}{\emph{IEEE transactions on pattern analysis and machine intelligence}} \bibinfo{volume}{42}, \bibinfo{number}{4} (\bibinfo{year}{2018}), \bibinfo{pages}{824--836}.
\newblock


\bibitem[\protect\citeauthoryear{{Merrill Lynch}}{{Merrill Lynch}}{1998}]%
        {MerrillLynch1998}
\bibfield{author}{\bibinfo{person}{{Merrill Lynch}}.} \bibinfo{year}{1998}\natexlab{}.
\newblock \bibinfo{booktitle}{\emph{Enterprise Information Portals: Industry Overview}}.
\newblock \bibinfo{type}{{T}echnical {R}eport}. \bibinfo{institution}{Merrill Lynch}.
\newblock
\urldef\tempurl%
\url{https://web.archive.org/web/20110724175845/http://ikt.hia.no/perep/eip_ind.pdf}
\showURL{%
\tempurl}
\newblock
\shownote{Accessed: 2025-02-16.}


\bibitem[\protect\citeauthoryear{Miller}{Miller}{2018}]%
        {miller2018open}
\bibfield{author}{\bibinfo{person}{Ren\'{e}e~J. Miller}.} \bibinfo{year}{2018}\natexlab{}.
\newblock \showarticletitle{Open data integration}.
\newblock \bibinfo{journal}{\emph{Proc. VLDB Endow.}} \bibinfo{volume}{11}, \bibinfo{number}{12} (\bibinfo{date}{Aug.} \bibinfo{year}{2018}), \bibinfo{pages}{2130–2139}.
\newblock
\showISSN{2150-8097}
\urldef\tempurl%
\url{https://doi.org/10.14778/3229863.3240491}
\showDOI{\tempurl}


\bibitem[\protect\citeauthoryear{Mirylenka, Christophides, Palpanas, Pefkianakis, and May}{Mirylenka et~al\mbox{.}}{2016}]%
        {mirylenka2016characterizing}
\bibfield{author}{\bibinfo{person}{Katsiaryna Mirylenka}, \bibinfo{person}{Vassilis Christophides}, \bibinfo{person}{Themis Palpanas}, \bibinfo{person}{Ioannis Pefkianakis}, {and} \bibinfo{person}{Martin May}.} \bibinfo{year}{2016}\natexlab{}.
\newblock \showarticletitle{Characterizing home device usage from wireless traffic time series}. In \bibinfo{booktitle}{\emph{19th International Conference on Extending Database Technology (EDBT)}}.
\newblock


\bibitem[\protect\citeauthoryear{Mitra, Craswell, et~al\mbox{.}}{Mitra et~al\mbox{.}}{2018}]%
        {mitra2018introduction}
\bibfield{author}{\bibinfo{person}{Bhaskar Mitra}, \bibinfo{person}{Nick Craswell}, {et~al\mbox{.}}} \bibinfo{year}{2018}\natexlab{}.
\newblock \showarticletitle{An introduction to neural information retrieval}.
\newblock \bibinfo{journal}{\emph{Foundations and Trends{\textregistered} in Information Retrieval}} \bibinfo{volume}{13}, \bibinfo{number}{1} (\bibinfo{year}{2018}), \bibinfo{pages}{1--126}.
\newblock


\bibitem[\protect\citeauthoryear{Mohoney, Pacaci, Chowdhury, Minhas, Pound, Renggli, Reyhani, Ilyas, Rekatsinas, and Venkataraman}{Mohoney et~al\mbox{.}}{2024}]%
        {mohoney2024incremental}
\bibfield{author}{\bibinfo{person}{Jason Mohoney}, \bibinfo{person}{Anil Pacaci}, \bibinfo{person}{Shihabur~Rahman Chowdhury}, \bibinfo{person}{Umar~Farooq Minhas}, \bibinfo{person}{Jeffery Pound}, \bibinfo{person}{Cedric Renggli}, \bibinfo{person}{Nima Reyhani}, \bibinfo{person}{Ihab~F Ilyas}, \bibinfo{person}{Theodoros Rekatsinas}, {and} \bibinfo{person}{Shivaram Venkataraman}.} \bibinfo{year}{2024}\natexlab{}.
\newblock \showarticletitle{Incremental IVF Index Maintenance for Streaming Vector Search}.
\newblock \bibinfo{journal}{\emph{arXiv preprint arXiv:2411.00970}} (\bibinfo{year}{2024}).
\newblock


\bibitem[\protect\citeauthoryear{Mohoney, Pacaci, Chowdhury, Mousavi, Ilyas, Minhas, Pound, and Rekatsinas}{Mohoney et~al\mbox{.}}{2023}]%
        {mohoney2023high}
\bibfield{author}{\bibinfo{person}{Jason Mohoney}, \bibinfo{person}{Anil Pacaci}, \bibinfo{person}{Shihabur~Rahman Chowdhury}, \bibinfo{person}{Ali Mousavi}, \bibinfo{person}{Ihab~F Ilyas}, \bibinfo{person}{Umar~Farooq Minhas}, \bibinfo{person}{Jeffrey Pound}, {and} \bibinfo{person}{Theodoros Rekatsinas}.} \bibinfo{year}{2023}\natexlab{}.
\newblock \showarticletitle{High-throughput vector similarity search in knowledge graphs}.
\newblock \bibinfo{journal}{\emph{Proceedings of the ACM on Management of Data}} \bibinfo{volume}{1}, \bibinfo{number}{2} (\bibinfo{year}{2023}), \bibinfo{pages}{1--25}.
\newblock


\bibitem[\protect\citeauthoryear{Mohoney, Sarda, Tang, Chowdhury, Pacaci, Ilyas, Rekatsinas, and Venkataraman}{Mohoney et~al\mbox{.}}{2025}]%
        {mohoney2025quake}
\bibfield{author}{\bibinfo{person}{Jason Mohoney}, \bibinfo{person}{Devesh Sarda}, \bibinfo{person}{Mengze Tang}, \bibinfo{person}{Shihabur~Rahman Chowdhury}, \bibinfo{person}{Anil Pacaci}, \bibinfo{person}{Ihab~F Ilyas}, \bibinfo{person}{Theodoros Rekatsinas}, {and} \bibinfo{person}{Shivaram Venkataraman}.} \bibinfo{year}{2025}\natexlab{}.
\newblock \showarticletitle{Quake: Adaptive Indexing for Vector Search}.
\newblock \bibinfo{journal}{\emph{arXiv preprint arXiv:2506.03437}} (\bibinfo{year}{2025}).
\newblock


\bibitem[\protect\citeauthoryear{Muja and Lowe}{Muja and Lowe}{2014}]%
        {muja2014scalable}
\bibfield{author}{\bibinfo{person}{Marius Muja} {and} \bibinfo{person}{David~G Lowe}.} \bibinfo{year}{2014}\natexlab{}.
\newblock \showarticletitle{Scalable nearest neighbor algorithms for high dimensional data}.
\newblock \bibinfo{journal}{\emph{IEEE transactions on pattern analysis and machine intelligence}} \bibinfo{volume}{36}, \bibinfo{number}{11} (\bibinfo{year}{2014}), \bibinfo{pages}{2227--2240}.
\newblock


\bibitem[\protect\citeauthoryear{Nerone, Holanda, De~Almeida, and Manegold}{Nerone et~al\mbox{.}}{2021}]%
        {nerone2021multidimensional}
\bibfield{author}{\bibinfo{person}{Matheus~Agio Nerone}, \bibinfo{person}{Pedro Holanda}, \bibinfo{person}{Eduardo~C De~Almeida}, {and} \bibinfo{person}{Stefan Manegold}.} \bibinfo{year}{2021}\natexlab{}.
\newblock \showarticletitle{Multidimensional adaptive \& progressive indexes}. In \bibinfo{booktitle}{\emph{2021 IEEE 37th International Conference on Data Engineering (ICDE)}}. IEEE, \bibinfo{pages}{624--635}.
\newblock


\bibitem[\protect\citeauthoryear{OpenAI}{OpenAI}{2025}]%
        {openai_embedding2023}
\bibfield{author}{\bibinfo{person}{OpenAI}.} \bibinfo{year}{2025}\natexlab{}.
\newblock \bibinfo{title}{OpenAI Embeddings}.
\newblock \bibinfo{howpublished}{Online}.
\newblock
\urldef\tempurl%
\url{https://platform.openai.com/docs/guides/embeddings}
\showURL{%
\tempurl}
\newblock
\shownote{Accessed: 2025-07-01.}


\bibitem[\protect\citeauthoryear{Pan, Wang, and Li}{Pan et~al\mbox{.}}{2024}]%
        {pan2024survey}
\bibfield{author}{\bibinfo{person}{James~Jie Pan}, \bibinfo{person}{Jianguo Wang}, {and} \bibinfo{person}{Guoliang Li}.} \bibinfo{year}{2024}\natexlab{}.
\newblock \showarticletitle{Survey of vector database management systems}.
\newblock \bibinfo{journal}{\emph{The VLDB Journal}} \bibinfo{volume}{33}, \bibinfo{number}{5} (\bibinfo{year}{2024}), \bibinfo{pages}{1591--1615}.
\newblock


\bibitem[\protect\citeauthoryear{Pavlovic, Sidlauskas, Heinis, and Ailamaki}{Pavlovic et~al\mbox{.}}{2018}]%
        {pavlovic2018quasii}
\bibfield{author}{\bibinfo{person}{Mirjana Pavlovic}, \bibinfo{person}{Darius Sidlauskas}, \bibinfo{person}{Thomas Heinis}, {and} \bibinfo{person}{Anastasia Ailamaki}.} \bibinfo{year}{2018}\natexlab{}.
\newblock \bibinfo{title}{QUASII: QUery-Aware Spatial Incremental Index}.
\newblock
\newblock
\urldef\tempurl%
\url{https://doi.org/10.5441/002/edbt.2018.29}
\showDOI{\tempurl}


\bibitem[\protect\citeauthoryear{Pennington, Socher, and Manning}{Pennington et~al\mbox{.}}{2014}]%
        {pennington2014glove}
\bibfield{author}{\bibinfo{person}{Jeffrey Pennington}, \bibinfo{person}{Richard Socher}, {and} \bibinfo{person}{Christopher~D Manning}.} \bibinfo{year}{2014}\natexlab{}.
\newblock \showarticletitle{Glove: Global vectors for word representation}. In \bibinfo{booktitle}{\emph{Proceedings of the 2014 conference on empirical methods in natural language processing (EMNLP)}}. \bibinfo{pages}{1532--1543}.
\newblock


\bibitem[\protect\citeauthoryear{pgvector contributors}{pgvector contributors}{2025}]%
        {pgvector}
\bibfield{author}{\bibinfo{person}{pgvector contributors}.} \bibinfo{year}{2025}\natexlab{}.
\newblock \bibinfo{title}{pgvector: Open-Source Vector Similarity Search for PostgreSQL}.
\newblock \bibinfo{howpublished}{\url{https://github.com/pgvector/pgvector}}.
\newblock
\newblock
\shownote{Accessed: 2025-03-01.}


\bibitem[\protect\citeauthoryear{Pinecone}{Pinecone}{2025}]%
        {pinecone}
\bibfield{author}{\bibinfo{person}{Pinecone}.} \bibinfo{year}{2025}\natexlab{}.
\newblock \bibinfo{title}{Pinecone - Vector Database}.
\newblock
\newblock
\urldef\tempurl%
\url{http://pinecone.io}
\showURL{%
\tempurl}
\newblock
\shownote{Accessed: 2025-03-01.}


\bibitem[\protect\citeauthoryear{Raza, Camerra, Murphy, Palpanas, and Picco}{Raza et~al\mbox{.}}{2015}]%
        {raza2015practical}
\bibfield{author}{\bibinfo{person}{Usman Raza}, \bibinfo{person}{Alessandro Camerra}, \bibinfo{person}{Amy~L Murphy}, \bibinfo{person}{Themis Palpanas}, {and} \bibinfo{person}{Gian~Pietro Picco}.} \bibinfo{year}{2015}\natexlab{}.
\newblock \showarticletitle{Practical data prediction for real-world wireless sensor networks}.
\newblock \bibinfo{journal}{\emph{IEEE Transactions on Knowledge and Data Engineering}} \bibinfo{volume}{27}, \bibinfo{number}{8} (\bibinfo{year}{2015}), \bibinfo{pages}{2231--2244}.
\newblock


\bibitem[\protect\citeauthoryear{Research}{Research}{2023}]%
        {soar_scann_blogpost}
\bibfield{author}{\bibinfo{person}{Google Research}.} \bibinfo{year}{2023}\natexlab{}.
\newblock \bibinfo{title}{SOAR: New Algorithms for Even Faster Vector Search with ScaNN}.
\newblock \bibinfo{howpublished}{Google Research Blog}.
\newblock
\urldef\tempurl%
\url{https://research.google/blog/soar-new-algorithms-for-even-faster-vector-search-with-scann/}
\showURL{%
\tempurl}
\newblock
\shownote{Accessed: February 22, 2025.}


\bibitem[\protect\citeauthoryear{Salton, Wong, and Yang}{Salton et~al\mbox{.}}{1975}]%
        {salton1975vector}
\bibfield{author}{\bibinfo{person}{Gerard Salton}, \bibinfo{person}{Anita Wong}, {and} \bibinfo{person}{Chung-Shu Yang}.} \bibinfo{year}{1975}\natexlab{}.
\newblock \showarticletitle{A vector space model for automatic indexing}.
\newblock \bibinfo{journal}{\emph{Commun. ACM}} \bibinfo{volume}{18}, \bibinfo{number}{11} (\bibinfo{year}{1975}), \bibinfo{pages}{613--620}.
\newblock


\bibitem[\protect\citeauthoryear{Samet}{Samet}{1984}]%
        {samet1984quadtree}
\bibfield{author}{\bibinfo{person}{Hanan Samet}.} \bibinfo{year}{1984}\natexlab{}.
\newblock \showarticletitle{The quadtree and related hierarchical data structures}.
\newblock \bibinfo{journal}{\emph{ACM Computing Surveys (CSUR)}} \bibinfo{volume}{16}, \bibinfo{number}{2} (\bibinfo{year}{1984}), \bibinfo{pages}{187--260}.
\newblock


\bibitem[\protect\citeauthoryear{Schuhknecht, Jindal, and Dittrich}{Schuhknecht et~al\mbox{.}}{2016}]%
        {schuhknecht2016experimental}
\bibfield{author}{\bibinfo{person}{Felix~Martin Schuhknecht}, \bibinfo{person}{Alekh Jindal}, {and} \bibinfo{person}{Jens Dittrich}.} \bibinfo{year}{2016}\natexlab{}.
\newblock \showarticletitle{An experimental evaluation and analysis of database cracking}.
\newblock \bibinfo{journal}{\emph{The VLDB Journal}}  \bibinfo{volume}{25} (\bibinfo{year}{2016}), \bibinfo{pages}{27--52}.
\newblock


\bibitem[\protect\citeauthoryear{Shasha}{Shasha}{1999}]%
        {shasha1999tuning}
\bibfield{author}{\bibinfo{person}{Dennis Shasha}.} \bibinfo{year}{1999}\natexlab{}.
\newblock \showarticletitle{Tuning time series queries in finance: Case studies and recommendations}.
\newblock \bibinfo{journal}{\emph{IEEE Data Eng. Bull.}} \bibinfo{volume}{22}, \bibinfo{number}{2} (\bibinfo{year}{1999}), \bibinfo{pages}{40--46}.
\newblock


\bibitem[\protect\citeauthoryear{Shieh and Keogh}{Shieh and Keogh}{2008}]%
        {shieh2008sax}
\bibfield{author}{\bibinfo{person}{Jin Shieh} {and} \bibinfo{person}{Eamonn Keogh}.} \bibinfo{year}{2008}\natexlab{}.
\newblock \showarticletitle{i SAX: indexing and mining terabyte sized time series}. In \bibinfo{booktitle}{\emph{Proceedings of the 14th ACM SIGKDD international conference on Knowledge discovery and data mining}}. \bibinfo{pages}{623--631}.
\newblock


\bibitem[\protect\citeauthoryear{Simhadri, Williams, Aum{\"u}ller, Douze, Babenko, Baranchuk, Chen, Hosseini, Krishnaswamny, Srinivasa, et~al\mbox{.}}{Simhadri et~al\mbox{.}}{2022}]%
        {simhadri2022results}
\bibfield{author}{\bibinfo{person}{Harsha~Vardhan Simhadri}, \bibinfo{person}{George Williams}, \bibinfo{person}{Martin Aum{\"u}ller}, \bibinfo{person}{Matthijs Douze}, \bibinfo{person}{Artem Babenko}, \bibinfo{person}{Dmitry Baranchuk}, \bibinfo{person}{Qi Chen}, \bibinfo{person}{Lucas Hosseini}, \bibinfo{person}{Ravishankar Krishnaswamny}, \bibinfo{person}{Gopal Srinivasa}, {et~al\mbox{.}}} \bibinfo{year}{2022}\natexlab{}.
\newblock \showarticletitle{Results of the NeurIPS’21 challenge on billion-scale approximate nearest neighbor search}. In \bibinfo{booktitle}{\emph{NeurIPS 2021 Competitions and Demonstrations Track}}. PMLR, \bibinfo{pages}{177--189}.
\newblock


\bibitem[\protect\citeauthoryear{Sivic and Zisserman}{Sivic and Zisserman}{2003}]%
        {sivic2003video}
\bibfield{author}{\bibinfo{person}{Sivic} {and} \bibinfo{person}{Zisserman}.} \bibinfo{year}{2003}\natexlab{}.
\newblock \showarticletitle{Video Google: A text retrieval approach to object matching in videos}. In \bibinfo{booktitle}{\emph{Proceedings ninth IEEE international conference on computer vision}}. IEEE, \bibinfo{pages}{1470--1477}.
\newblock


\bibitem[\protect\citeauthoryear{Sun, Simcha, Dopson, Guo, and Kumar}{Sun et~al\mbox{.}}{2024}]%
        {sun2024soar}
\bibfield{author}{\bibinfo{person}{Philip Sun}, \bibinfo{person}{David Simcha}, \bibinfo{person}{Dave Dopson}, \bibinfo{person}{Ruiqi Guo}, {and} \bibinfo{person}{Sanjiv Kumar}.} \bibinfo{year}{2024}\natexlab{}.
\newblock \showarticletitle{SOAR: improved indexing for approximate nearest neighbor search}.
\newblock \bibinfo{journal}{\emph{Advances in Neural Information Processing Systems}}  \bibinfo{volume}{36} (\bibinfo{year}{2024}).
\newblock


\bibitem[\protect\citeauthoryear{Tang, Yang, Zhang, Luo, Fan, Cao, Madden, and Halevy}{Tang et~al\mbox{.}}{2024}]%
        {tangsymphony}
\bibfield{author}{\bibinfo{person}{Nan Tang}, \bibinfo{person}{Chenyu Yang}, \bibinfo{person}{Zhengxuan Zhang}, \bibinfo{person}{Yuyu Luo}, \bibinfo{person}{Ju Fan}, \bibinfo{person}{Lei Cao}, \bibinfo{person}{Sam Madden}, {and} \bibinfo{person}{Alon Halevy}.} \bibinfo{year}{2024}\natexlab{}.
\newblock \showarticletitle{Symphony: Towards trustworthy question answering and verification using RAG over multimodal data lakes}.
\newblock \bibinfo{journal}{\emph{IEEE Data Eng. Bull}} \bibinfo{volume}{48}, \bibinfo{number}{4} (\bibinfo{year}{2024}), \bibinfo{pages}{135--146}.
\newblock


\bibitem[\protect\citeauthoryear{Van~Gysel, de~Rijke, and Kanoulas}{Van~Gysel et~al\mbox{.}}{2016}]%
        {van2016learning}
\bibfield{author}{\bibinfo{person}{Christophe Van~Gysel}, \bibinfo{person}{Maarten de Rijke}, {and} \bibinfo{person}{Evangelos Kanoulas}.} \bibinfo{year}{2016}\natexlab{}.
\newblock \showarticletitle{Learning latent vector spaces for product search}. In \bibinfo{booktitle}{\emph{Proceedings of the 25th ACM international on conference on information and knowledge management}}. \bibinfo{pages}{165--174}.
\newblock


\bibitem[\protect\citeauthoryear{{Vespa.ai}}{{Vespa.ai}}{2024}]%
        {vespa_hybrid_search}
\bibfield{author}{\bibinfo{person}{{Vespa.ai}}.} \bibinfo{year}{2024}\natexlab{}.
\newblock \bibinfo{title}{Hybrid Search}.
\newblock
\newblock
\urldef\tempurl%
\url{https://docs.vespa.ai/en/tutorials/hybrid-search.html}
\showURL{%
\tempurl}
\newblock
\shownote{Accessed: 2024-03-01.}


\bibitem[\protect\citeauthoryear{Wang, Wang, Zeng, Tu, Gan, and Li}{Wang et~al\mbox{.}}{2012}]%
        {wang2012scalable}
\bibfield{author}{\bibinfo{person}{Jing Wang}, \bibinfo{person}{Jingdong Wang}, \bibinfo{person}{Gang Zeng}, \bibinfo{person}{Zhuowen Tu}, \bibinfo{person}{Rui Gan}, {and} \bibinfo{person}{Shipeng Li}.} \bibinfo{year}{2012}\natexlab{}.
\newblock \showarticletitle{Scalable k-nn graph construction for visual descriptors}. In \bibinfo{booktitle}{\emph{2012 IEEE Conference on Computer Vision and Pattern Recognition}}. IEEE, \bibinfo{pages}{1106--1113}.
\newblock


\bibitem[\protect\citeauthoryear{Wang, Wang, Jia, Li, Zeng, Zha, and Hua}{Wang et~al\mbox{.}}{2013}]%
        {wang2013trinary}
\bibfield{author}{\bibinfo{person}{Jingdong Wang}, \bibinfo{person}{Naiyan Wang}, \bibinfo{person}{You Jia}, \bibinfo{person}{Jian Li}, \bibinfo{person}{Gang Zeng}, \bibinfo{person}{Hongbin Zha}, {and} \bibinfo{person}{Xian-Sheng Hua}.} \bibinfo{year}{2013}\natexlab{}.
\newblock \showarticletitle{Trinary-projection trees for approximate nearest neighbor search}.
\newblock \bibinfo{journal}{\emph{IEEE transactions on pattern analysis and machine intelligence}} \bibinfo{volume}{36}, \bibinfo{number}{2} (\bibinfo{year}{2013}), \bibinfo{pages}{388--403}.
\newblock


\bibitem[\protect\citeauthoryear{Wang, Yi, Guo, Jin, Xu, Li, Wang, Guo, Li, Xu, et~al\mbox{.}}{Wang et~al\mbox{.}}{2021}]%
        {wang2021milvus}
\bibfield{author}{\bibinfo{person}{Jianguo Wang}, \bibinfo{person}{Xiaomeng Yi}, \bibinfo{person}{Rentong Guo}, \bibinfo{person}{Hai Jin}, \bibinfo{person}{Peng Xu}, \bibinfo{person}{Shengjun Li}, \bibinfo{person}{Xiangyu Wang}, \bibinfo{person}{Xiangzhou Guo}, \bibinfo{person}{Chengming Li}, \bibinfo{person}{Xiaohai Xu}, {et~al\mbox{.}}} \bibinfo{year}{2021}\natexlab{}.
\newblock \showarticletitle{Milvus: A purpose-built vector data management system}. In \bibinfo{booktitle}{\emph{Proceedings of the 2021 International Conference on Management of Data}}. \bibinfo{pages}{2614--2627}.
\newblock


\bibitem[\protect\citeauthoryear{Wang, Zhang, Sebe, Shen, et~al\mbox{.}}{Wang et~al\mbox{.}}{2017}]%
        {wang2017survey}
\bibfield{author}{\bibinfo{person}{Jingdong Wang}, \bibinfo{person}{Ting Zhang}, \bibinfo{person}{Nicu Sebe}, \bibinfo{person}{Heng~Tao Shen}, {et~al\mbox{.}}} \bibinfo{year}{2017}\natexlab{}.
\newblock \showarticletitle{A survey on learning to hash}.
\newblock \bibinfo{journal}{\emph{IEEE transactions on pattern analysis and machine intelligence}} \bibinfo{volume}{40}, \bibinfo{number}{4} (\bibinfo{year}{2017}), \bibinfo{pages}{769--790}.
\newblock


\bibitem[\protect\citeauthoryear{Wang, Wang, Wang, Palpanas, and Wang}{Wang et~al\mbox{.}}{2023}]%
        {wang2023dumpy}
\bibfield{author}{\bibinfo{person}{Zeyu Wang}, \bibinfo{person}{Qitong Wang}, \bibinfo{person}{Peng Wang}, \bibinfo{person}{Themis Palpanas}, {and} \bibinfo{person}{Wei Wang}.} \bibinfo{year}{2023}\natexlab{}.
\newblock \showarticletitle{Dumpy: A compact and adaptive index for large data series collections}.
\newblock \bibinfo{journal}{\emph{Proceedings of the ACM on Management of Data}} \bibinfo{volume}{1}, \bibinfo{number}{1} (\bibinfo{year}{2023}), \bibinfo{pages}{1--27}.
\newblock


\bibitem[\protect\citeauthoryear{Wang, Wang, Wang, Palpanas, and Wang}{Wang et~al\mbox{.}}{2024}]%
        {wang2024dumpyos}
\bibfield{author}{\bibinfo{person}{Zeyu Wang}, \bibinfo{person}{Qitong Wang}, \bibinfo{person}{Peng Wang}, \bibinfo{person}{Themis Palpanas}, {and} \bibinfo{person}{Wei Wang}.} \bibinfo{year}{2024}\natexlab{}.
\newblock \showarticletitle{DumpyOS: A data-adaptive multi-ary index for scalable data series similarity search}.
\newblock \bibinfo{journal}{\emph{The VLDB Journal}} \bibinfo{volume}{33}, \bibinfo{number}{6} (\bibinfo{year}{2024}), \bibinfo{pages}{1887--1911}.
\newblock


\bibitem[\protect\citeauthoryear{Weaviate}{Weaviate}{2025}]%
        {weaviate}
\bibfield{author}{\bibinfo{person}{Weaviate}.} \bibinfo{year}{2025}\natexlab{}.
\newblock \bibinfo{title}{Weaviate - Vector Database}.
\newblock
\newblock
\urldef\tempurl%
\url{http://weaviate.io}
\showURL{%
\tempurl}
\newblock
\shownote{Accessed: 2025-03-01.}


\bibitem[\protect\citeauthoryear{Weber, Schek, and Blott}{Weber et~al\mbox{.}}{1998}]%
        {DBLP:conf/vldb/WeberSB98}
\bibfield{author}{\bibinfo{person}{Roger Weber}, \bibinfo{person}{Hans{-}J{\"{o}}rg Schek}, {and} \bibinfo{person}{Stephen Blott}.} \bibinfo{year}{1998}\natexlab{}.
\newblock \showarticletitle{A Quantitative Analysis and Performance Study for Similarity-Search Methods in High-Dimensional Spaces}. In \bibinfo{booktitle}{\emph{VLDB'98, Proceedings of 24rd International Conference on Very Large Data Bases, August 24-27, 1998, New York City, New York, {USA}}}.
\newblock


\bibitem[\protect\citeauthoryear{Xu, Yang, Zhang, Pan, Chen, Shen, Zhou, and Du}{Xu et~al\mbox{.}}{2025}]%
        {Xu25-tribase}
\bibfield{author}{\bibinfo{person}{Qian Xu}, \bibinfo{person}{Juan Yang}, \bibinfo{person}{Feng Zhang}, \bibinfo{person}{Junda Pan}, \bibinfo{person}{Kang Chen}, \bibinfo{person}{Youren Shen}, \bibinfo{person}{Amelie~Chi Zhou}, {and} \bibinfo{person}{Xiaoyong Du}.} \bibinfo{year}{2025}\natexlab{}.
\newblock \showarticletitle{Tribase: A Vector Data Query Engine for Reliable and Lossless Pruning Compression using Triangle Inequalities}.
\newblock \bibinfo{journal}{\emph{Proc. ACM Manag. Data}} \bibinfo{volume}{3}, \bibinfo{number}{1}, Article \bibinfo{articleno}{82} (\bibinfo{date}{Feb.} \bibinfo{year}{2025}), \bibinfo{numpages}{28}~pages.
\newblock
\urldef\tempurl%
\url{https://doi.org/10.1145/3709743}
\showDOI{\tempurl}


\bibitem[\protect\citeauthoryear{Xu, Liang, Li, Xu, Chen, Zhang, Li, Yang, Yang, Yang, et~al\mbox{.}}{Xu et~al\mbox{.}}{2023}]%
        {xu2023spfresh}
\bibfield{author}{\bibinfo{person}{Yuming Xu}, \bibinfo{person}{Hengyu Liang}, \bibinfo{person}{Jin Li}, \bibinfo{person}{Shuotao Xu}, \bibinfo{person}{Qi Chen}, \bibinfo{person}{Qianxi Zhang}, \bibinfo{person}{Cheng Li}, \bibinfo{person}{Ziyue Yang}, \bibinfo{person}{Fan Yang}, \bibinfo{person}{Yuqing Yang}, {et~al\mbox{.}}} \bibinfo{year}{2023}\natexlab{}.
\newblock \showarticletitle{SPFresh: Incremental In-Place Update for Billion-Scale Vector Search}. In \bibinfo{booktitle}{\emph{Proceedings of the 29th Symposium on Operating Systems Principles}}. \bibinfo{pages}{545--561}.
\newblock


\bibitem[\protect\citeauthoryear{Zardbani, Mamoulis, Idreos, and Karras}{Zardbani et~al\mbox{.}}{2023}]%
        {zardbani2023adaptive}
\bibfield{author}{\bibinfo{person}{Fatemeh Zardbani}, \bibinfo{person}{Nikos Mamoulis}, \bibinfo{person}{Stratos Idreos}, {and} \bibinfo{person}{Panagiotis Karras}.} \bibinfo{year}{2023}\natexlab{}.
\newblock \showarticletitle{Adaptive indexing of objects with spatial extent}.
\newblock \bibinfo{journal}{\emph{Proceedings of the VLDB Endowment}} \bibinfo{volume}{16}, \bibinfo{number}{9} (\bibinfo{year}{2023}), \bibinfo{pages}{2248--2260}.
\newblock


\bibitem[\protect\citeauthoryear{Zhang, Shin, R{\'e}, Cafarella, and Niu}{Zhang et~al\mbox{.}}{2016}]%
        {zhang2016extracting}
\bibfield{author}{\bibinfo{person}{Ce Zhang}, \bibinfo{person}{Jaeho Shin}, \bibinfo{person}{Christopher R{\'e}}, \bibinfo{person}{Michael Cafarella}, {and} \bibinfo{person}{Feng Niu}.} \bibinfo{year}{2016}\natexlab{}.
\newblock \showarticletitle{Extracting databases from dark data with deepdive}. In \bibinfo{booktitle}{\emph{Proceedings of the 2016 International Conference on Management of Data}}. \bibinfo{pages}{847--859}.
\newblock


\bibitem[\protect\citeauthoryear{Zhang, Xu, Chen, Sui, Xie, Cai, Chen, He, Yang, Yang, et~al\mbox{.}}{Zhang et~al\mbox{.}}{2023}]%
        {zhang2023vbase}
\bibfield{author}{\bibinfo{person}{Qianxi Zhang}, \bibinfo{person}{Shuotao Xu}, \bibinfo{person}{Qi Chen}, \bibinfo{person}{Guoxin Sui}, \bibinfo{person}{Jiadong Xie}, \bibinfo{person}{Zhizhen Cai}, \bibinfo{person}{Yaoqi Chen}, \bibinfo{person}{Yinxuan He}, \bibinfo{person}{Yuqing Yang}, \bibinfo{person}{Fan Yang}, {et~al\mbox{.}}} \bibinfo{year}{2023}\natexlab{}.
\newblock \showarticletitle{$\{$VBASE$\}$: Unifying Online Vector Similarity Search and Relational Queries via Relaxed Monotonicity}. In \bibinfo{booktitle}{\emph{17th USENIX Symposium on Operating Systems Design and Implementation (OSDI 23)}}. \bibinfo{pages}{377--395}.
\newblock


\bibitem[\protect\citeauthoryear{Zoumpatianos, Idreos, and Palpanas}{Zoumpatianos et~al\mbox{.}}{2014}]%
        {zoumpatianos2014indexing}
\bibfield{author}{\bibinfo{person}{Kostas Zoumpatianos}, \bibinfo{person}{Stratos Idreos}, {and} \bibinfo{person}{Themis Palpanas}.} \bibinfo{year}{2014}\natexlab{}.
\newblock \showarticletitle{Indexing for interactive exploration of big data series}. In \bibinfo{booktitle}{\emph{Proceedings of the 2014 ACM SIGMOD international conference on Management of data}}. \bibinfo{pages}{1555--1566}.
\newblock


\end{thebibliography}

\end{document}